\documentclass[aps,prd,twocolumn,nopacs,floatfix,amsmath,nofootinbib,superscriptaddress,amssymb,floatfix]{revtex4}
\usepackage{graphicx,color,dcolumn,booktabs,bm}
\usepackage{longtable,lscape}
\usepackage{pdfpages}
\usepackage{txfonts}
\usepackage{overpic}
\usepackage{amssymb}
\usepackage{makecell}
\usepackage{indentfirst}
\usepackage{feynmf}
\usepackage{slashed}
\usepackage{cases}
\usepackage{color}
\usepackage{multirow}
\usepackage{longtable,lscape}
\usepackage{threeparttable}
\usepackage{epstopdf}
\usepackage{makecell}
\def\lrpartial{\buildrel\leftrightarrow\over\partial}
\usepackage{ulem}
\usepackage{enumerate}
\makeatletter

\newcommand{\Rmnum}[1]{\expandafter\@slowromancap\romannumeral #1@}
\makeatother
\usepackage{threeparttable}
\usepackage{graphicx,color,dcolumn,booktabs,bm}
\usepackage[colorlinks,
            citecolor=blue,
            anchorcolor=red,
            menucolor=red,
            linkcolor=red,
            filecolor=red,
            runcolor=red,
            urlcolor=blue,
            frenchlinks=red]{hyperref}
\usepackage{epstopdf}
\usepackage{array}

\begin{document}
\title{Masses and radiative decay widths of the $D_{s0}^*(2317)$ and $D_{s1}^{\prime}(2460)$ and their bottom analogs}

\author{Zi-Le Zhang}
\affiliation{School of Physical Science and Technology, Lanzhou University, Lanzhou 730000, China}
\affiliation{Research Center for Hadron and CSR Physics, Lanzhou University and Institute of Modern Physics of CAS, Lanzhou 730000, China}

\author{Zhan-Wei Liu}\email{liuzhanwei@lzu.edu.cn}
\affiliation{School of Physical Science and Technology, Lanzhou University, Lanzhou 730000, China}
\affiliation{Research Center for Hadron and CSR Physics, Lanzhou University and Institute of Modern Physics of CAS, Lanzhou 730000, China}
\affiliation{Lanzhou Center for Theoretical Physics, Key Laboratory of Theoretical Physics of Gansu Province,Key Laboratory of Quantum Theory and Applications of MoE,and MoE Frontiers Science Center for Rare Isotopes, Lanzhou University, Lanzhou 730000, China}

\author{Si-Qiang Luo}
\affiliation{School of Physical Science and Technology, Lanzhou University, Lanzhou 730000, China}
\affiliation{Research Center for Hadron and CSR Physics, Lanzhou University and Institute of Modern Physics of CAS, Lanzhou 730000, China}
\affiliation{Lanzhou Center for Theoretical Physics, Key Laboratory of Theoretical Physics of Gansu Province,Key Laboratory of Quantum Theory and Applications of MoE,and MoE Frontiers Science Center for Rare Isotopes, Lanzhou University, Lanzhou 730000, China}

\author{Ping Chen}
\affiliation{School of Physical Science and Technology, Lanzhou University, Lanzhou 730000, China}
\affiliation{Research Center for Hadron and CSR Physics, Lanzhou University and Institute of Modern Physics of CAS, Lanzhou 730000, China}

\author{Zhi-Hui Guo}
\affiliation{Department of Physics and Hebei Key Laboratory of Photophysics Research and Application,Hebei Normal University, Shijiazhuang 050024, China}

\begin{abstract}
We study the mass spectra and radiative decays of $D_{s0}^*(2317)$ and $D_{s1}^{\prime}(2460)$ in an unquenched framework. In addition to coupled channel effects between the $c\bar{s}$ cores and $D^{(*)}K$ channels, $D^{(*)}K$-$D^{(*)}K$ self interactions are also considered in this work and we succeed to reproduce their mass spectra. Furthermore, we study the radiative decays of the $D_{s0}^*(2317)$ and $D_{s1}^{\prime}(2460)$ by simultaneously including the compound structures of conventional $c\bar{s}$ cores and $D^{(*)}K$ components. We also calculate their bottom analogs with heavy quark symmetry. Our study offers useful insights into the important unquenched effects in the formation of $D_{s0}^*(2317)$, $D_{s1}^{\prime}(2460)$ and the bottom counterparts. 
\end{abstract}

\maketitle

\section{Introduction}\label{sec1}

The rich hadron spectroscopy provides a good opportunity to decode the behavior of QCD in the nonperturbative region \cite{Liu:2024uxn,Chen:2016spr,Chen:2022asf,Dong:2021juy,Meng:2022ozq}. With the abundance of hadron data, many exotic states have been discovered. Their decay behavior will help us disclose the nonperturbative strong interactions furthermore.

In 2003, the {\it BABAR} Collaboration first observed the $D_{s0}^*(2317)$ through the $D_s^+\pi^0$ invariant mass spectrum and estimated that its mass is near $2.32$ GeV and its width is small \cite{BaBar:2003oey}. Shortly afterward the CLEO Collaboration confirmed the existence of $D_{s0}^*(2317)$ and found another state $D_{s1}^{\prime}(2460)$ in the $D_s^+\pi^0$ channel  \cite{CLEO:2003ggt}. They were further confirmed later by Belle and {\it BABAR} \cite{Belle:2003kup,Belle:2003guh,BaBar:2004yux,BaBar:2003cdx,BaBar:2006eep}. Their isospin and spin parity quantum number $I(J^P)$ are determined as $0(0^+)$ and $0(1^+)$~\cite{ParticleDataGroup:2024cfk}, and their masses and widths are~\cite{ParticleDataGroup:2024cfk}
\begin{equation}
    \begin{split}
        m_{D_{s0}^*}(2317)&=2317.7\pm0.5\;{\rm MeV},\\
        \Gamma_{D_{s0}^*}(2317)&<3.8\;{\rm MeV},\\
        m_{D_{s1}^{\prime}}(2460)&=2459.5\pm0.6\;{\rm MeV},\\
        \Gamma_{D_{s1}^{\prime}}(2460)&<3.5\;{\rm MeV}.\nonumber
    \end{split}
\end{equation}
However, such experimental values of the masses of $D_{s0}^*(2317)$ and $D_{s1}^{\prime}(2460)$ are clearly smaller than the predictions from the conventional quark model~\cite{Godfrey:1985xj,Godfrey:2015dva,Godfrey:1986wj,Ebert:2009ua}, which is usually referred as the low mass problem. This low mass problem also occurs in the well-known exotic hadrons $\Lambda(1405)$, $X(3872)$, and $\Lambda_c(2940)$ \cite{Silvestre-Brac:1991qqx,vanBeveren:2003kd,Kalashnikova:2005ui,Li:2009ad,Danilkin:2010cc,Luo:2019qkm}. Given the ongoing debate and various theoretical interpretations, the nature of the $D_{s0}^*(2317)$ and $D_{s1}^{\prime}(2460)$ and their bottom analogs remains an intriguing topic for further investigation. 

Indeed, the discovery of the $D_{s0}^*(2317)$ and $D_{s1}^{\prime}(2460)$ states has sparked numerous theoretical works attempting to interpret their nature. In the modified quark model, the mass spectra of $D_s$ mesons were studied by including the screening effects \cite{Gao:2022bsb,Song:2015nia}. To explain the low mass problem, the $cq\bar{s}\bar{q}$ tetraquarks with open charms were investigated by using the constituent quark model, QCD sum rule, and so on \cite{Cheng:2003kg,Dmitrasinovic:2005gc,Maiani:2004vq,Wang:2006uba,Chen:2004dy,Kim:2005gt,Nielsen:2005ia}. These two charmed mesons can also be the mixture of conventional $P$-wave quark-antiquark and four-quark components \cite{Vijande:2006hj}.

Given that their masses are close to the $D^{(*)}K$ threshold, they are considered to be $S$-wave isoscalar $D^{(*)}K$ molecules in many phenomenological studies~\cite{Kolomeitsev:2003ac,Szczepaniak:2003vy,vanBeveren:2003kd,Barnes:2003dj,Hofmann:2003je,Gamermann:2006nm,Guo:2006fu,Guo:2006rp,Flynn:2007ki,Faessler:2007gv,Guo:2009ct,Xie:2010zza,Liu:2011mi,Cleven:2010aw,Wu:2011yb,Guo:2015dha,Albaladejo:2016hae,Du:2017ttu,Guo:2018tjx,Albaladejo:2018mhb,Wu:2019vsy,Kong:2021ohg,Wang:2012bu,Huang:2021fdt,Ikeno:2023ojl,Liu:2023uly,Wu:2022wgn,Liu:2022dmm,Chen:2016ypj,Chen:2022svh,Liu:2024xbw,Kim:2023htt,Montesinos:2024uhq}. The lattice QCD simulations of the $D_{s0}^*(2317)$ and $D_{s1}^{\prime}(2460)$ have also revealed the molecular composition \cite{Mohler:2011ke,Liu:2012zya,Mohler:2013rwa,Lang:2014yfa,Bali:2017pdv,Alexandrou:2019tmk}. The coupled channel effects between the $c\bar{s}$ cores and the $D^{(*)}K$ channels are found to be important for these two charmed mesons \cite{Simonov:2004ar,Luo:2021dvj,Hao:2022vwt,Ni:2021pce,Yang:2023tvc,Coito:2011qn,Hwang:2004cd}. In addition, the $D^{(*)}K$ interactions were also considered in Refs. \cite{Ortega:2016mms, Yang:2021tvc}. The $D_{s0}^{*}(2317)$ is found to be dominated by the $DK$ channel with the Hamiltonian effective field theory in Ref.~\cite{Yang:2021tvc}, while the primary component of $D_{s0}^{*}(2317)$ is advocated to be the $c\bar{s}$ core in the resonating group method~\cite{Ortega:2016mms}.

Since heavy quark flavor symmetry establishes a connection between the open charm and bottom sectors, there has been considerable interest in predicting the bottom analogs of $D_{s0}^*(2317)$ and $D_{s1}^{\prime}(2460)$, i.e., $B_{s0}^*(1P)$ and $B_{s1}^{\prime}(1P)$. However, their bottom analogs are still not observed in experiments. The $B_s$ spectrum were studied with the conventional quark model \cite{Godfrey:2016nwn,DiPierro:2001dwf,Lu:2016bbk,Ebert:2009ua,Sun:2014wea,li:2021hss}. In Ref. \cite{Hao:2022ibj}, the screened nonrelativistic quark model was used to investigate the masses of the $B_s$ mesons. The $B_{s0}^*(1P)$ and $B_{s1}^{\prime}(1P)$ were proposed as the two-antiquark and four-quark mixtures in Ref. \cite{Vijande:2007ke}. In addition, the $B^{(*)}\bar{K}$ hadron molecule was advocated in Refs.~\cite{Kolomeitsev:2003ac,Guo:2006fu,Guo:2006rp,Cleven:2010aw,Colangelo:2012xi,Altenbuchinger:2013vwa,Sun:2018zqs,Ni:2021pce,Kim:2023htt}. The unquenched picture was adopted in Refs.~\cite{Albaladejo:2016ztm,Yang:2022vdb,Ortega:2016pgg}. The lattice QCD simulations were also carried out for these two states~\cite{Gregory:2010gm,Lang:2015hza,Hudspith:2023loy}.

Apart from the mass spectra, the radiative decays are also crucial to revealing the internal structures of these states. The radiative decays of $X(3872)$ were measured by different experimental groups \cite{Belle:2011wdj,BaBar:2008flx,LHCb:2014jvf,BESIII:2020nbj,LHCb:2024tpv} and studied within various theoretical contexts~\cite{Badalian:2012jz,Dong:2009uf,Cincioglu:2016fkm,Chen:2024xlw}. 
The E1 and M1 processes of charmonia and bottomia have been investigated in Refs.~\cite{Godfrey:2015dia,Segovia:2016xqb,Wang:2018rjg,Li:2012vc,Deng:2016ktl,Deng:2016stx}, which can be compared with the experimental values listed in PDG \cite{ParticleDataGroup:2024cfk}. The radiative decays of $D^*(2010)\to D\gamma$ were also measured experimentally \cite{CLEO:1992xqa,BESIII:2014rqs}. 

The radiative transitions for conventional $D_s$ or $B_s$ mesons were explored in Refs. \cite{Godfrey:2003kg,Godfrey:2016nwn,Lu:2016bbk,Godfrey:2015dva,Li:2019tbn,Radford:2009bs,Chen:2020jku,Green:2016occ,Godfrey:2005ww,Colangelo:2005hv,Lutz:2007sk}, and those for the conventional quark-antiquark and four-quark mixtures were studied in Refs. \cite{Vijande:2006hj,Vijande:2007ke}. Within the molecular picture,  the processes of $D_{s0}^*(2317)\to D_{s}^*\gamma$, $D_{s1}^{\prime}(2460)\to D_{s}^*\gamma$, $D_{s1}^{\prime}(2460)\to D_{s0}^*(2317)\gamma$ were studied in Refs. \cite{Cleven:2014oka,Fu:2021wde,Faessler:2007gv,Faessler:2007us,Xiao:2016hoa,Feng:2012zzf,Feng:2012zze} and $B_{s0}^*(1P)\to B_{s}^*\gamma$, $B_{s1}^{\prime}(1P)\to B_{s}^*\gamma$, $B_{s1}^{\prime}(1P)\to B_{s}\gamma$ were addressed in Refs. \cite{Faessler:2008vc,Wang:2019ehs,Cleven:2014oka,Fu:2021wde}. In this work, we will push forward the calculation of the radiative decays of $D_s$ and $B_s$ excitations by simultaneously including the conventional two-quark cores and $D^{(*)}K$/$B^{(*)}K$ components. 
We consider not only the coupled-channel effects between the bare $c\bar{s}$ cores and $D^{(*)}K$ channels but also the $D^{(*)}K$-$D^{(*)}K$ interactions. 
By solving the coupled-channel Schr{\"o}dinger equations, the masses and the wave functions can be obtained. With such wave functions, we further deal with their radiative decays. 

This article is organized as follows. After the introduction, we present the general formalism of the  coupled-channel Schr{\"o}dinger equation for the masses of the compound states in Sec.~\ref{sec2}, and the involved interactions are given in detail in Sec.~\ref{sec3}. We analyze the unquenched effect for $D_{s0}^*(2317)$, $D_{s1}^{\prime}(2460)$, and their bottom partners in Sec.~\ref{sec4}. Their radiative decays are then studied in Sec.~\ref{sec5}. A short summary follows in Sec.~\ref{sec6}.

\section{The coupled-channel Schr{\"o}dinger equation} \label{sec2}

In this section, we will present the coupled-channel Schr{\"o}dinger equation, which includes not only the coupling between the bare state and the hadron-hadron channel but also the self interaction of the two-hadron system.

It is plausible that the $D_{s0}^*(2317)$ could contain a single $c\bar{s}$ core plus the $DK$ cloud. To uncover this state, we need to set up the coupled-channel Schr{\"o}dinger equation by including one bare state, which can be found in our previous work \cite{Zhang:2022pxc}.

The $D_{s1}^{\prime}(2460)$ is a $J^P=1^+$ state and thus is related to $1P$ $c\bar{s}$ cores in the traditional quark model. There are two independent such cores $|c\bar{s},1^+,j_\ell=1/2\rangle$ and $|c\bar{s},1^+,j_\ell=3/2\rangle$ where $j_\ell$ is the angular momentum for the light quark freedom. To understand the $D_{s1}^{\prime}(2460)$, we need to solve the coupled-channel Schr{\"o}dinger equation including two bare states. We will elaborate this formalism in the following.

The physical particle involving two bare states can be expressed as~\cite{Weinberg:1965zz,Guo:2017jvc,Tornqvist:1995kr,Kalashnikova:2005ui,Danilkin:2009hr,Eichten:1978tg,Lu:2017hma,Anwar:2018yqm,Ortega:2021fem,Ortega:2021yis,Ortega:2016pgg,Ortega:2009hj} 
\begin{eqnarray}
\begin{split}\label{eq:phy2}
\left|\Psi\right\rangle=&c_{\alpha_1} \left|\Psi_{\alpha_1}\right\rangle+c_{\alpha_2} \left|\Psi_{\alpha_2}\right\rangle+\int {\rm d}^3\textbf{p}~\phi_{BC}(\textbf{p}) \left|BC,\textbf{p} \right\rangle,
\end{split}
\end{eqnarray}
with the normalization condition 
\begin{equation}\label{eq:Norm}
  |c_{\alpha_1}|^2+|c_{\alpha_2}|^2+\int \left|\phi_{BC}(\textbf{p})\right|^2{\rm d}^3{\bf p}=1,
\end{equation}
where the $\alpha_1$, $\alpha_2$, and $BC$ denote the first bare state, second bare state, and hadron-hadron channel, respectively. The full coupled-channel Schr{\"o}dinger equation which contains two bare states can be then written as
\begin{equation}\label{eq:coupchaSch}
\left(\begin{array}{cccc}
	\hat{H}_0&\hat{H}_{0}&\hat{H}_{I}\\
	\hat{H}_{0}&\hat{H}_0&\hat{H}_{I}\\
	\hat{H}_{I}&\hat{H}_{I}&\hat{H}_{BC}
\end{array}\right)
\left(\begin{array}{c}
c_{\alpha_1}|\Psi_{\alpha_1}\rangle\\
c_{\alpha_2}|\Psi_{\alpha_2}\rangle\\
\phi_{BC}(\textbf{p}) \left|BC,\textbf{p} \right\rangle
\end{array}\right)=M\left(\begin{array}{c}
c_{\alpha_1}|\Psi_{\alpha_1}\rangle\\
c_{\alpha_2}|\Psi_{\alpha_2}\rangle\\
\phi_{BC}(\textbf{p}) \left|BC,\textbf{p} \right\rangle
\end{array}\right),
\end{equation}	
where $\hat H_0$ is the Hamiltonian of the bare $c\bar s$ cores and can be obtained from the conventional potential model, $\hat H_I$ is the Hamiltonian describing the coupled-channel effect between the bare states and the $D^* K$ channel. As a novelty, we also consider $\hat{H}_{BC}$ for the $D^* K$-$D^* K$ interaction. 

From Eq. (\ref{eq:coupchaSch}), one obtains
\begin{equation}\label{eq:EXtwobareEQ1}
\begin{split}
	&M_{1}c_{\alpha_1}+M^{\prime}c_{\alpha_2}+\int \phi_{BC}({\bf p})H_{\alpha_1\to BC}^*({\bf p}){\rm d}^3{\bf p}=M c_{\alpha_1},\\
	&M^{\prime}c_{\alpha_1}+M_{2}c_{\alpha_2}+\int \phi_{BC}({\bf p})H_{\alpha_2\to BC}^*({\bf p}){\rm d}^3{\bf p}=M c_{\alpha_2},\\
	&c_{\alpha_1}H_{\alpha_1\to BC}({\bf p})+c_{\alpha_2}H_{\alpha_2\to BC}({\bf p})
	\\&\qquad+\int \langle BC,{\bf p}|\hat{H}_{BC}\phi_{BC}({\bf p}^\prime)|BC,{\bf p^{\prime}}\rangle {\rm d}^3{\bf p^\prime}=M \phi_{BC}({\bf p})\, ,
\end{split}
\end{equation}
where 
\begin{equation}
\begin{split}\label{eq:TranM}
 &M_1=\langle\Psi_{\alpha_1}|\hat{H}_0|\Psi_{\alpha_1}\rangle,\\
 &M_2=\langle\Psi_{\alpha_2}|\hat{H}_0|\Psi_{\alpha_2}\rangle,\\
    &M^{\prime}=\langle\Psi_{\alpha_1}|\hat{H}_0|\Psi_{\alpha_2}\rangle=\langle\Psi_{\alpha_2}|\hat{H}_0|\Psi_{\alpha_1}\rangle,\\
    &H_{\alpha_1\to BC}({\bf p})=\langle BC,{\bf p}|\hat{H}_{I}|{\Psi_{\alpha_1}}\rangle,\\
    &H_{\alpha_2\to BC}({\bf p})=\langle BC,{\bf p}|\hat{H}_{I}|{\Psi_{\alpha_2}}\rangle.
\end{split}
\end{equation}
The matrix elements sandwiching the $\hat{H}_{BC}$ can be explicitly expressed as
\begin{equation}\label{eq:expandHBC}
\begin{split}
&\int \langle BC,{\bf p}|\hat{H}_{BC}\phi_{BC}({\bf p}^\prime)|BC,{\bf p^\prime}\rangle {\rm d}^3{\bf p^\prime}\\
=&E_{BC}({\bf p})\phi_{BC}({\bf p})+\int V_{BC\to BC}({\bf p},{\bf p}^\prime)\phi_{BC}({\bf p}^{\prime}) {\rm d}^3{\bf p}^{\prime},
\end{split}
\end{equation}
where $E_{BC}({\bf p})=m_B+m_C+\frac{p^{2}}{2m_B}+\frac{p^{2}}{2m_C}$ is the free energy of the intermediate $BC$ channel and $V_{BC\to BC}({\bf p},{\bf p}^\prime)$ stands for the hadron-hadron interaction in momentum space.

To simplify the notations of Eq.~\eqref{eq:EXtwobareEQ1}, we define
\begin{equation}
    \begin{split}
      d_1=&\frac{M-M_{2}}{M^2-MM_{1}-{M^{\prime}}^2-MM_{2}+M_{1}M_{2}},\\
      d_2=&\frac{M^{\prime}}{M^2-MM_{1}-{M^{\prime}}^2-MM_{2}+M_{1}M_{2}},\\
      d_3=&\frac{M-M_{1}}{M^2-MM_{1}-{M^{\prime}}^2-MM_{2}+M_{1}M_{2}},
    \end{split}
\end{equation}
which give
\begin{equation}\label{eq:EXtwobareEQc}
\begin{split}
c_{\alpha_1}=&\int \phi_{BC}({\bf p})\big[ d_1 H_{\alpha_1\to BC}^*({\bf p})+d_2H_{\alpha_2\to BC}^*({\bf p})\big]{\rm d}^3{\bf p},\\
c_{\alpha_2}=&\int \phi_{BC}({\bf p})[d_2H_{\alpha_1\to BC}^*({\bf p})+d_3H_{\alpha_2\to BC}^*({\bf p})]{\rm d}^3{\bf p}\ .
\end{split}
\end{equation}
Then Eq. (\ref{eq:EXtwobareEQ1}) can be rewritten as 
\begin{equation}
\begin{split}\label{eq:FullChannel}
&
E_{BC}({\bf p})\phi_{BC}({\bf p})
+\int {\rm d}^3{\bf p^\prime}\phi_{BC}({\bf p^\prime})\Big[
V_{BC\to BC}({\bf p},{\bf p^\prime})
\\&\qquad 
+H_{\alpha_1\to BC}({\bf p})\left(d_1 H_{\alpha_1\to BC}^*({\bf p^\prime})+d_2H_{\alpha_2\to BC}^*({\bf p^\prime})\right)
\\&\qquad 
+H_{\alpha_2\to BC}({\bf p})\left(d_2H_{\alpha_1\to BC}^*({\bf p^\prime})+d_3H_{\alpha_2\to BC}^*({\bf p^\prime})\right)
%\\&\quad
\Big] \\
=&M\phi_{BC}({\bf p}).
\end{split}
\end{equation}
Up to this point, we have provided a simple and effective method for solving the unquenched coupled-channel equations that include both two bare states and direct hadron-hadron channel.

We point out that if the relevant term of one of the bare states is removed, Eq.~(\ref{eq:FullChannel}) is automatically reduced to the coupled-channel equation with only one bare state in our previous work \cite{Zhang:2022pxc}. Furthermore, if we remove the relevant terms of both two bare states, the equation becomes a Schr{\"o}dinger equation for the pure molecular state. We can also easily adjust the above formula to account for the situation when the interaction $V_{BC\to BC}({\bf p},{\bf p}^\prime)$ of the direct hadron-hadron channel is dropped. 

For obtaining the solution of Eq. (\ref{eq:FullChannel}), we use the complete-basis expansion method, where a set of the complete bases can be chosen as the harmonic oscillator basis, the Gaussian basis, and so on. In our study, we use the Gaussian basis to expand the wave function of bound hadron-hadron channel $\phi_{BC}(\textbf{p})$ in  Eq.~(\ref{eq:FullChannel}) \cite{Hiyama:2003cu,Hiyama:2012sma}
\begin{eqnarray}
	\phi_{BC}({\bf p})=\sum_{i=1}^{N_{max}}C_{il}\phi_{ilm}^p({\bf p}),
\end{eqnarray}
where $C_{il}$ is the coefficient of the corresponding basis, and $\phi_{ilm}^p({\bf p})$ denotes the Gaussian basis. In the coordinate space, the latter is given by 
\begin{eqnarray}\label{eq:Gr}
\phi_{nlm}^r(\nu_n,\textbf{r})=N_{nl}r^le^{-v_nr^2}Y_{lm}(\hat{\textbf{r}}),
\end{eqnarray}
where $N_{nl}$ is the normalization constant. By taking the Fourier transformation, the Gaussian basis in momentum space can be written as 
\begin{eqnarray}
\phi_{nlm}^p(\nu_n,\textbf{p})=(-i)^l\phi_{nlm}^r\left(\frac{1}{4\nu_n},\textbf{p}\right).\label{eq:Gp}
\end{eqnarray}
In Eqs.~(\ref{eq:Gr}) and (\ref{eq:Gp}), $v_n$ corresponds to the Gaussian ranges, i.e.,  
\begin{eqnarray}
	v_n=1/r_n^2,\quad r_n=r_1a^{n-1}\quad (n=1,2\dots N_{\rm max}).
\end{eqnarray}
 In practice, we vary $a$, $r_1$ and $N_{\rm max}$ to make the eigenvalues corresponding to the ground states as low as possible while keeping the solutions stable enough,  which gives $a= 1.6238$, $r_1=0.0051$ fm and $N_{\rm max}=20$. 

After introducing the above Gaussian basis, all the relevant Hamiltonian matrix elements can be expressed in simple forms: 
\begin{equation}\label{eq:matrixelement}
\begin{split}
T_{fi}=&\int \textrm{d}^3\textbf{p}^\prime \phi_{flm}^{p*}(\nu_f,\textbf{p}^{\prime})E_{BC}(\textbf{p}^{\prime})\phi_{ilm}^p(\nu_i,\textbf{p}^{\prime}),\\
{\cal M}_{fi}=&\int \textrm{d}^3\textbf{p}^{\prime}\textrm{d}^3\textbf{p}\Big[H_{\alpha_1\to BC}({\bf p}^\prime)\Big(d_1 H_{\alpha_1\to BC}^*({\bf p})\\&+d_2 H_{\alpha_2\to BC}^*({\bf p})\Big)+H_{\alpha_2\to BC}({\bf p}^\prime)\\&\times\Big(d_2H_{\alpha_1\to BC}^*({\bf p})+d_3H_{\alpha_2\to BC}^*({\bf p})\Big)\Big]\\&\times
\phi_{flm}^{p*}(\nu_f,\textbf{p}^{\prime})\phi_{ilm}^p(\nu_i,\textbf{p}),\\
V_{fi}=&\int \textrm{d}^3\textbf{p}^{\prime}\textrm{d}^3\textbf{p} \phi_{flm}^{p*}(\nu_f,\textbf{p}^{\prime})V_{BC\to BC}(\textbf{p},\textbf{p}^{\prime})\phi_{ilm}^p(\nu_i,\textbf{p}),\\
%=&\int \textrm{d}^3\textbf{r} V_{BC\to BC}(\textbf{r}) \phi_{flm}^{r*}(\nu_f,\textbf{r})\phi_{ilm}^r(\nu_i,\textbf{r}),\\
N_{fi}=&\int \textrm{d}^3 \textbf{r}^{\prime}\phi_{flm}^{r*}(\nu_f,\textbf{r}^{\prime})\phi_{ilm}^r(\nu_i,\textbf{r}^{\prime}).
\end{split}
\end{equation}
With the above matrix elements, Eq. (\ref{eq:FullChannel}) can be recast into the following general eigenvalue equation
\begin{eqnarray}\label{eq:EigM}
	\sum_{i=1}^{N_{max}}C_{il}(T_{fi}+{\cal M}_{fi}+V_{fi})=M\sum_{i=1}^{N_{max}}C_{il}N_{fi}\ .
\end{eqnarray}
Because both sides of Eq. (\ref{eq:EigM}) depend on $M$, we are in fact dealing with a special eigenvalue-equation problem. We scan the possible values of $M$ in the left-side matrix $T+{\cal M}+V$ within a reasonable range until the eigenvalue of Eq.~(\ref{eq:EigM}) gets equal to the input $M$, and this mass is exactly the solution of Eq.~\eqref{eq:EigM}. The corresponding eigenvector gives the coefficient $C_{il}$.

\section{Detailed interactions}\label{sec3}

In this section, we provide the quark-model descriptions of the bare $c\bar s$/$\bar b s$ cores in Sec.~\ref{sec:c/bsint}, the effective Lagrangian calculations of the $D^{(*)}K$/$B^{(*)}\bar{K}$ self-interactions in Sec.~\ref{sec:DKint} and the couplings between the bare states and the $D^{(*)}K/B^{(*)}\bar{K}$ in Sec.~\ref{sec:DKbareCoup}, respectively. To implement these detailed interactions into the coupled-channel Schr{\"o}dinger equations, one can then study the unquenched effects of the $D_{s0}^*(2317)$, $D_{s1}^{\prime}(2460)$, and their bottom analogs.

\subsection{Quark-model description of the bare cores}\label{sec:c/bsint}
We employ a nonrelativistic potential model to calculate the mass spectrum of the bare charmed-strange and bottom-strange states. The Hamiltonian $\hat{H}_0$ in Eq. (\ref{eq:coupchaSch}) can be represented as~\cite{Luo:2021dvj}
\begin{equation}\label{H0}
\hat{H}_0=\sum\limits_{i=1}\left(m_i+\frac{p_i^2}{2m_i}\right)+\sum\limits_{i<j}V_{ij},
\end{equation}
where $m_i$ and $p_i$ are the mass and momentum of the $i$-th constituent quark, respectively. $V_{ij}$ is the quark-quark/quark-antiquark interaction, which consists of the one-gluon-exchange potentials and confining potentials 
\begin{equation}\label{Vij}
V_{ij}=H_{ij}^{\rm conf}+H_{ij}^{\rm hyp}+H_{ij}^{\rm so(cm)}+H_{ij}^{\rm so(tp)}.
\end{equation}
Here the first term denotes the spin-independent Cornell potential
\begin{equation}\label{conf1}
H_{ij}^{\rm conf}=-\frac{4}{3}\frac{\alpha_s}{r_{ij}}+br_{ij}+C,
\end{equation}
where $\alpha_s$, $b$, and $C$ stand for the coupling constant of one-gluon exchange, the strength of linear confinement, and the mass-renormalized constant, respectively. The hyperfine spin-spin interaction reads 
\begin{equation}
H_{ij}^{\rm hyp}=\frac{4\alpha_s}{3m_im_j}\left(\frac{8\pi}{3}{\bf s}_i\cdot{\bf s}_j\tilde{\delta}(r)+\frac{1}{r_{ij}^3}S({\bf r}_{ij},{\bf s}_i,{\bf s}_j)\right),
\end{equation}
where
\begin{equation}
\tilde{\delta}(r)=\frac{\sigma^3}{\pi^{3/2}}{\rm e}^{-\sigma^2r^2}
\end{equation}
is a Gaussian smearing function including a smearing parameter $\sigma$, and
\begin{equation}
S({\bf r}_{ij},{\bf s}_i,{\bf s}_j)=\frac{3{\bf s}_i\cdot{\bf r}_{ij}{\bf s}_j\cdot{\bf r}_{ij}}{r_{ij}^2}-{\bf s}_i\cdot{\bf s}_j
\end{equation}
is a tensor operator. The color-magnetic term and Thomas-precession piece of the spin-orbit interactions could be expressed as
\begin{equation}
H_{ij}^{\rm so(cm)}=\frac{4\alpha_s}{3r_{ij}^3}\left(\frac{1}{m_i}+\frac{1}{m_j}\right)\left(\frac{{\bf s}_i}{m_i}+\frac{{\bf s}_j}{m_j}\right)\cdot {\bf L}\,,
\end{equation}
and
\begin{equation}
H_{ij}^{\rm so(tp)}=-\frac{1}{2r_{ij}}\frac{\partial H_{ij}^{\rm conf}}{\partial r_{ij}}\left(\frac{{\bf s}_i}{m_i^2}+\frac{{\bf s}_j}{m_j^2}\right)\cdot {\bf L}\,,
\end{equation}
respectively.

As the $D_s$ and $B_s$ are typical heavy-light systems, it is convenient to construct the wave function in the heavy quark symmetry, i.e.,
\begin{equation}\label{eq:PsiJMmeson}
\Psi_{JM}\sim [[s_q\phi_{NL}({\bf r})]_{j_\ell}s_Q]_{JM},
\end{equation}
where $s_q$ and $s_Q$ are the spins of the light and heavy quarks, respectively. $\phi_{NL}({\bf r})$ is the spatial wave function, and $N$ and $L$ are radial and orbital quantum numbers, in order.

By reproducing the masses of the low-lying well-established $\pi$, $K$, $D$, $D_s$, and $B_s$ mesons, we can determine the parameters in the quenched potential model, and the explicit results are: $m_{u/d}=0.336$ GeV, $m_{s}=0.540$ GeV, and $m_{c}=1.660$ GeV, $m_{b}=4.730$ GeV, $\alpha_s=0.570$, $b=0.144$ GeV$^2$, $\sigma =1.028$~GeV, and $C_{D_s}=-0.447$ GeV, $C_{B_s}=-0.122$ GeV.

Using these parameters and the interactions above, we obtain the relevant bare masses in Table \ref{table:M0}, which are close to the results in Ref.~\cite{Ebert:2009ua}. 

\renewcommand\tabcolsep{0.55cm}
\renewcommand\arraystretch{1.5}
\begin{table}[!htbp]
	\caption{The quenched masses in the nonrelativistic potential model. Here, $M_0$ is the bare mass of $D_{s0}^*(2317)$.  $M_1$ and $M_2$ are the bare masses of the $j_{\ell}=1/2$ and $3/2$ components in the $D_{s1}^{\prime}(2460)$, respectively, and $M^{\prime}$ is their transition matrix element as defined in Eq. (\ref{eq:TranM}). The masses are given in units of MeV.    }\label{table:M0}
	\centering
	\begin{tabular}{lcccc}\toprule[1.0pt]\toprule[1.0pt]
	Core&$M_0$&$M_1$&$M_2$&$M^{\prime}$\\ \midrule[1.pt]
    $c\bar{s}$&2447&2537&2526&4\\
    $s\bar{b}$&5821&5862&5861&2\\
		\bottomrule[1.0pt]	\bottomrule[1.0pt]
	\end{tabular}
\end{table}

\subsection{Two-hadron interactions in effective Lagrangian}\label{sec:DKint}

\begin{figure}[!htbp]
	\centering
	\includegraphics[width=0.85\linewidth]{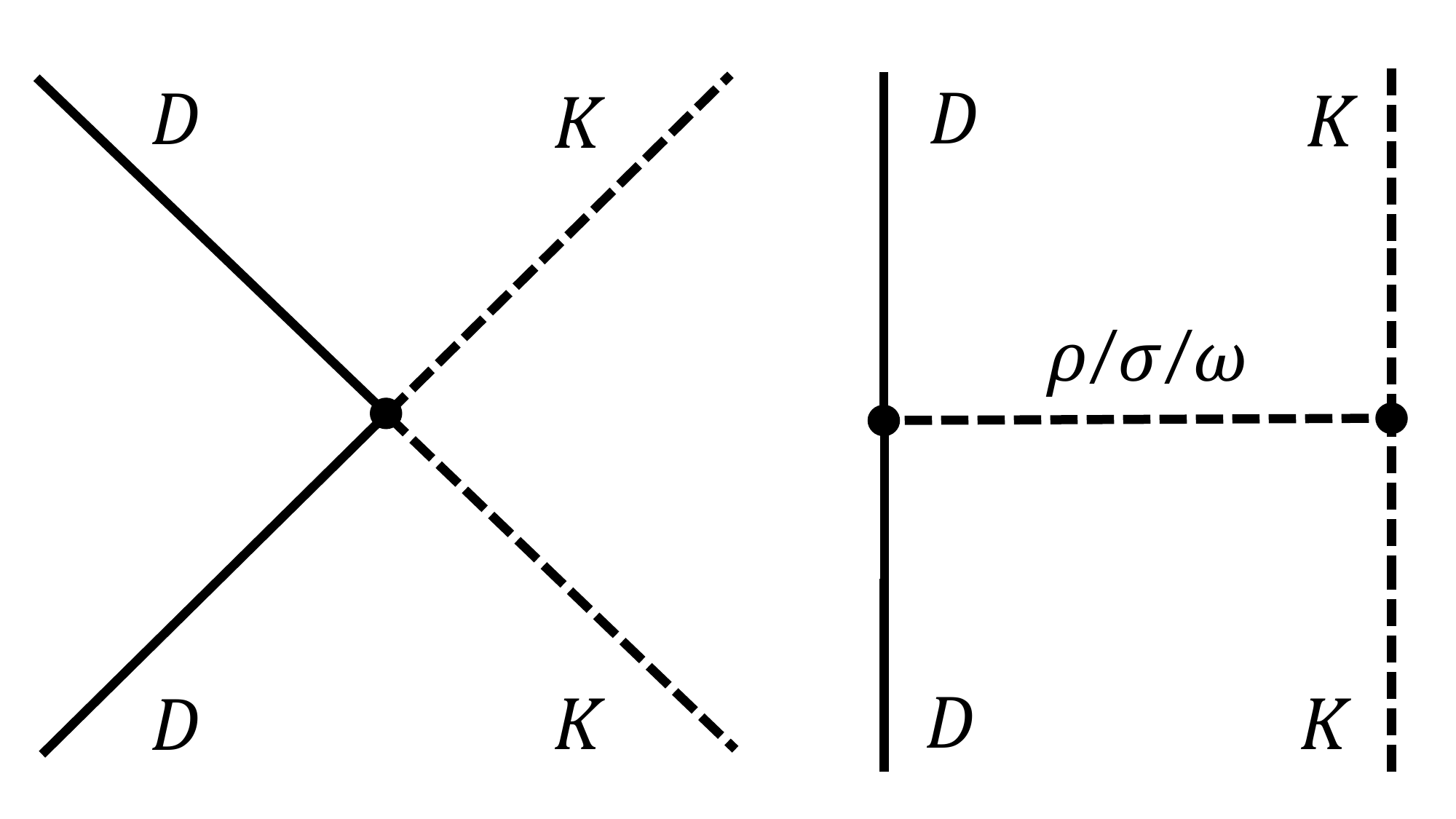}
	\caption{The Feynman diagrams for the $D^{(*)}K$/$B^{(*)}\bar{K}$ interaction. In the heavy quark limit, the interactions of $D^* K$ and $B^{(*)}\bar{K}$ share the same forms as $DK$. }\label{fig:feynman}
\end{figure}

We employ the effective Lagrangian method to calculate the hadron-hadron interactions $V_{BC\to BC}$. Here we simultaneously consider the contact term and the exchange of light-mesons ($\sigma$, $\rho$, $\omega$) as illustrated in Fig. \ref{fig:feynman} to describe the interaction of the $S$-wave $D^{(*)}K$/$B^{(*)}\bar{K}$ system. 

In the following, we first give the effective Lagrangians and Feynman diagrams of the $DK$ system. Next, we provide the $DK$ effective potentials in momentum space which include the contact term, 1$\sigma$-exchange, 1$\rho$-exchange, and 1$\omega$-exchange diagram contributions. Effective potentials in coordinate space are then obtained via the Fourier transformation. 

The Lagrangian with the $DK$ contact term takes the form~\cite{Wise:1992hn,Manohar:2000dt}
\begin{eqnarray}
\mathcal{L}_{\mathcal{D}\mathcal{D}\mathcal{P}\mathcal{P}}=\frac{i}{2f_K^2}\bar{\mathcal{D}}(\mathcal{P} v \cdot \partial \mathcal{P}-v \cdot \partial \mathcal{P} \mathcal{P})\mathcal{D}.
\end{eqnarray}
Here, $v=(1,0,0,0)$ stands for the 4-velocity of $D$ meson, and $\mathcal{D}$ denotes the charmed meson fields $\mathcal{D}^{T}=(D^{0},~D^{+},D_s^+)$. $f_K$ is the kaon decay constant with the value of $f_K=0.113$ MeV. The pseudoscalar field $\mathcal{P}$ is
\begin{eqnarray}
\mathcal{P}= \left(\begin{array}{ccc}
\frac{\pi^0}{\sqrt{2}}+\frac{\eta}{\sqrt{6}} &\pi^+&K^{+} \nonumber\\
\pi^- &-\frac{\pi^0}{\sqrt{2}}+\frac{\eta}{\sqrt{6}}&K^{0}\\
K^{-}&\bar{K}^{0}&-\frac{2}{\sqrt{6}}\eta
\end{array}\right)\,.
\end{eqnarray}

The Lagrangians with the $DD\sigma$ and $DDV$ couplings are given by~\cite{Yan:1992gz,Burdman:1992gh,Casalbuoni:1996pg,Falk:1992cx}
\begin{eqnarray}
&&\mathcal{L}_{{\mathcal{D}}{\mathcal{D}}\sigma} = -2g_s{\mathcal{D}}_b^{\dag}{\mathcal{D}}_b\sigma\,,\\
&&\mathcal{L}_{{\mathcal{D}}{\mathcal{D}}{\mathcal{V}}} = -\sqrt{2}\beta g_V{\mathcal{D}}_b{\mathcal{D}}_a^{\dag} v\cdot{\mathcal{V}}_{ba}\,,
\end{eqnarray}
where the vector meson matrix $\mathcal{V}$ is  
\begin{eqnarray}
\mathcal{V}= \left(\begin{array}{ccc}
\frac{\rho^0}{\sqrt{2}}+\frac{\omega}{\sqrt{2}} &\rho^+&K^{*+} \nonumber\\
\rho^- &-\frac{\rho^0}{\sqrt{2}}+\frac{\omega}{\sqrt{2}}&K^{*0}\\
K^{*-}&\bar{K}^{*0}&\phi
\end{array}\right).
\end{eqnarray}
The above equations give a vanishing $DD\phi$ coupling since it is Okubo-Zweig-Iizuka (OZI) suppressed. The Lagrangian with the couplings of $K\bar{K}\sigma$, $K\bar{K}\omega$ and $KK\rho$ can be written as~\cite{Lin:1999ad,Nagahiro:2008mn}
\begin{eqnarray}
\mathcal{L}_{\sigma KK} &=& g_{\sigma }m_K\bar{K} K\sigma,\\
\mathcal {L}_{\rho KK} &=& ig_{\rho KK}\left(\bar K\partial^{\mu}K
       - \partial^{\mu}\bar{K}K\right)\bm{\tau}\cdot\bm{\rho_{\mu}},\\
\mathcal {L}_{\omega KK} &=& ig_{\omega KK}\left(\bar K\partial^{\mu}K
       - \partial^{\mu}\bar{K}K\right)\omega_{\mu},
\end{eqnarray}
where $g_s=g_{\sigma}$ is estimated by the quark model \cite{Wang:2019aoc}. The $\sigma$ meson coupling constant $g_{s}$ can be calculated by the quark model: $g_{s}=g_{\sigma NN}/3$ with $g_{\sigma NN}^2/(4\pi)=5.69$ \cite{Isola:2003fh}. Following Ref. \cite{Chen:2011cj}, we take $g_{\rho {K}^{(*)}{K}^{(*)}} =3.425$ and $g_{\omega {K}^{(*)}{K}^{(*)}} = 4.396$. Using the vector meson dominance, $\beta$ can be fixed as 0.9, and $g_V=m_{\rho}/f_{\pi}=5.8$ \cite{Sun:2011uh}.

With these Lagrangians, we can calculate the Feynman amplitudes $\mathcal{M}_{DK\to DK}({\bf q})$ corresponding to the Feynman diagrams in Fig.~\ref{fig:feynman}. Then the effective potential related to such amplitudes can be given by the Breit approximation \cite{Breit:1929zz},
\begin{equation}
\mathcal{V}_{DK\to DK}({\bf q})=-\frac{\mathcal{M}_{DK\to DK}({\bf q})}{\sqrt{\Pi_i 2m_i \Pi_f 2m_f}},
\end{equation}
where $m_i$ and $m_f$ are the masses of the initial and final states, respectively. The flavor wave function of the isoscalar $D^{(*)}K$ system with $\left|I=0,I_3=0\right\rangle$ is
\begin{equation}\label{eq:flavor}
  \frac{1}{\sqrt{2}}\left(\left|D^{(*)+}K^{0}\right\rangle+\left|D^{(*)0}K^+\right\rangle\right) .
\end{equation}
In momentum space, these effective potentials of the Feynman diagrams in Fig. \ref{fig:feynman} are found to be
\begin{equation}
\begin{split}\label{eq:VXepress}
&V_{\rm{Contact}}=-\frac{m_K}{2f_K^2},\\
&V_{\rho}=-\frac{3}{2}\beta g_V g_{\rho KK}\frac{1}{{\textbf{q}}^2+m_{\rho}},\\
&V_{\omega}=-\frac{1}{2}\beta g_V g_{\omega KK} \frac{1}{{\bf q}^2+m_{\omega}^2},\\
&V_{\sigma}=-g_s g_{\sigma}\frac{1}{{\bf q}^2+m_{\sigma}^2},\\
&\mathcal{V}_{\rm Total}=V_{\rm{Contact}}+V_{\rho}+V_{\omega}+V_{\sigma}\,,
 \end{split}
\end{equation}
where the approximation of $q^0=0$ is taken.

\begin{figure}[htbp]
	\centering
	\includegraphics[width=0.95\linewidth]{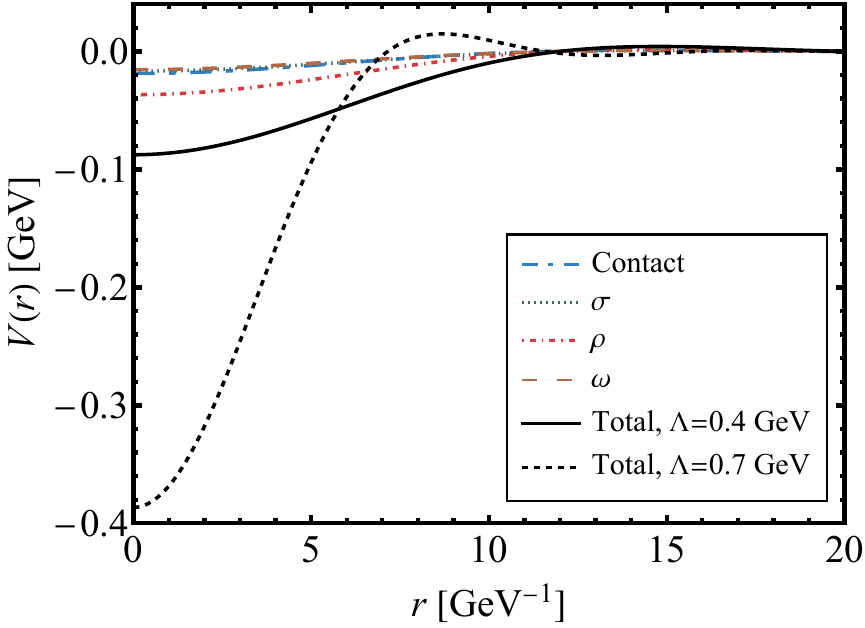}
	\caption{The $I=0$ interaction potentials of $D^{(*)}K$/$B^{(*)}\bar{K}$ systems in coordinate space. Notice that with the heavy quark symmetry, the $D^{(*)}K$ and $B^{(*)}\bar{K}$ potentials are equal. \label{fig:potential}}
\end{figure}

Based on the effective potential $\mathcal{V}_{BC\to BC}({\bf q})$ in momentum space, its counterpart $\mathcal{V}_{BC\to BC}({\bf r})$ in coordinate space can be obtained via the following Fourier transformation
\begin{eqnarray}\label{eq:FT}
	V_{BC\to BC}({\bf r})=\int\frac{\textrm{d}^3{\bf q}}{(2\pi)^3}{\rm e}^{i{\bf q}\cdot\textbf{r}}\mathcal{V}_{BC\to BC}({\bf q})\mathcal{F}({\bf q})\,,
\end{eqnarray}
where $\mathcal{F}(\textbf{q})={\rm e}^{-\textbf{q}^{2n}/\Lambda^{2n}}$ is the form factor with Gauss form to get rid off divergences showing up in $|\textbf{q}|\to \infty$ limit and to also renormalize the potential \cite{Xu:2017tsr}. 
The matrix element $V_{fi}$ in Eq.~(\ref{eq:matrixelement}) can be then conveniently calculated via
\begin{equation}
V_{fi}=\int \textrm{d}^3\textbf{r} \phi_{flm}^{r*}(\nu_f,\textbf{r})V_{BC\to BC}(\textbf{r})\phi_{ilm}^r(\nu_i,\textbf{r})\, .
\end{equation}

So far we have focused on the $DK$ effective potential. In the heavy-quark limit, the interactions of $D^*K$ and $DK$ system are in fact exactly the same. In Fig.~\ref{fig:potential}, we illustrate their potentials in coordinate space. It is evident that the contribution from each term, including the contact interaction and the exchanges of $\rho$, $\sigma$ and $\omega$, is attractive. The effects of the $\sigma$ and $\omega$ exchanges are rather weak, with similar strengths. The strongest attractive interaction is provided by the $\rho$ exchange and it is dominant over the others. Furthermore, the potential with the cutoff $\Lambda=0.7$ GeV is found to be more attractive than that with $\Lambda=0.4$~GeV.

In the heavy quark limit, the $S$-wave $B^{(*)}\bar{K}$ interaction shares the same form as that of the $D^{(*)}K$ system~\cite{Chen:2016ypj,Chen:2022svh}. Nevertheless, the binding properties of the two systems are different because of their different reduced masses $\mu_{AB}=M_A M_B/(M_A+M_B)$.

\subsection{Interaction between the bare core and hadron-hadron channel}\label{sec:DKbareCoup}

For the interaction between the bare states and $D^{(*)}K$/$B^{(*)}\bar{K}$ in Eq.~(\ref{eq:FullChannel}), we employ the quark-pair-creation model, whose Hamiltonian is given by~\cite{Micu:1968mk,LeYaouanc:1972vsx}
\begin{equation}
	\hat{H}_I=g\int \textrm{d}^3x\bar{\psi}(x)\psi(x).
\end{equation}
Here, $g=2m_q\gamma$, $m_q$ is the mass of the creation quark, and the dimensionless parameter $\gamma$ describes the strength of the quark and antiquark pair creation from the vacuum, which can be determined phenomenologically by the OZI allowed decay widths of charmonia. 

In the nonrelativistic limit, $\hat{H}_I$ is equivalent to \cite{Chen:2016iyi}
\begin{equation}
\begin{split}
	\hat{H}_I=&-3\gamma\sum_{m}\left<1,m;1,-m|0,0\right>\int \textrm{d}^3\textbf{p}_{\mu}d^3\textbf{p}_{\nu}\delta\left(\textbf{p}_{\mu}+\textbf{p}_{\nu}\right)\\&\times\mathcal{Y}_1^m\left(\frac{\textbf{p}_{\mu}-\textbf{p}_{\nu}}{2}\right)\omega^{\left(\mu,\nu\right)}\phi^{\left(\mu,\nu\right)}\chi_{-m}^{\left(\mu,\nu\right)}b_{\mu}^{\dagger}\left(\textbf{p}_{\mu}\right)d_{\nu}^{\dagger}\left(\textbf{p}_{\nu}\right)\, ,
\end{split}
\end{equation}
where $\omega$, $\phi$, $\chi$, and $\mathcal{Y}_1^m$ 
denote the color, flavor, spin, and orbital angular momentum functions of the quark pair, respectively. $b_{\mu}^{\dagger}$ and $d_{\nu}^{\dagger}$ are quark and antiquark creation operators, in order. In this work, we determine the dimensionless parameter $\gamma$ to be $5.7$ from the fit to the total decay width of $D_{s2}^*(2573)$.

Using the wave functions to describe the bare hadrons in previous Sec.~\ref{sec:c/bsint}, we can obtain the transition amplitudes of bare cores to hadron-hadron channels in Eq. (\ref{eq:TranM}) with the quark-pair-creation model.

\section{Spectra from coupled Schr{\"o}dinger equations}\label{sec4}

Based on the formalism described in Sec.~\ref{sec2} and the detailed interaction in Sec.~\ref{sec3}, we are ready to study the dynamical couplings between the $S$-wave $D^{(*)}K$ channels and the bare $c\bar s$ cores, and explore the unquenched effects for the $D_{s0}^*(2317)$ and $D_{s1}^{\prime}(2460)$ mesons. Within the same framework, we will also investigate their bottom analogs.

\subsection{Quenched results from the bare $c\bar s$ cores}

In Sec. \ref{sec3}, we enumerate the results under the quenched picture. We can see that the bare mass of $|c\bar{s},0^+,j_\ell=1/2\rangle$ is obtained to be $2447$ MeV, about 80 MeV above the $DK$ threshold and $130$ MeV larger than the measured mass of $D_{s0}^*(2317)$. The bare mass of $|c\bar{s},1^+,j_\ell=1/2\rangle$ is obtained to be $2537$ MeV, which is about $32$ MeV above the $D^*K$ threshold and $77$ MeV larger than the measured mass of $D_{s1}^{\prime}(2460)$. Thus we can clearly see that the near-threshold behavior and the prominent low-mass problem exist in these two states.

From Table \ref{table:M0}, the mixing interaction $M'$ between the $|c\bar{s},1^+,j_\ell=1/2\rangle$ and $|c\bar{s},1^+,j_\ell=3/2\rangle$ is weak. In the heavy quark limit, $j_\ell$ is a good quantum number and different states with different $j_\ell$ will not mix.

\subsection{Adding the couplings between $c\bar s$ core and $D^{(*)}K$}\label{sec.ivb}

Because the masses of the physical states $D_{s0}^*(2317)$ and $D_{s1}^{\prime}(2460)$ are clearly lower than the bare $c\bar{s}$ cores, they cannot be explained as the $c\bar{s}$ pairs in the simple quenched quark model. In this part, we explore the situation by including the coupling between $c\bar{s}$ and $D^{(*)}K$ but excluding the $D^{(*)}K$-$D^{(*)}K$ interaction, and give the results in the middle part of Table \ref{table:MResult}.

According to the values in this table, the resulting masses decrease due to the couplings between $c\bar{s}$ core and $D^{(*)}K$ channel, and they are now closer to the measured values. In the $D_{s0}^*$ case, the probabilities of the bare $c\bar{s}$ core and $DK$ are $42\%$ and $58\%$, respectively. In the $D_{s1}^{\prime}$ case, the proportions of $D^*K$ and the bare states with $j_{\ell}=1/2$ and $j_{\ell}=3/2$ are $37.4\%$, $62.1\%$, and $0.5\%$, in order. We see that the component of $|c\bar{s},1^+,j_\ell=1/2\rangle$ is dominant for $D_{s1}^{\prime}$. 

Although the masses are shifted down, one cannot still fully reproduce the experimental masses of $D_{s0}^*(2317)$ and $D_{s1}^{\prime}(2460)$. Therefore, additional effects need to be considered.

\renewcommand\tabcolsep{0.4cm}
\renewcommand\arraystretch{1.5}
\begin{table*}[!htbp]
	\caption{The masses in the quenched quark model, with coupled-channel effects between the bare cores and the $D^{(*)}K/B^{(*)}\bar{K}$, and with $D^{(*)}K$-$D^{(*)}K$/$B^{(*)}\bar{K}$-$B^{(*)}\bar{K}$ interactions furthermore. Here, $r_{\rm RMS}$ refers to the root-mean-square radius of the $D^{(*)}K/B^{(*)}\bar{K}$ component, and $P({\rm bare})$ represents the probability of bare $c\bar{s}/\bar{b}s$ core in the eigenstate.}\label{table:MResult}
	\centering
	\begin{tabular}{cc|ccc|ccc}\toprule[1.0pt]\toprule[1.0pt]
		Cases&Quenched&\multicolumn{3}{c|}{Without hadron-hadron interaction (approximation)}&\multicolumn{3}{c}{With hadron-hadron interaction (full result)}\\\midrule[1.pt]
		States&$M$ (MeV)&$M$ (MeV)&$r_{\rm RMS}$ (fm)&$P({\rm bare})$ $(\%)$&$M$ (MeV)&$r_{\rm RMS}$ (fm)&$P({\rm bare})$ $(\%)$\\\midrule[1.pt]
		$D_{s0}^*(2317)$&2447&2349&1.84&42.0&2321&1.33&34.4\\
		$D_{s1}^{\prime}(2460)$&2537&2478&1.48&62.6&2456&1.30&47.8\\
		$B_{s0}^*(1P)$&5821&5741&1.30&58.2&5714&1.17&44.5\\
		$B_{s1}^{\prime}(1P)$&5862&5787&1.26&61.7&5762&1.16&47.2\\
		\bottomrule[1.0pt]	\bottomrule[1.0pt]
	\end{tabular}
\end{table*}	

\subsection{Further adding the $D^{(*)}K$-$D^{(*)}K$ interaction}

In addition to the effects already discussed in Sec.~\ref{sec.ivb},  we further take into account the $D^{(*)}K$-$D^{(*)}K$ interactions in Eq. (\ref{eq:FullChannel}) to solve the unquenched effect of $D_{s0}^*(2317)$ and $D_{s1}^{\prime}(2460)$. We list the results after including the $D^{(*)}K$ interaction with the cutoff $\Lambda=0.41$ GeV in the right part of Table \ref{table:MResult}. The results show that the $D^{(*)}K$-$D^{(*)}K$ interaction can indeed further lower the masses down. Now the resulting masses from our study are very close to the experimental ones of $D_{s0}^*(2317)$ and $D_{s1}^{\prime}(2460)$. 

\begin{figure}[!htbp]
  	\centering
  	\includegraphics[width=1.0\linewidth]{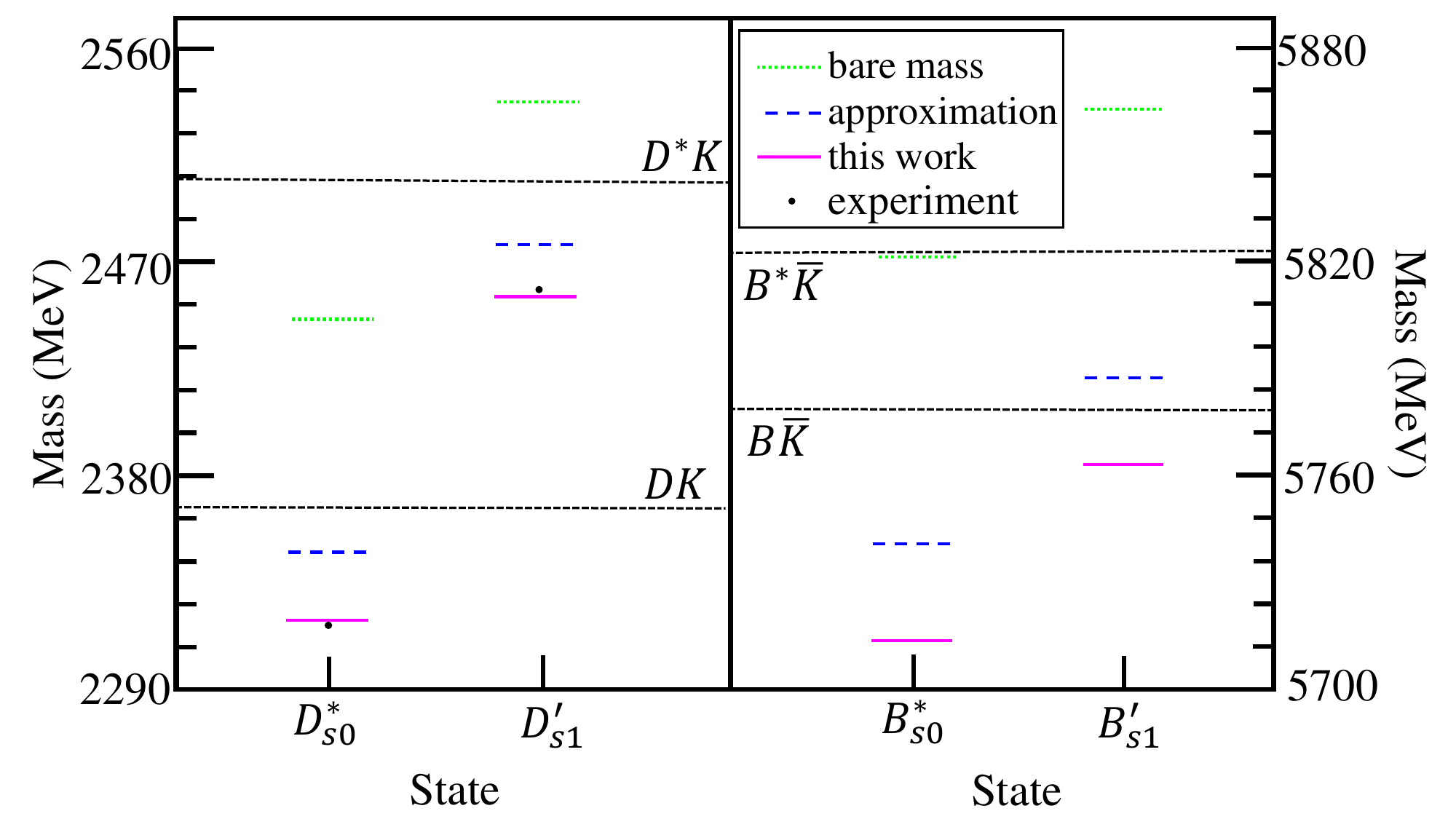}
  	\caption{Comparison of the masses in Table \ref{table:MResult} and experimental measurements. See Table~\ref{table:MResult} for the meaning of notations. The results labeled as this work correspond to the full result in our study. }\label{fig:compare}
  \end{figure}

For easy comparisons, the spectra of the quenched quark model and those with the full coupled-channel effects as shown in Table \ref{table:MResult} are visualized in the left part of Fig.~\ref{fig:compare}. The masses in the quenched quark model are much larger than the experimental ones. With the continuous improvement of the unquenched effects, the theoretical masses gradually overlap with the experimental ones. Both the $c\bar{s}$-$D^{(*)}K$ couplings and $D^{(*)}K$-$D^{(*)}K$ interactions are important to understand the inner structures of $D_{s0}^*(2317)$ and $D_{s1}^{\prime}(2460)$. 

Using Eqs. (\ref{eq:EXtwobareEQc}) and (\ref{eq:Gr}), we can obtain the $D^{(*)}K$ radial wave function and bare $c\bar{s}$ core probability $|c_{\alpha_i}|^2$ which can further help us to reveal the role of $D^{(*)}K$ channels and bare $c\bar{s}$ cores in physical state $D_{s0}^*(2317)$ and $D_{s1}^{\prime}(2460)$. From Table \ref{table:MResult}, we can see the $0^+$ state, i.e., $D_{s0}^*(2317)$,  is a little fatter with larger radius and contains lesser bare $c\bar{s}$ core than the $1^+$ one, which is consistent with the conclusions from earlier
studies~\cite{MartinezTorres:2014kpc,Albaladejo:2018mhb,Tan:2021bvl}. The $D^{(*)}K$ possibility in the physical states $D_{s0}^*(2317)$ and $D_{s1}^{\prime}(2460)$ are about $65.6\%$ and $52.2\%$, respectively. The predominantly $D^{(*)}K$ structure of these two states has an indirect robust theoretical support in Refs. \cite{MartinezTorres:2014kpc,Albaladejo:2016lbb,Du:2017zvv}.

\subsection{Closer inspection of binding energy and radial wave function}

If $D_{s0}^*(2317)$ and $D_{s1}^{\prime}(2460)$ are pure $D^{(*)}K$ molecules, their binding energies would be about 50 MeV. They would be very deeply bound states compared to the deuteron, which indicates the attractive $D^{(*)}K$ interactions must be extremely larger than the $pn$ attraction! However, the $D^{(*)}K$-$D^{(*)}K$ interactions in our framework are not extraordinarily strong. If neglecting the $c\bar{s}$ core, the pure $DK$ and $D^{*}K$ systems can only form bound states with binding energies around $2$ MeV and $1$ MeV, respectively. This means the interaction strength for $D^{(*)}K$ in this work is comparable to that for $pn$ in the deuteron.

Of course, we can artificially increase the cutoff $\Lambda$ in order to obtain very large $D^{(*)}K$ interactions, so that one can obtain the masses of pure $D^{(*)}K$ molecules close to the experimental values. When the cutoff $\Lambda$ reaches $0.73$ GeV, the masses of the pure $0^+$ $DK$ and $1^+$ $D^{*}K$ become $2318$ MeV and $2458$ MeV, respectively. In this case, the interaction strength for $D^{(*)}K$ is very different from that of the $pn$ system. However, we do not exactly know how different their interaction strengths should be in the real world, and thus we also show the results for the pure molecule assumption in the following discussion.

\begin{figure*}[!htbp]
		\centering
		\includegraphics[width=0.45\linewidth]{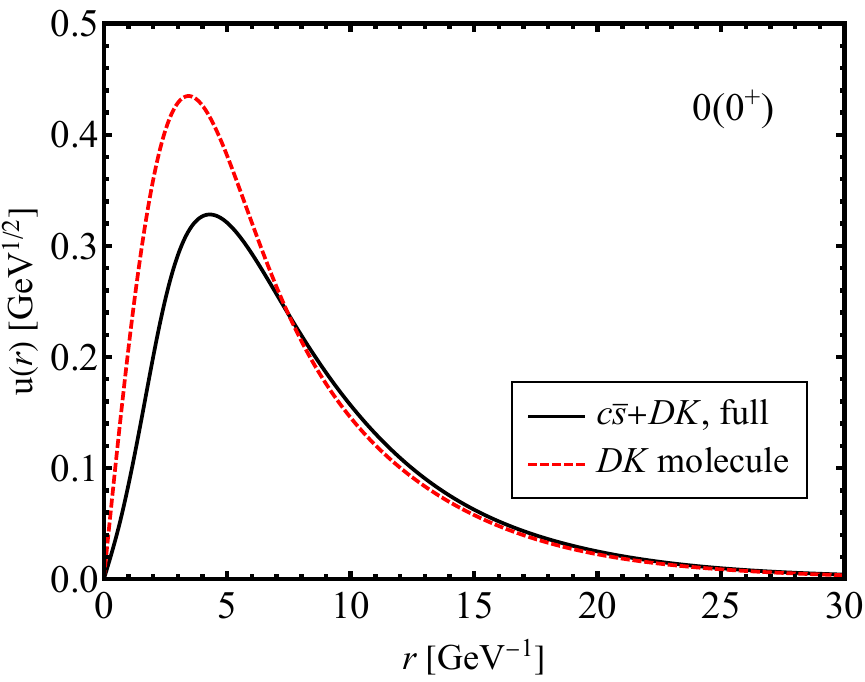}~~~~
        \includegraphics[width=0.45\linewidth]{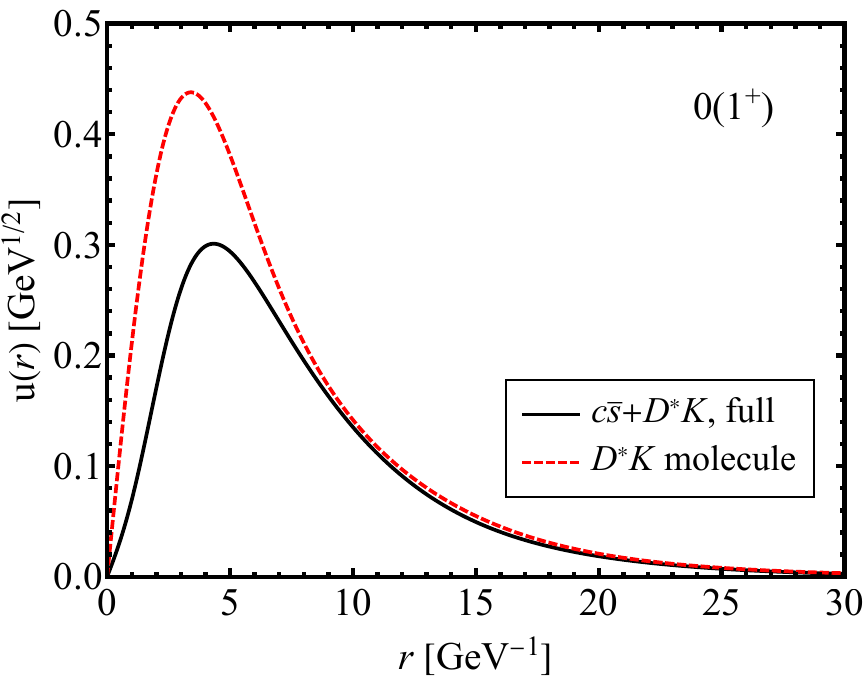}
		\caption{The $D^{(*)}K$ radial wave functions $u(r)=r\phi_{D^{(*)}K}(r)$ for the pure $D^{(*)}K$ molecule (with red dashed lines), and the dressed state with $D^{(*)}K$ interaction (with black solid lines). }\label{wavecs0}
	\end{figure*}

In addition to the spectra, the bare state will also bring a correction to the $D^{(*)}K$ radial wave functions. In Fig.~\ref{wavecs0}, we give the $D^{(*)}K$ radial wave functions in two cases: the pure $D^{(*)}K$ molecule (the red dot line) and the physical state dressed with $D^{(*)}K$ coupled channel involving hadron-hadron interaction (the black solid line). The results indicate that the bare states can obviously affect $D^{(*)}K$ radial wave functions and bind the $D^{(*)}K$ tighter. The radial wave functions will be used later in the calculation of the radiative decay of these two states.

\subsection{The prediction for the bottom analogs}
With the method of describing the $D_{s0}^*(2317)$ and $D_{s1}^{\prime}(2460)$, we use the same coupling constants and cutoff parameters to further give the predictions of the spectra of their bottom analogs. Here, we consider their bottom-strange analogs as the $\bar{b}s$ cores plus $S$-wave $B^{(*)}\bar{K}$ channels. 

According to the values of Table \ref{table:MResult}, the two $B_s(1P)$ states have approximate root-mean-square radius. The probability of bare core in $B_{s0}^*(1P)$ is bigger than that in $D_{s0}^*(2317)$.

From Fig. \ref{fig:compare}, the masses of the two $B_s(1P)$ states clearly decrease after considering the $B^{(*)}\bar{K}$ interactions. When using the cutoff that reproduces the masses of $D_{s0}^*(2317)$ and $D_{s1}^{\prime}(2460)$, we can predict $M_{B_{s0}^*}=5714$ MeV and $M_{B_{s1}^{\prime}}=5762$ MeV. 
  
\section{Radiative decay}\label{sec5}

To shed more light on the properties of $D_{s0}^*(2317)$, $D_{s1}^{\prime}(2460)$, we further explore their electromagnetic properties. We systematically study the electromagnetic properties of the bare cores, the pure $D^{(*)}K$/$B^{(*)}\bar{K}$ molecular states, and their physical mixtures. 

\subsection{The E1 transitions for the bare cores}\label{sub:RWbare}

The E1 transitions of the bare $D_s$/$B_s$ states play important roles in their electromagnetic properties. Under the assumption that they are simple $c\bar s$ or $\bar b s$ states, the partial widths associated with the E1 transitions can be obtained using the following expression~\cite{Brodsky:1968ea,Kwong:1988ae}
\begin{align}\label{eq:E1}
\Gamma(i \to f+\gamma)=\frac{4}{3} \langle e_{Q} \rangle^{2} \alpha \omega^{3} C_{f i} \delta_{S S^{\prime}}|\langle f|r| i\rangle|^{2},
\end{align}
with
\begin{align}
C_{f i}&=max\left(L, L^{\prime}\right)\left(2 J^{\prime}+1\right)\left\{\begin{array}{ccc}
L^{\prime} & J^{\prime} & S \\
J & L & 1
\end{array}\right\}^{2},
\end{align}
and
\begin{equation}
 e_{Q} =  \left\{\begin{array}{ll}(m_{s} e_{c}-m_{c} e_{\bar{s}})/(m_{c}+m_{s}) & \quad \mbox{for}\; D_s\;\mbox{meson}\\
(m_{b} e_{s}-m_{s} e_{\bar{b}})/(m_{b}+m_{s}) & \quad\mbox{for}\;B_s\;\mbox{meson}
\end{array} \right..
\label{Eq:eQ}
\end{equation}
$J$, $L$, and $S$ are the total angular momentum, orbital angular momentum, and spin of initial state meson, while $J^{\prime}$ and $L^{\prime}$ are the total and orbital angular momenta of final state meson, correspondingly. 
$\omega$ is the emitted photon energy, and $\langle f\left|r\right| i\rangle$ denotes the transition matrix element which can be expressed precisely as $\int _0^{\infty} R_{n^{\prime}l^{\prime}}(r)rR_{n,l}(r)r^2dr$, where the radial wave function $R_{n,l}(r)$ is obtained from the nonrelativistic potential model in Sec. \ref{sec:c/bsint}.   

If the $D_{s0}^*(2317)$ and $D_{s1}^{\prime}(2460)$ are assumed to be conventional $c\bar{s}$ mesons with only quark-antiquark cores, the widths for $D_{s0}^*\to D_s^*\gamma$, $D_{s1}^{\prime}\to D_s^*\gamma$, and $D_{s1}^{\prime}\to D_s\gamma$ are predicted to be 12.7, 14.3, and 6.0 keV, respectively. For the bottom analogs, the widths for $B_{s0}^*\to B_s^*\gamma$, $B_{s1}^{\prime}\to B_s^*\gamma$, and $B_{s1}^{\prime}\to B_s\gamma$ are 73.3, 77.2, and 44.9 keV, correspondingly. The bottom mesons have larger widths because of the larger values of $e_Q$ in Eq. (\ref{Eq:eQ}).

\subsection{The radiative decay for pure molecular state}\label{sub:RWbound}

If the $D_{s0}^*(2317)$ and $D_{s1}^{\prime}(2460)$ are pure $D^{(*)}K$ molecular states, their radiative decays can be totally different from those of conventional quark-antiquark mesons. We study the processes that the pure $D^{(*)}K$ molecules decay to $D_s^{(*)}\gamma$ in this subsection.

We denote the molecular state consisting of hadron $A$ and hadron $B$ as $[AB]$, and the radiative decay widths $\Gamma_{[AB]\to C\gamma}$ can be calculated as follows. The decay amplitude is \cite{Luo:2023hnp,Zhang:2006ix,Chen:2024xlw}
\begin{equation}\label{eq:decayABtoCG}
\begin{split}
\mathcal{M}^{JM}_{[AB]\to C\gamma}=&\int\frac{\rm d^3\textbf{p}^{\prime}}{(2\pi)^{3/2}}\,\frac{\sqrt{2m_{[AB]}}}{\sqrt{2m_A}\sqrt{2m_B}}\hat\phi^{JM}_{[AB]}({\bf p^{\prime}})\\&\otimes\hat{\mathcal{M}}_{AB\to C\gamma}({\bf p},{\bf p}^\prime),\\
\end{split}
\end{equation}

\begin{eqnarray}\label{phiJM}
\hat\phi^{JM}_{[AB]}({\bf p}^{\prime})&=&\bigg\{\phi_{[AB]|^3L_J\rangle}(|\textbf{p}^{\prime}|)\,\,C^{S,m_S}_{1m_1,1m_2}C^{J,M}_{Sm_S,Lm_L}\,Y_{L,m_L}(\theta,\phi) \nonumber\\
&&\qquad \bigg|\forall ~ L,S,m_L,m_1,m_2\bigg\}.
\end{eqnarray}
where $\hat{\mathcal{M}}_{AB\to C\gamma}({\bf p},{\bf p}^\prime)$ is the scattering amplitude of $AB\to C\gamma$ process, and $\hat\phi^{JM}_{[AB]}({\bf p}^{\prime})$ denotes the wave function of the molecular state $[AB]$ in momentum space which can be obtained via Eq. (\ref{eq:EigM}).

Then the decay width can be expressed as
\begin{equation}\label{eq:decayGAtoCD}
\Gamma_{[AB]\to C\gamma}=\frac{1}{2J+1}\frac{|\textbf{p}|}{32\pi^2m^2_{\left[AB\right]}}\sum_{M}\int \left|\mathcal{M}^{JM}_{[AB]\to C\gamma}\right|^2{\rm d}\Omega_{\textbf{p}}\,,
\end{equation}
where the factor $1/(2J+1)$ comes from the spin average over the initial polarization vector. In Eqs.~(\ref{eq:decayABtoCG})-(\ref{eq:decayGAtoCD}), ${\bf p}^\prime$ and ${\bf p}$ are the momentum of $A$ and $C$, respectively. 

\begin{figure*}[!htbp]
  	\centering
	\includegraphics[width=0.98\linewidth]{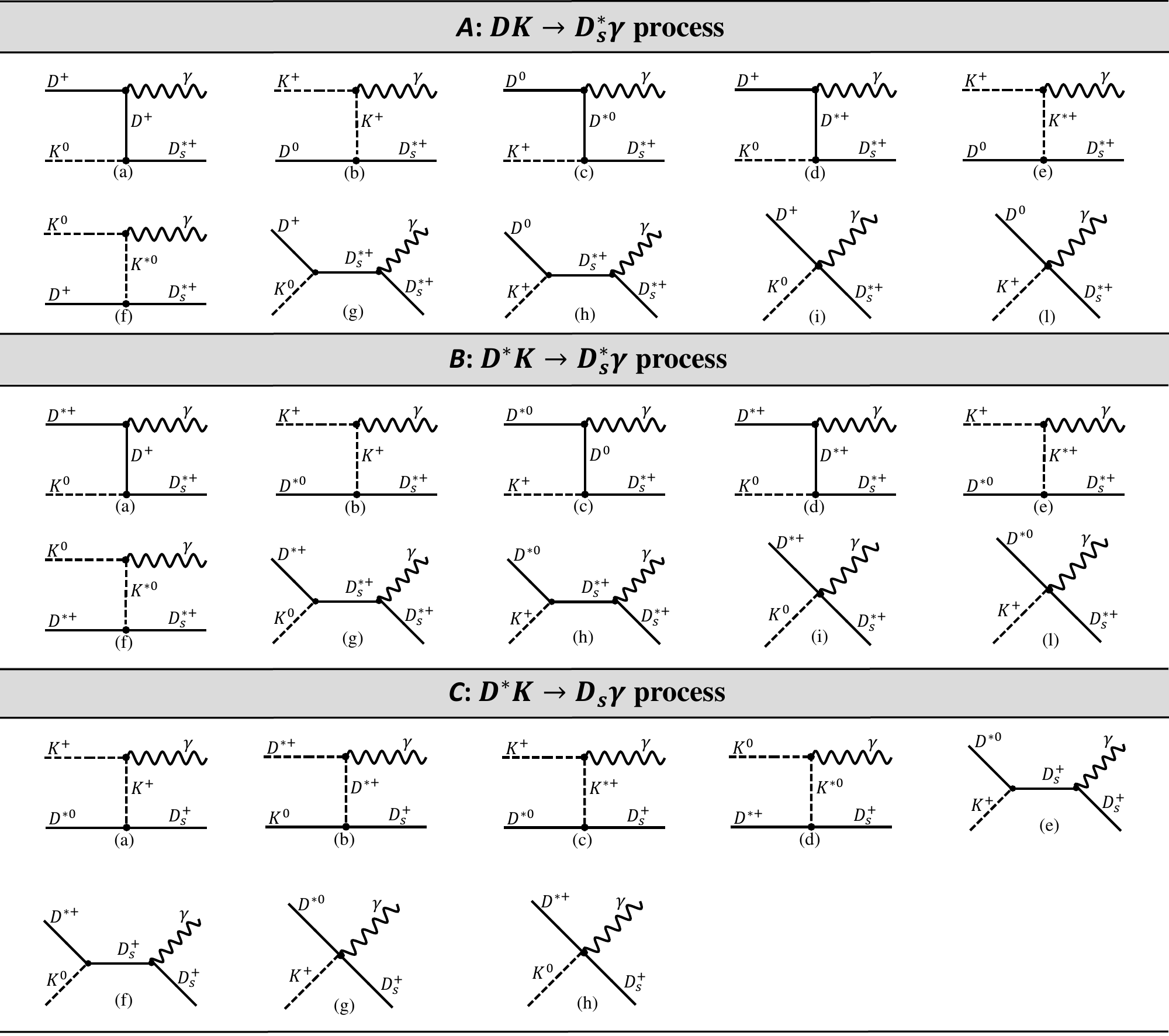}
  	\caption{Feynman diagrams for the radiative transition $D^{(*)}K\to D_s^*\gamma$.}\label{RDFeynmanD}
  \end{figure*}

To give the radiative decay width $\Gamma_{[DK]\to D_s^*\gamma}$, we need to calculate the scattering amplitude $\hat{\mathcal{M}}_{AB\to C\gamma}({\bf p},{\bf p}^\prime)$. The corresponding Feynman diagrams are shown in Fig.~\ref{RDFeynmanD}. 

The effective Lagrangians responsible for $\mathcal{D}^{(*)} \mathcal{D}^{(*)} V$ and $\mathcal{D}^{(*)} \mathcal{D}^{(*)} P$ interactions are~\cite{Chen:2010re,Chen:2014sra,Faessler:2007gv}
\begin{equation}
\begin{split}
\mathcal{L}_{\mathcal{D}^{(\ast)}\mathcal{D}^{(\ast)}\mathcal{P}}=&ig_{\mathcal{D}^*\mathcal{D}P}\mathcal{D}^{*\mu\dagger}\mathcal{D}\lrpartial_{\mu}\mathcal{P}\\&- g_{\mathcal{D}^*\mathcal{D}^*P}\epsilon_{\mu\nu\alpha\beta} \mathcal{D}^{*\nu\dagger}(\partial^{\beta}\mathcal{D}^{*\alpha})\partial^{\mu}\mathcal{P}+H.c. ,\label{Eq:LDDP}\\
\end{split}
\end{equation}
\begin{equation}
\begin{split}
 \mathcal{L}_{\mathcal{D}^{(*)} \mathcal{D}^{(*)} \mathcal{V}}=&-ig_{\mathcal{DDV}} \mathcal{D}_i^\dagger \lrpartial_{\mu} \mathcal{D}^j (\mathcal{V}^\mu)^i_j-2f_{\mathcal{D}^* \mathcal{D} \mathcal{V}} \epsilon_{\mu\nu\alpha\beta}\\
&\times(\partial^\mu \mathcal{V}^\nu)^i_j (\mathcal{D}_i^\dagger\lrpartial{}^{\!\alpha} \mathcal{D}^{*\beta j}-\mathcal{D}_i^{*\beta\dagger}\lrpartial{}{\!^\alpha} \mathcal{D}^j)\\
&+ ig_{\mathcal{D}^*\mathcal{D}^*\mathcal{V}} \mathcal{D}^{*\nu\dagger}_i \lrpartial_{\!\mu} \mathcal{D}^{*j}_\nu(\mathcal{V}^\mu)^i_j\\
&+4if_{\mathcal{D}^*\mathcal{D}^*\mathcal{V}} \mathcal{D}^{*\dagger}_{i\mu}(\partial^\mu \mathcal{V}^\nu-\partial^\nu \mathcal{V}^\mu)^i_j \mathcal{D}^{*j}_\nu,\label{Eq:LDDV}
\end{split}
\end{equation}
where $\mathcal{V}$ and $\mathcal{P}$ denote the matrices of the vector octet and pseudoscalar octet, respectively,  $A{\stackrel{\leftrightarrow}{\partial^\mu}}B =A(\partial_{\mu}B)-(\partial_{\mu}A)B$, and $\mathcal{D}^{(*)T}=(D^{(*)0},D^{(*)+},D_s^{(*)+})$. The values of coupling constants are given by~\cite{Kaymakcalan:1983qq,Oh:2000qr,Casalbuoni:1996pg,Chen:2010re,Chen:2014sra}
\begin{eqnarray}
     g_{D^*DP}&=&\frac{g}{f_{\pi}}\sqrt{m_{D^*}m_D},\quad g_{D^*D^*P}= \frac{2g}{f_{\pi}},\quad
     f_{D^*DV}=\frac{\lambda g_V}{\sqrt{2}}, \nonumber\\ f_{D^*D^*V}&=&\frac{\lambda g_V}{\sqrt{2}}m_{D^*},\quad g_{DDV}=g_{D^*D^*V}=\frac{\beta g_V}{\sqrt{2}},
  \end{eqnarray}
 where the gauge couplings are: $\lambda=0.56$ \cite{Chen:2010re} and $g=0.16$~\cite{Wang:2006ida}.

The effective Lagrangians involving the electromagnetic interaction vertices are \cite{Faessler:2008vc,Dong:2009uf}
\begin{eqnarray}
\begin{split}
 \mathcal{L}_{\mathcal{DD}\gamma}&=&i e A_{\mu}D^{-} {\stackrel{\leftrightarrow}{\partial^\mu}}D^{+}+i e A_{\mu}D_s^{-} {\stackrel{\leftrightarrow}{\partial^\mu}}D_s^{+},\label{Eq:DDgamma} \\  
 \end{split}
\end{eqnarray}
\begin{eqnarray}
\begin{split}
\mathcal{L}_{\mathcal{D}^{*} \mathcal{D} \gamma} =&\bigg\{\frac{e}{4}g_{{D^{*+}D^+}\gamma} \varepsilon^{\mu \nu \alpha \beta} F_{\mu\nu} {D}^{*+}_{\alpha \beta} {D}^- \\
&+\frac{e}{4}g_{{D^{*0}{D}^0}\gamma} \varepsilon^{\mu \nu \alpha \beta} F_{\mu\nu} D^{*0}_{\alpha \beta} \bar{{D}}^0\bigg\}+ H.c,\label{Eq:DsDgamma}
\end{split}
\end{eqnarray}
\begin{eqnarray}
\begin{split}
\mathcal{L}_{\mathcal{D}^{*} \mathcal{D}^{*} \gamma} =&-ie A_{\mu} \bigg\{g^{\alpha
\beta} D_{\alpha}^{*-} {\stackrel{\leftrightarrow}{\partial^\mu}} D_{\beta}^{*+} - g^{\mu
\beta} D_{\alpha}^{*-} \partial^{\alpha} D_{\beta}^{*+}\\
&+ g^{\mu\alpha }\partial^{\beta} D_{\alpha}^{*-}  D_{\beta}^{*+}\bigg\}\\
&-i e A_{\mu} \bigg\{g^{\alpha \beta} D_{s\alpha}^{*-} {\stackrel{\leftrightarrow}{\partial^\mu}} D_{s\beta}^{*+} -g^{\mu
\beta} D_{s\alpha}^{*-} \partial^{\alpha} D_{s\beta}^{*+} \\
&+ g^{\mu\alpha }\partial^{\beta} D_{s\alpha}^{*-}  D_{s\beta}^{*+}\bigg\},\label{Eq:DsDsgamma}\\
\end{split}
\end{eqnarray}

\begin{eqnarray}
\begin{split}
\mathcal{L}_{\mathcal{K}^{*} \mathcal{K} \gamma}=&\bigg\{\frac{e}{4}g_{{K^{*+}K^+}\gamma} \varepsilon^{\mu \nu \alpha \beta} F_{\mu\nu} {K}^{*+}_{\alpha \beta} {K}^-\\
&+\frac{e}{4}g_{{K^{*0}{K}^0}\gamma} \varepsilon^{\mu \nu \alpha \beta} F_{\mu\nu} K^{*0}_{\alpha \beta}\bar{{K}}^0\bigg\}+ H.c,\label{Eq:KsKgamma}\\
\end{split}
\end{eqnarray}
\begin{eqnarray}
\mathcal{L}_{\mathcal{KK}\gamma}=i e A_{\mu}K^{-} {\stackrel{\leftrightarrow}{\partial^\mu}}K^{+},\label{Eq:KKgamma}
\end{eqnarray}
with $F_{\mu \nu} =\partial_{\mu}A_{\nu}-\partial_{\nu} A_{\mu}$ and $D^{*0,+}_{\mu \nu} = \partial_{\mu} D^{*0,+}_{\nu} -\partial_{\nu} D^{*0,+}_{\mu}$. According to the experimental widths $\Gamma(D^{*+}\to D^+\gamma)=1.54$ keV \cite{ParticleDataGroup:2024cfk} and $\Gamma(D^{*0}\to D^0\gamma)=26.04$~keV \cite{Dong:2008gb,ParticleDataGroup:2024cfk}, the coupling constants $g_{D^{*} D \gamma}$ are fixed as
\begin{eqnarray}
|g_{D^{*+} D^{+} \gamma}|&=&0.5\;\mathrm{GeV}^{-1}, \\
|g_{D^{*0}D^{0}\gamma}|&=&2.0\;\mathrm{GeV}^{-1}.
\end{eqnarray}
By using the experimental measurements of $\Gamma(K^{*+}\to K^+\gamma)=50.29\;\rm{keV}$ and $\Gamma(K^{*0}\to K^0\gamma)=116.19\;\rm{keV}$ \cite{ParticleDataGroup:2024cfk,Faessler:2008vc}, the $K^{*}K\gamma$ coupling constants are determined as  
\begin{eqnarray}
|g_{K^{*+} K^{+} \gamma}|&=&0.836\;\mathrm{GeV}^{-1},\\
|g_{K^{*0}K^{0}\gamma}|&=&1.267\;\mathrm{GeV}^{-1}.
\end{eqnarray}
In our calculation, we take $g_{D^{*+}D^{+}\gamma}=0.5$ GeV$^{-1}$, $g_{D^{*0}D^{0}\gamma} = 2.0$ GeV$^{-1}$, $g_{K^{*+}K^{+} \gamma}=0.836$ GeV$^{-1}$, and $g_{K^{*0}K^{0}\gamma}=-1.267$~GeV$^{-1}$ as suggested in Ref.~\cite{Faessler:2008vc}. The interaction of four-point vertex $\mathcal{D}^{(*)} \mathcal{D}^{(*)} K\gamma $ can be obtained by replacing $\partial_{\mu}$ with $\partial_{\mu}+i e QA_{\mu}$ in Eq.~(\ref{Eq:LDDP}).

%The electromagnetic interactions of $D^0D^0\gamma$ and $D^{*0}D^{*0}\gamma$ do not exist.  

\iffalse
\begin{figure*}[!htbp]
 	\centering
	\includegraphics[width=0.95\linewidth]{DK.pdf}
  	\caption{Feynman diagrams for the radiative transition $DK\to D_s^*\gamma$.}\label{DK}
\end{figure*}
\fi 
 With these effective Lagrangians, we can write down the scattering amplitudes for the $DK\to D_s^*\gamma$ process shown in Fig. \ref{RDFeynmanD}, 
  
  \begin{eqnarray}
  \begin{split}
  i\mathcal{M}_a^{A}=&i^2[ie\epsilon_{\gamma}^{*\mu}(-iq_\mu-ip_{1\mu)}][ig_{D^*DP}\epsilon_{D_s^*}^{*\nu}(-ip_{2\nu}+iq_{\nu})]\\&\times \frac{i}{q^2-m_{D^+}^2},
  \end{split}
     \end{eqnarray}
     
     \begin{eqnarray}
     \begin{split}
  i\mathcal{M}_b^{A}=&i^2[ie\epsilon_{\gamma}^{*\mu}(-ip_{2\mu}-iq_{\mu})][ig_{D^*DP}\epsilon_{D_s^*}^{*\nu}(-iq_{\nu}+ip_{1\nu})]\\&\times \frac{i}{q^2-M_{K^+}^2},
  \end{split}
     \end{eqnarray}
     
     \begin{eqnarray}
     \begin{split}
  i\mathcal{M}_c^{A}=&i^2\big[\frac{1}{4}eg_{D^{*0}D^{0}\gamma}\varepsilon^{\mu \nu\alpha \beta}\epsilon_{\gamma}^{*\eta}(ip_{3\mu}g_{\nu\eta}-ip_{3\nu}g_{\mu\eta})\\&\times (iq_{\alpha}g_{\beta\tau}-iq_{\beta}g_{\alpha\tau})\big] \big[-g_{D^*D^*P}\varepsilon_{\delta\phi\sigma\gamma}(-ip_2^{\delta})\\&\times \epsilon_{D_s^*}^{*\phi}(-iq^{\gamma})\big] i\frac{-g^{\tau\sigma}+q^{\tau}q^{\sigma}/M_{D^{*0}}^2}{q^2-M_{D^{*0}}^2},
  \end{split}
     \end{eqnarray}
     
     \begin{eqnarray}
     \begin{split}
  i\mathcal{M}_d^{A}=&i^2\big[\frac{e}{4}eg_{D^{*+}D^{+}\gamma}\varepsilon^{\mu \nu\alpha \beta}\epsilon_{\gamma}^{*\eta}(ip_{3\mu}g_{\nu\eta}-ip_{3\nu}g_{\mu\eta})\\&\times(iq_{\alpha}g_{\beta\tau}-iq_{\beta}g_{\alpha\tau})\big]\big[-g_{D^*D^*P}\varepsilon_{\delta\phi\sigma\gamma}(-ip_2^{\delta})\\&\times\epsilon_{D_s^*}^{*\phi}(-iq^{\gamma})\big] i\frac{-g^{\tau\sigma}+q^{\tau}q^{\sigma}/M_{D^{*+}}^2}{q^2-M_{D^{*+}}^2},
  \end{split}
     \end{eqnarray}
     
     \begin{eqnarray}
     \begin{split}
  i\mathcal{M}_e^{A}=&i^2\big[2f_{D^*DV}\varepsilon_{\delta\phi\sigma\gamma}(-iq^{\delta})(-ip_4^{\sigma}-ip_1^{\sigma})\epsilon_{D_s^*}^{*\gamma}\big]\\&\times \big[\frac{1}{4}e g_{K^{*+}K^+\gamma}\varepsilon^{\mu\nu\alpha \beta}\epsilon_{\gamma}^{*\eta}(ip_{3\mu}g_{\nu\eta}-ip_{3\nu}g_{\mu\eta})\\&\times(iq_{\alpha}g_{\beta\tau}-iq_{\beta}g_{\alpha\tau})\big] i\frac{-g^{\phi \tau}+q^{\phi}q^{\tau}/M_{K^{*+}}^2}{q^2-M_{K^{*+}}^2},
  \end{split}
     \end{eqnarray}
     
     \begin{eqnarray}
     \begin{split}
  i\mathcal{M}_f^{A}=&i^2\big[2f_{D^*DV}\varepsilon_{\delta\phi\sigma\gamma}(-iq^{\delta})(-ip_4^{\sigma}-ip_1^{\sigma})\epsilon_{D_s^*}^{*\gamma}\big]\\&\times \big[\frac{1}{4}e g_{K^{*0}K^0\gamma}\varepsilon^{\mu\nu\alpha \beta}\epsilon_{\gamma}^{*\eta}(ip_{3\mu}g_{\nu\eta}-ip_{3\nu}g_{\mu\eta})\\&\times (iq_{\alpha}g_{\beta\tau}-iq_{\beta}g_{\alpha\tau})\big] i\frac{-g^{\phi \tau}+q^{\phi}q^{\tau}/M_{K^{*0}}^2}{q^2-M_{K^{*0}}^2},
  \end{split}
     \end{eqnarray}
     
     \begin{eqnarray}
     \begin{split}
  i\mathcal{M}_{g}^{A}=&i^2[ig_{D^*DP}(-ip_{2\mu}+ip_{1\mu})](-ie)\\&\times \big[g_{\alpha\beta}(-iq_{\nu}-ip_{4\nu})+g_{\nu\beta}(iq_{\alpha})+g_{\alpha\nu}(ip_{4\beta})\big]\\&\times \epsilon_{\gamma}^{*\nu}\epsilon_{D_s^*}^{*\alpha} i\frac{-g^{\mu\beta}+q^{\mu}q^{\beta}/M_{D_s^{*+}}^2}{q^2-m_{D_s^{*+}}^2},
  \end{split}
     \end{eqnarray}
     
     \begin{eqnarray}
     \begin{split}
	i\mathcal{M}_{h}^{A}=&i^2[ig_{D^*DP}(-ip_{2\mu}+ip_{1\mu})](-ie)\\&\times \big[g_{\alpha\beta}(-iq_{\nu}-ip_{4\nu})+g_{\nu\beta}(iq_{\alpha})+g_{\alpha\nu}(ip_{4\beta})\big]\\&\times\epsilon_{\gamma}^{*\nu}\epsilon_{D_s^*}^{*\alpha} i\frac{-g^{\mu\beta}+q^{\mu}q^{\beta}/M_{D_s^{*+}}^2}{q^2-m_{D_s^{*+}}^2},
  \end{split}
     \end{eqnarray}
  
  \begin{eqnarray}
      	i\mathcal{M}_{i}^{A}&=&i^3eg_{D^*DP}\epsilon_{\gamma}^{*\mu}\epsilon_{D_s^*\mu}^*,
  \end{eqnarray}
  
   \begin{eqnarray}
      	i\mathcal{M}_{l}^{A}&=&-i^3eg_{D^*DP}\epsilon_{\gamma}^{*\mu}\epsilon_{D_s^*\mu}^*,
  \end{eqnarray}
where $\epsilon_\gamma$ and $\epsilon_{D_s^*}$ are the polarization vectors of photon and $D_s^*$ meson, respectively \cite{Chen:2024xlw}, and $q$ is the momentum of the exchanged meson.
 
The total scattering amplitude for the $DK\to D_s^*\gamma$ process can be expressed as
\begin{eqnarray}
\begin{split}
    \mathcal{M}_{DK\to D_s^*\gamma}^{\rm Total}=\sum_{n=a}^{l}\mathcal{M}_{n}^{A}.
\end{split}
\end{eqnarray}
Considering the flavor wave function in Eq. (\ref{eq:flavor}), the decay amplitude in Eq. (\ref{eq:decayABtoCG}) need to be multiplied by the $1/\sqrt{2}$. To describe the off-shell effects of the exchanged mesons and account for the inner structures of the interaction vertices, we introduce the following form factor in the amplitudes~\cite{Chen:2024xlw}
\begin{eqnarray}\label{eq:FFS}
\mathcal{F}(\textbf{q})={\rm e}^{-\textbf{q}^{2n}/\Lambda^{\prime 2n}}.
\end{eqnarray}

Meanwhile, we also calculate the decay process of $[D^*K]\to D_s^* \gamma$ similar to the $[DK]\to D_s^* \gamma$ one. Their corresponding Feynman amplitudes in Fig.~\ref{RDFeynmanD} take the forms
 
  \begin{eqnarray}
  \begin{split}
     i\mathcal{M}_a^{B}=&i^2\big[\frac{1}{4}eg_{D^{*+}D^+\gamma}\varepsilon^{\mu \nu\alpha \beta}\epsilon_{\gamma}^{*\eta}\epsilon_{D^*}^{\tau}(ip_{3\mu}g_{\nu\eta}-ip_{3\nu}g_{\mu\eta})\\&\times(-ip_{1\alpha}g_{\beta\tau}+ip_{1\beta}g_{\alpha\tau})\big] \big[ig_{D^*DP}\epsilon_{D_{s}^{*}}^{*\phi}\\&\times(-ip_{2\phi}+iq_{\phi})\big]\frac{i}{q^2-M_{D^+}^2},
     \end{split}
     \end{eqnarray}
     
     \begin{eqnarray}
     \begin{split}
     i\mathcal{M}_b^{B}=&i^2\big[ie\epsilon_{\gamma}^{*\mu}(-ip_{2\mu}-iq_{\mu})\big]\\&\times\big[-g_{D^*D^*P}\epsilon_{\eta \tau \rho \sigma }(-iq^{\eta})\epsilon_{D_s^*}^{*\tau}\varepsilon_{D^*}^{\rho}(-ip_1^{\sigma})\big]\\&\times\frac{i}{q^2-M_{K^+}^2},
     \end{split}
     \end{eqnarray}
     
     \begin{eqnarray}
     \begin{split}
     i\mathcal{M}_c^{B}=&i^2\big[\frac{1}{4}eg_{D^{*0}D^0\gamma}\varepsilon^{\mu\nu\alpha \beta}\epsilon_{\gamma}^{*\eta}\epsilon_{D^*}^{\tau}(ip_{3\mu}g_{\nu\eta}-ip_{3\nu}g_{\mu\eta})\\&\times(-ip_{1\alpha}g_{\beta\tau}+ip_{1\beta}g_{\alpha\tau})\big]\big[ig_{D^*DP}\epsilon_{D_s^*}^{*\phi}\\&\times(-ip_{2\phi}+iq_{\phi})\big]\frac{i}{q^2-M_{D^+}^2},
     \end{split}
     \end{eqnarray}
     
     \begin{eqnarray}
     \begin{split}
     i\mathcal{M}_d^{B}=&i^2(-ie)\epsilon^{*\mu}_{\gamma}\epsilon^{\beta}_{D^*}\big[g_{\alpha\beta}(-iq_{\mu}-ip_{1\mu})+ig_{\mu\beta}(ip_{1\alpha})\\&+g_{\mu\alpha}(iq_{\beta})\big]\big[-g_{D^*D^*P}\varepsilon_{\eta\tau \rho\sigma}(-ip_2^{\eta})\epsilon_{D_s^*}^{*\tau}(-iq^{\sigma})\big]\\&\times i\frac{-g^{\alpha\rho}+q^{\alpha}q^{\rho}/{M_{D^{*+}}^2}}{q^2-M_{D^{*+}}^2},
     \end{split}
     \end{eqnarray}
     
     \begin{eqnarray}
     \begin{split}
     i\mathcal{M}_e^{B}=&i^2\big[ig_{D^*D^*V}\epsilon_{D_s^*}^{*\rho}\epsilon_{D^*}^{\phi}g_{\rho\phi}(-ip_{1\sigma}-ip_{4\sigma})\\&+4if_{D^*D^*V}\epsilon_{D^*}^{\phi}\epsilon_{D_s^*}^{*\rho}(-iq_{\rho}g_{\phi\sigma}+iq_{\phi}g_{\rho\sigma})\big]\\&\times\big[\frac{1}{4} e g_{K^*K\gamma} \varepsilon^{\mu \nu\alpha\beta}\epsilon_{\gamma}^{*\eta}(ip_{3\mu}g_{\nu\eta}-ip_{3\nu}g_{\mu\eta})\\&\times(iq_{\alpha}g_{\beta\tau}-iq_{\beta}g_{\alpha\tau})\big]i\frac{-g^{\sigma\tau}+q^{\sigma}q^{\tau}/M_{K^{*+}}^2}{q^2-M_{K^{*+}}^2},
     \end{split}
     \end{eqnarray}
     
     \begin{eqnarray}
     \begin{split}
     i\mathcal{M}_f^{B}=&i^2\big[ig_{D^*D^*V}\epsilon_{D_s^*}^{*\rho}\epsilon_{D^*}^{\phi}g_{\rho\phi}(-ip_{1\sigma}-ip_{4\sigma})\\&+4if_{D^*D^*V}\epsilon_{D^*}^{\phi}\epsilon_{D_s^*}^{*\rho}(-iq_{\rho}g_{\phi\sigma}+iq_{\phi}g_{\rho\sigma})\big]\\&\times\big[\frac{1}{4} e g_{K^*K\gamma}\varepsilon^{\mu \nu\alpha \beta}\epsilon_{\gamma}^{*\eta}(ip_{3\mu}g_{\nu\eta}-ip_{3\nu}g_{\mu\eta})\\&\times(iq_{\alpha}g_{\beta\tau}-iq_{\beta}g_{\alpha\tau})\big]i\frac{-g^{\sigma\tau}+q^{\sigma}q^{\tau}/M_{K^{*+}}^2}{q^2-M_{K^{*+}}^2},
     \end{split}
     \end{eqnarray}
     
     \begin{eqnarray}
     \begin{split}
     i\mathcal{M}_{g}^{B}=&i^2\big[-g_{D^*D^*P}\varepsilon_{\eta\tau\rho\sigma}(-ip_2^{\eta})\epsilon_{D^*}^{\rho}(-ip_1^{\sigma})\big]\\&\times(-ie)[g_{\alpha\beta}(-ip_{4\mu}-iq_{\mu})+g_{\mu\beta}(iq_{\alpha})\\&+g_{\mu\alpha}(ip_{4\beta})]\epsilon_{\gamma}^{*\mu}\epsilon_{D_s^*}^{*\alpha}i\frac{-g^{\tau\beta}+q^{\tau}q^{\beta}/M_{D_s^{*+}}^2}{q^2-M_{D_s^{*+}}^2},
     \end{split}
     \end{eqnarray}
     
     \begin{eqnarray}
     \begin{split}
	i\mathcal{M}_{h}^{B}=&i^2\big[-g_{D^*D^*P}\varepsilon_{\eta\tau\rho\sigma}(-ip_2^{\eta})\epsilon_{D^*}^{\rho}(-ip_1^{\sigma})\big]\\&\times(-ie)[g_{\alpha\beta}(-ip_{4\mu}-iq_{\mu})+g_{\mu\beta}(iq_{\alpha})\\&+g_{\mu\alpha}(ip_{4\beta})]\epsilon_{\gamma}^{*\mu}\epsilon_{D_s^*}^{*\alpha}i\frac{-g^{\tau\beta}+q^{\tau}q^{\beta}/M_{D_s^{*+}}^2}{q^2-M_{D_s^{*+}}^2},
	\end{split}
  \end{eqnarray}
  
   \begin{eqnarray}
   \begin{split}
      i\mathcal{M}_{i}^{B}=i^2e g_{D^*D^*P}\varepsilon_{\mu\nu\alpha\beta}(-i p_2^{\mu})\epsilon_{D_s^*}^{\nu}\epsilon_{D^*}^{\alpha}\epsilon_{\gamma}^{\beta},
    \end{split}
  \end{eqnarray}  
  
  \begin{eqnarray}
   \begin{split}
       i\mathcal{M}_{l}^{B}=&i^2e g_{D^*D^*P}\varepsilon_{\mu\nu\alpha\beta}\epsilon_{\gamma}^{\mu}\epsilon_{D_s^*}^{\nu}\epsilon_{D^*}^{\alpha}(-i p_{1}^{\beta}).
       \end{split}
  \end{eqnarray}

In addition, we also calculate the decay process of $[D^*K]\to D_s \gamma$. The Feynman diagrams for the $D^*K\to D_s \gamma$ process are shown in Fig.~\ref{RDFeynmanD}, and their corresponding Feynman amplitudes are

\begin{eqnarray}
   \begin{split}
       i\mathcal{M}_a^C=&i^2\big[ig_{D^*DP}(ip_{4\mu}+iq_{\mu})\epsilon_{D^*}^{\mu}\big]\big[ie\epsilon_{\gamma}^{*\nu}(-ip_{2\nu}-iq_{\nu})\big]\\
       &\times \frac{i}{q^2-m_{K^+}^2},
   \end{split}
\end{eqnarray}
 
\begin{eqnarray}
   \begin{split}
       i\mathcal{M}_b^C=&i^2\big[ig_{D^*DP}(ip_{4\mu}+ip_{2\mu})]\big](-ie)\epsilon^{*\rho}_{\gamma}\epsilon^{\beta}_{D^*}\\\times &\big[g_{\alpha\beta}(-iq_{\rho}-ip_{1\rho})+ig_{\rho\beta}(ip_{1\alpha})+g_{\rho\alpha}(iq_{\beta})\big]\\
       \times &i\frac{-g^{\mu \alpha}+q^{\mu}q^{\alpha}/{M_{D^{*+}}^2}}{q^2-M_{D^{*+}}^2},
   \end{split}
\end{eqnarray}

\begin{eqnarray}
     \begin{split}
  i\mathcal{M}_c^{C}=&i^2\big[-2f_{D^*DV}\varepsilon_{\delta\phi\sigma\gamma}(-iq^{\delta})(-ip_4^{\sigma}-ip_1^{\sigma})\epsilon_{D^*}^{\gamma}\big]\\&\times \big[\frac{1}{4}e g_{K^{*+}K^+\gamma}\varepsilon^{\mu\nu\alpha \beta}\epsilon_{\gamma}^{*\eta}(ip_{3\mu}g_{\nu\eta}-ip_{3\nu}g_{\mu\eta})\\&\times(iq_{\alpha}g_{\beta\tau}-iq_{\beta}g_{\alpha\tau})\big] i\frac{-g^{\phi \tau}+q^{\phi}q^{\tau}/M_{K^{*+}}^2}{q^2-M_{K^{*+}}^2},
  \end{split}
     \end{eqnarray}
     
\begin{eqnarray}
     \begin{split}
  i\mathcal{M}_d^{C}=&i^2\big[-2f_{D^*DV}\varepsilon_{\delta\phi\sigma\gamma}(-iq^{\delta})(-ip_4^{\sigma}-ip_1^{\sigma})\epsilon_{D^*}^{\gamma}\big]\\&\times \big[\frac{1}{4}e g_{K^{*0}K^0\gamma}\varepsilon^{\mu\nu\alpha \beta}\epsilon_{\gamma}^{*\eta}(ip_{3\mu}g_{\nu\eta}-ip_{3\nu}g_{\mu\eta})\\&\times(iq_{\alpha}g_{\beta\tau}-iq_{\beta}g_{\alpha\tau})\big] i\frac{-g^{\phi \tau}+q^{\phi}q^{\tau}/M_{K^{*0}}^2}{q^2-M_{K^{*0}}^2},
  \end{split}
     \end{eqnarray}

\begin{eqnarray}
   \begin{split}
       i\mathcal{M}_e^C=&i^2\big[ig_{D^*DP}(ip_{2\mu}+iq_{\mu})\epsilon_{D^*}^{\mu}\big]\big[ie\epsilon_{\gamma}^{*\nu}(-ip_{4\nu}-iq_{\nu})\big]\\
       &\times \frac{i}{q^2-m_{D_s^+}^2},
   \end{split}
\end{eqnarray}

\begin{eqnarray}
   \begin{split}
       i\mathcal{M}_f^C=&i^2\big[ig_{D^*DP}(ip_{2\mu}+iq_{\mu})\epsilon_{D^*}^{\mu}\big]\big[ie\epsilon_{\gamma}^{*\nu}(-ip_{4\nu}-iq_{\nu})\big]\\
       &\times \frac{i}{q^2-m_{D_s^+}^2},
   \end{split}
\end{eqnarray}

\begin{eqnarray}
   \begin{split}
       i\mathcal{M}_g^C=-2ie g_{D^*DP}\epsilon_{\gamma}^{* \mu}\epsilon_{D^*\mu},
   \end{split}
\end{eqnarray}

\begin{eqnarray}
   \begin{split}
       i\mathcal{M}_h^C=-ie g_{D^*DP}\epsilon_{\gamma}^{* \mu}\epsilon_{D^*\mu}.
   \end{split}
\end{eqnarray}

We have checked that gauge invariance is preserved in our amplitudes. That is, the total amplitude $\epsilon_{\gamma}^{*\mu}\mathcal{M}_{\mu}$ vanishes when we replace $\epsilon_{\gamma}^{*\mu}$ with the photon momentum $p_3^{\mu}$, i.e.,  $p_3^{\mu}\mathcal{M}_{\mu}=0$.

Based on heavy quark symmetry, our study can be easily extended to the $B^{(*)}\bar{K}$ system. Of course, the coupling constants should be replaced with the bottomed ones  \cite{Oh:2000qr,Xiao:2016mho,Becirevic:2009yb,Faessler:2008vc}
 \begin{eqnarray}
     &&g_{B^*BP}=\frac{g}{f_{\pi}}\sqrt{m_{B^*}m_B},\quad g_{B^*B^*P}= \frac{2g}{f_{\pi}},\quad
     f_{B^*BV}=\frac{\lambda g_V}{\sqrt{2}}, \nonumber\\ 
     &&f_{B^*B^*V}=\frac{\lambda g_V}{\sqrt{2}}m_{B^*},\quad g_{BBV}=g_{B^*B^*V}=\frac{\beta g_V}{\sqrt{2}}, \nonumber\\
   &&g_{B^{*+}B^+ \gamma}=0.5\;{\rm GeV}^{-1};\quad  g_{B^{*0}B^0 \gamma}=-1\;{\rm GeV}^{-1}.
\end{eqnarray}

  \begin{figure}[!htbp]
  	\centering
	\includegraphics[width=0.49\linewidth]{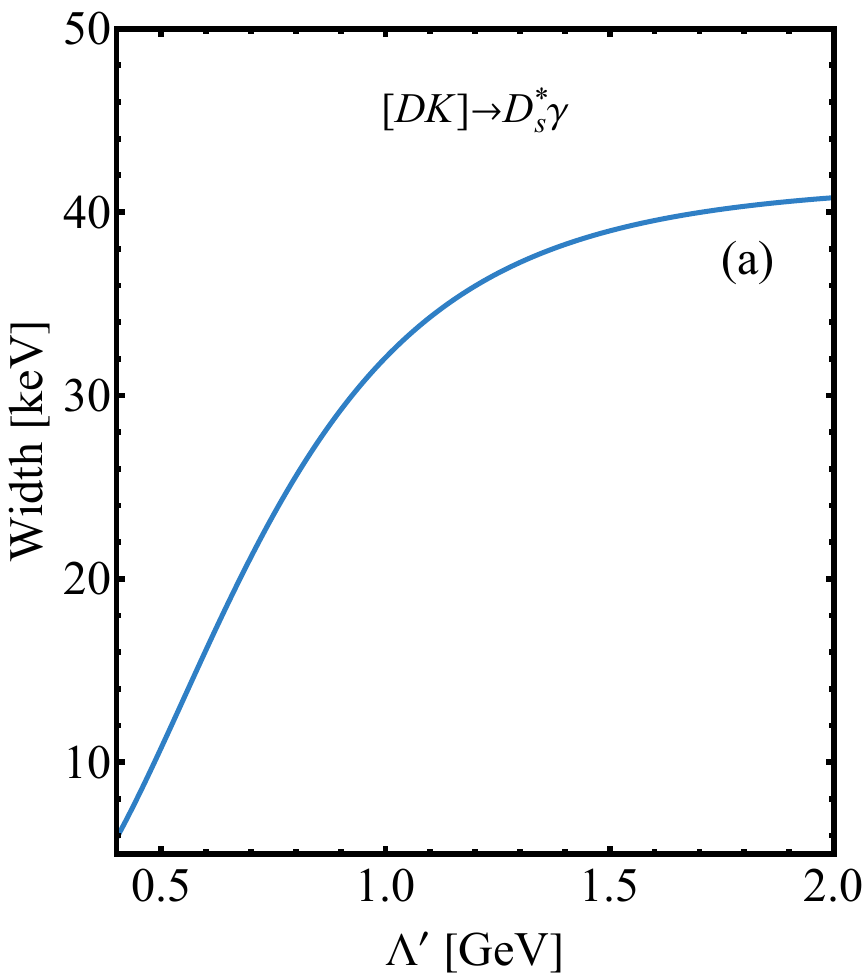}
		\includegraphics[width=0.49\linewidth]{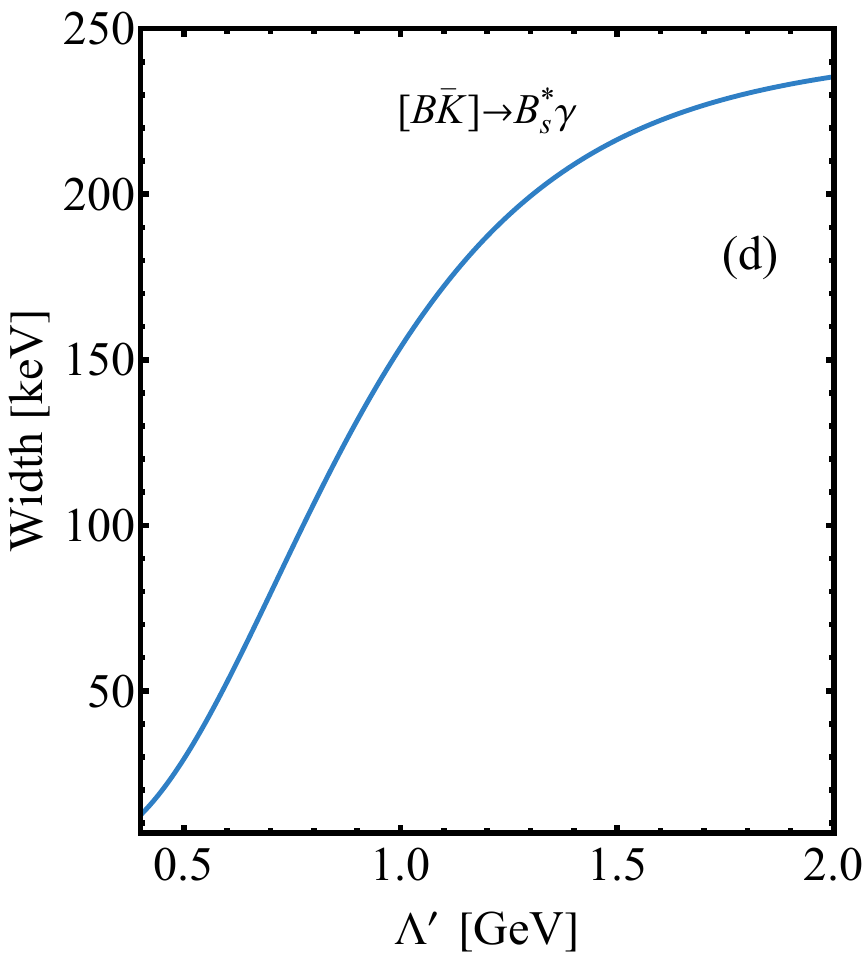}
	\includegraphics[width=0.49\linewidth]{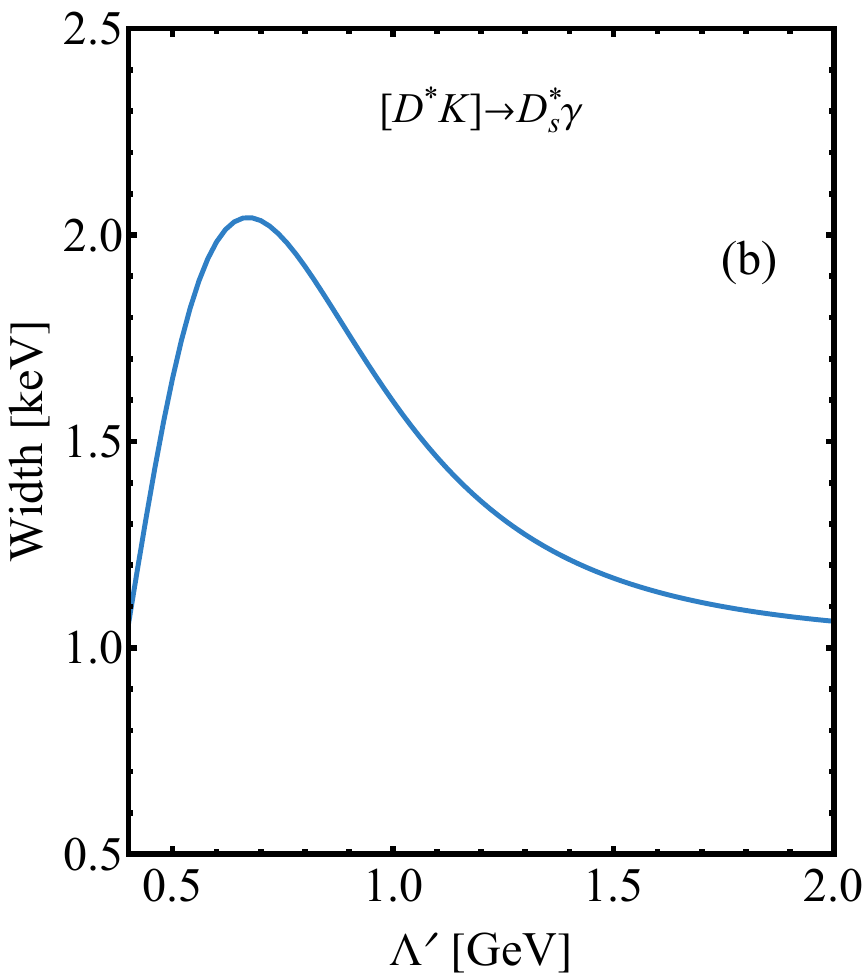}
	\includegraphics[width=0.49\linewidth]{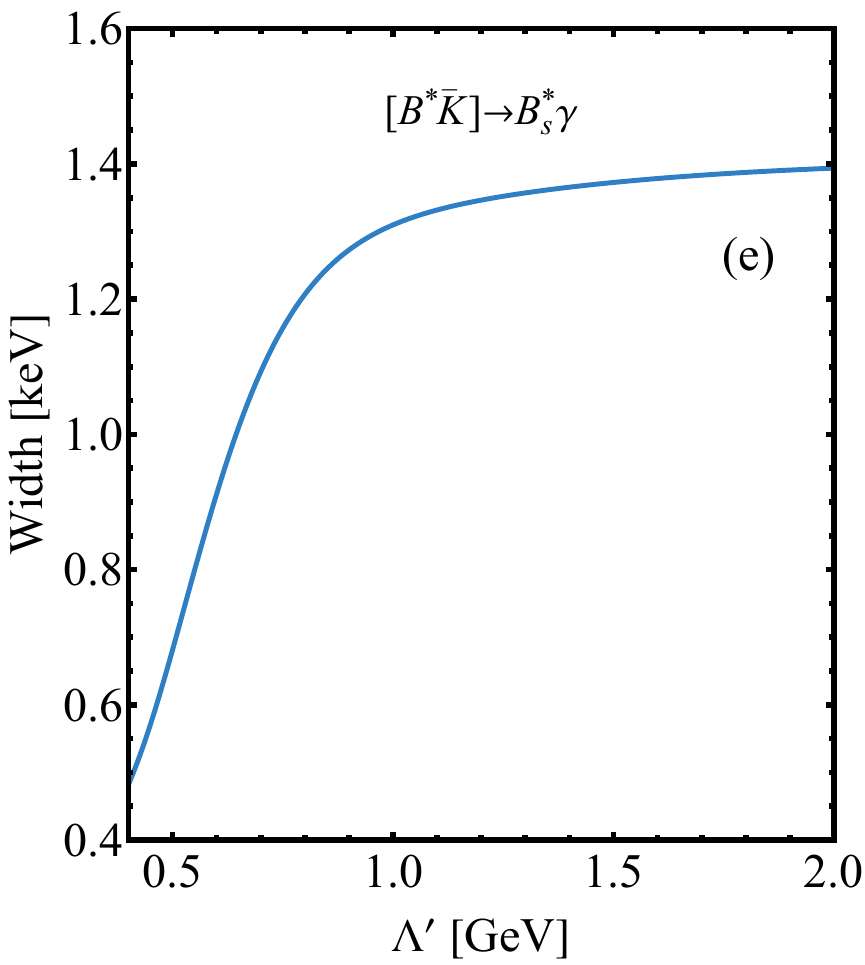}
	\includegraphics[width=0.49\linewidth]{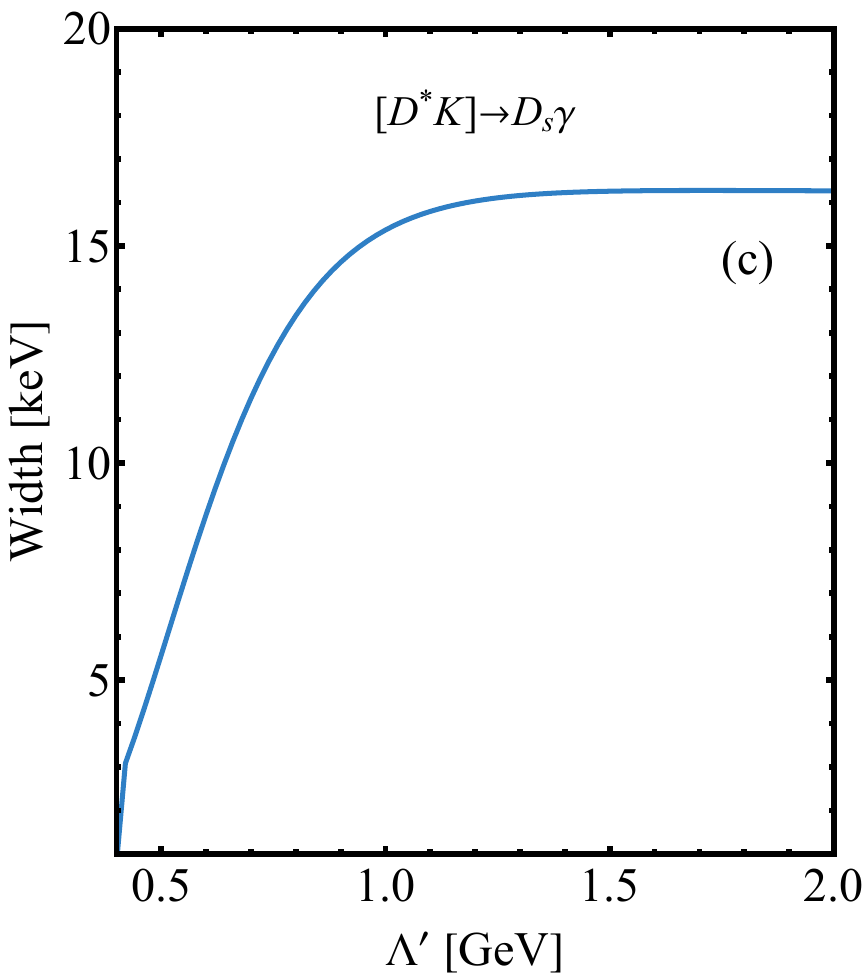}
    \includegraphics[width=0.49\linewidth]{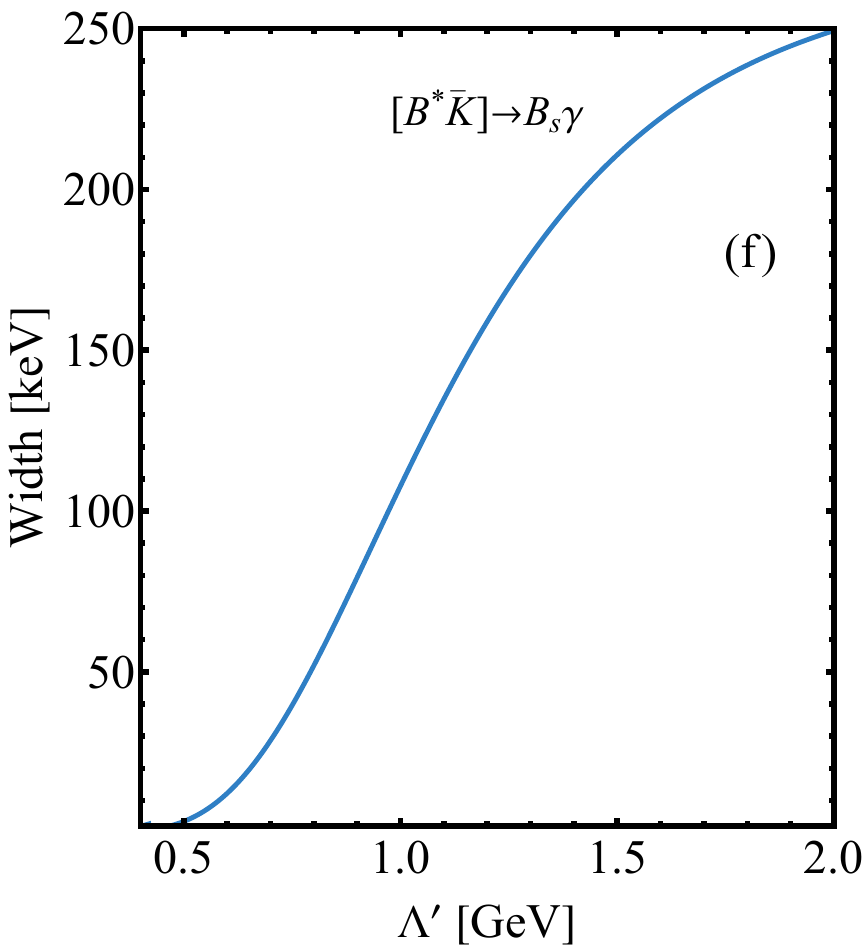}
  	\caption{
  The dependence of the decay widths on the cutoff $\Lambda^{\prime}$ in the pure molecule picture.}\label{fig:widthpuremolecular}
  \end{figure}
  
\begin{figure}[!htbp]
  	\centering
		\includegraphics[width=0.49\linewidth]{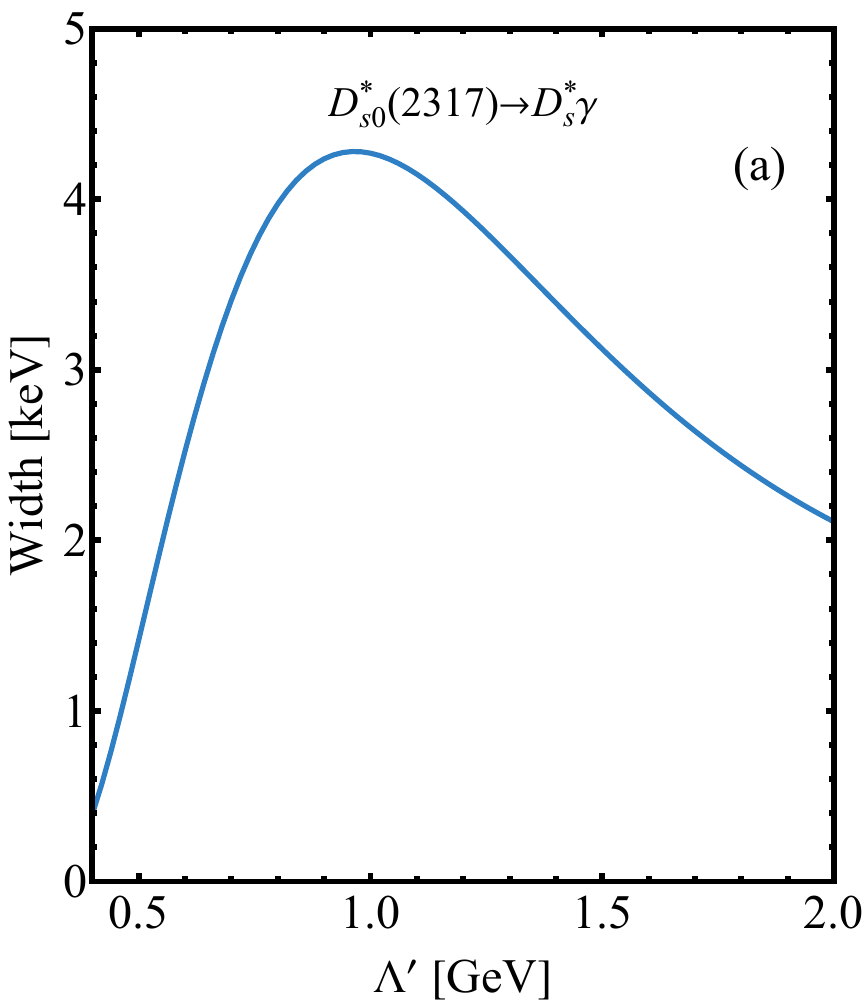}
		\includegraphics[width=0.49\linewidth]{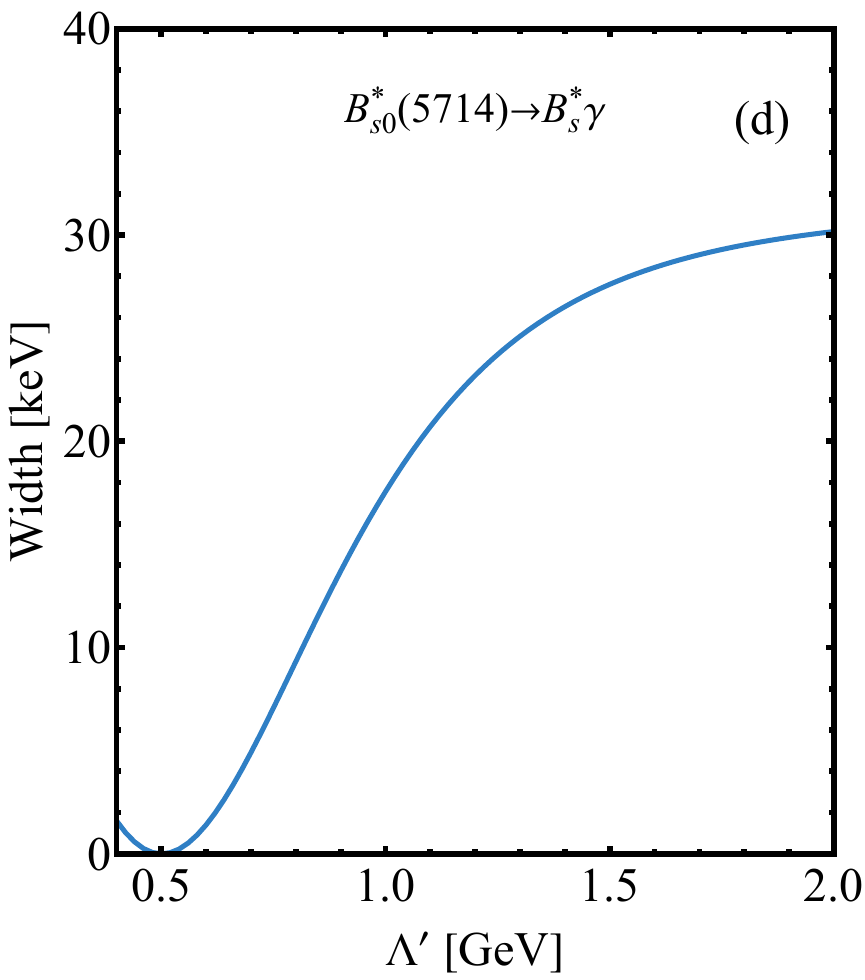}
	\includegraphics[width=0.49\linewidth]{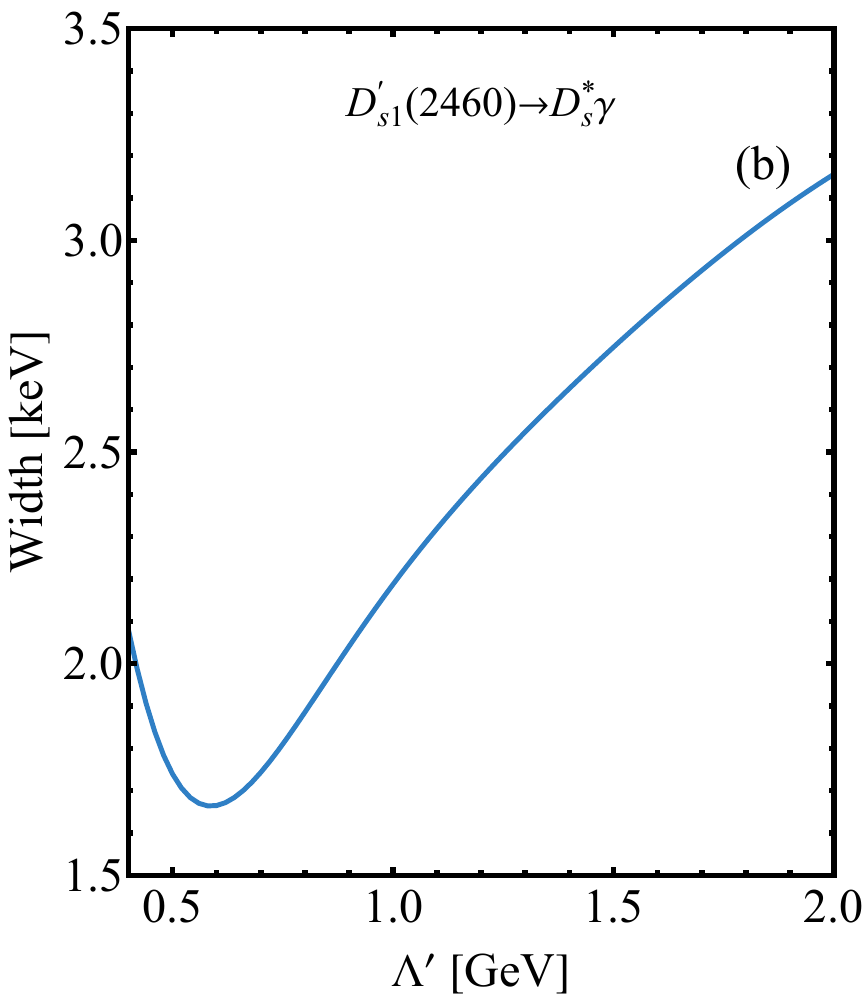}
	\includegraphics[width=0.49\linewidth]{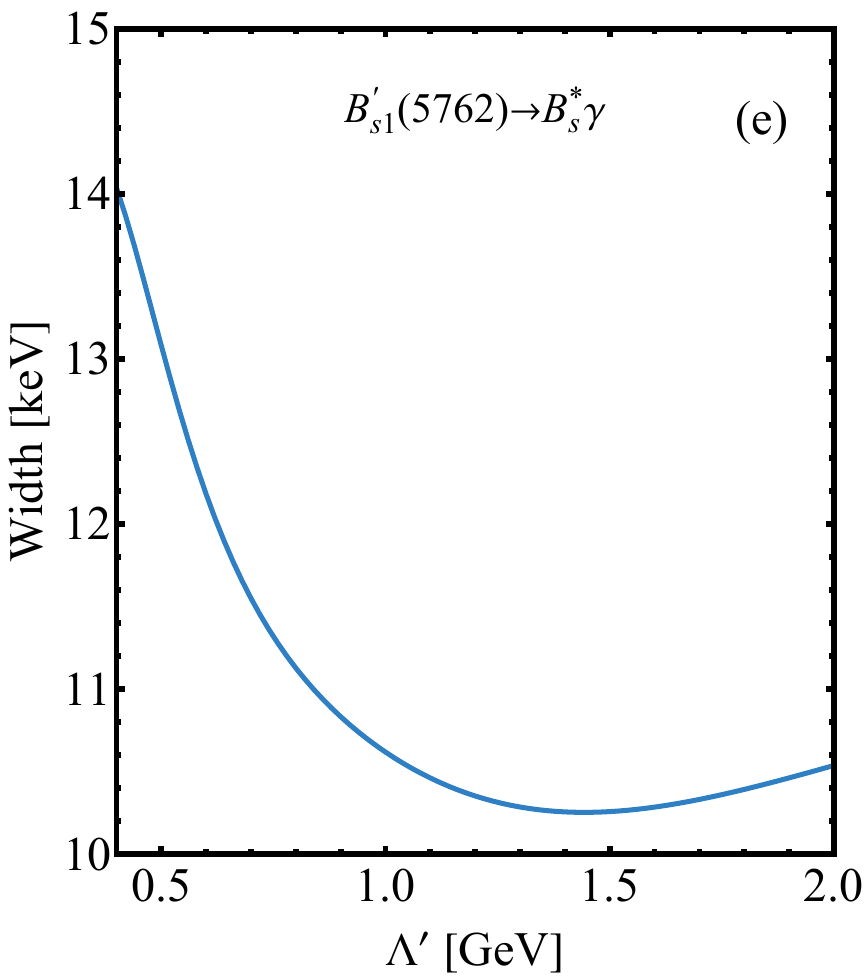}
	\includegraphics[width=0.49\linewidth]{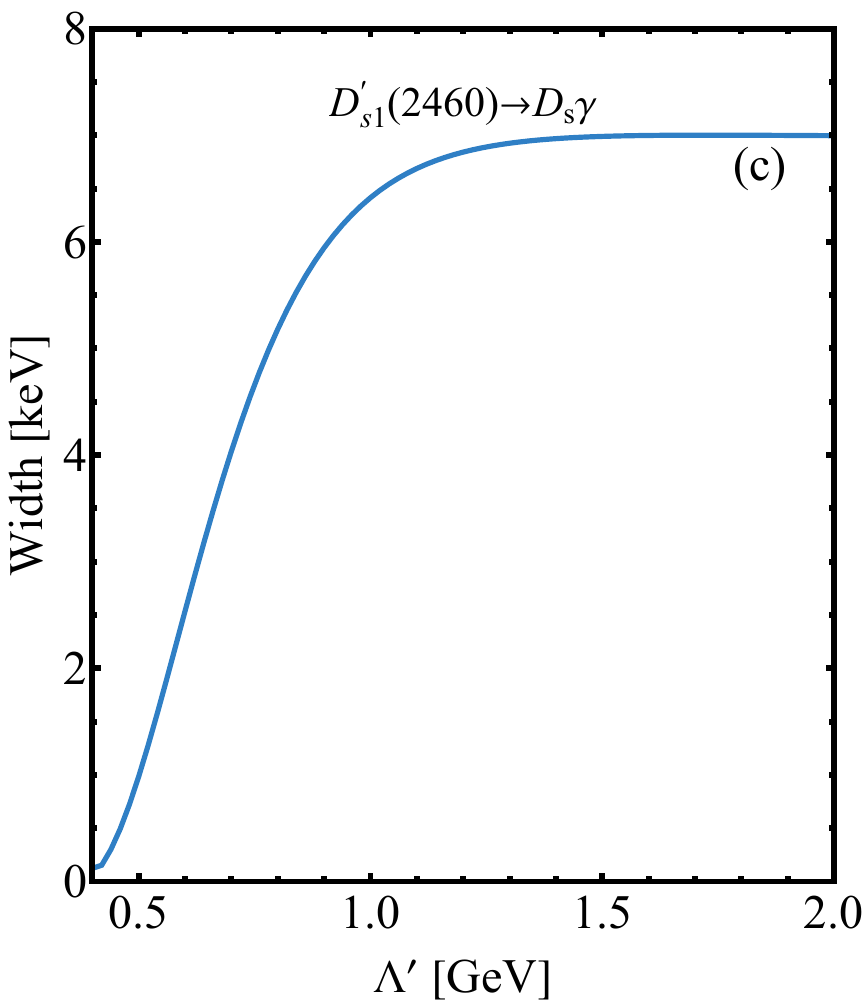}
    \includegraphics[width=0.49\linewidth]{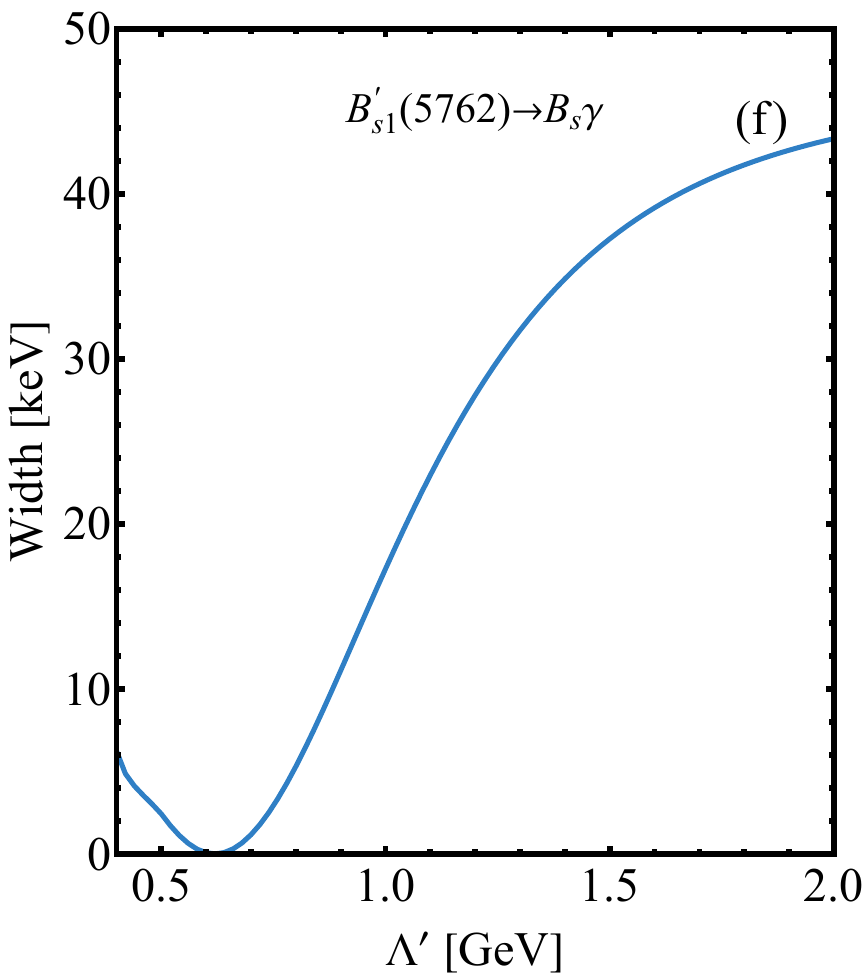}
  	\caption{ The dependence of the decay widths on the cutoff $\Lambda^{\prime}$ with physical states as the mixtures of bare cores and molecules.}\label{fig:widthfull}
  \end{figure}

With the above preparations, we can calculate the radiative decay widths of the pure molecular states and the results are shown in Fig. \ref{fig:widthpuremolecular}. By taking the cutoff $\Lambda^{\prime}$ in the range 0.4$-$2.0 GeV, the decay width of $[DK]\to D^*_s\gamma$ is about 4$-$45 keV, and those of $[D^*K]\to D^*_s\gamma$ and $[D^*K]\to D_s\gamma$ are 0.4$-$2.1 keV and 1$-$16 keV, respectively. For the bottom sector, the decay widths of $[B\bar K]\to B_s^*\gamma$, $[B^*\bar K]\to B_s^*\gamma$, and $[B^*\bar K]\to B_s\gamma$ are about 8$-$238 keV, 0.2 keV$-$1.4 keV, and 4$-$250 keV, correspondingly. The ratio between $[DK]\to D_{s}^*\gamma$ and $[BK]\to B_{s}^*\gamma$ is $\frac{[BK]\to B_{s}^*\gamma}{[DK]\to D_{s}^*\gamma}\approx 2\sim3$, which is close to the prediction of Ref.~ \cite{Faessler:2008vc}.

\subsection{The radiative decay for the physical state}

In the real situation, the physical state is a composite structure consisting of the bare quark-antiquark cores $N$ and hadron-hadron channels $[AB]$, and its radiative decay width can be then expressed as 
\begin{equation}
\begin{split}
    \Gamma_{N+[AB]\to C\gamma}^{\rm Total}=&\frac{1}{2J+1}\frac{\left|\bf p\right|}{32\pi^2m^2}\sum_M\int {\rm d}\Omega_{\textbf{p}}\\&\times\left|\mathcal{M}_{{N}\to C\gamma}+\mathcal{M}_{{[AB]}\to C\gamma}\right|^2\\
    =&\frac{1}{2J+1}\frac{\left|\bf p\right|}{32\pi^2m^2}\sum_M\int {\rm d}\Omega_{\textbf{p}}\bigg(\left|\mathcal{M}_{{N}\to C\gamma}\right|^2\\&+\left|\mathcal{M}_{{[AB]}\to C\gamma}\right|^2+\mathcal{M}_{{N}\to C\gamma}^{\dagger}\mathcal{M}_{{[AB]}\to C\gamma}+H.c.\bigg),\\
    \end{split}
\end{equation}
where $m$ is the mass of the physical state consisting of $N$ and $[AB]$. $\mathcal{M}_{{N}\to C\gamma}$ and $\mathcal{M}_{{[AB]}\to C\gamma}$ are the radiative transition amplitudes which already include the contributions from the probability amplitude or the wave function. 

In Fig. \ref{fig:widthfull}, we show the radiative decay widths with the physical states as the mixings of bare cores and hadron-hadron channels. The ratio $\Gamma_{D_{s1}^{\prime}(2460)\to D_{s}\gamma}/\Gamma_{D_{s1}^{\prime}(2460)\to D_{s}^*\gamma}$ is bigger than 2 when $\Lambda^{\prime}$ is above 0.7 GeV, which is consistent with the current experimental data \cite{ParticleDataGroup:2024cfk}. By comparing with Fig. \ref{fig:widthpuremolecular} obtained by only including the pure molecular components, we notice that the widths change significantly when the bare cores are introduced, especially for $D_{s0}^*(2317)\to D_s^*\gamma$, $D_{s1}^{\prime}(2460)\to D_s\gamma$, $B_{s0}^*(5714)\to B_s^*\gamma$, $B_{s1}^{\prime}(5762)\to B_s^*\gamma$, and $B_{s1}^{\prime}(5762)\to B_s\gamma$.

\renewcommand\tabcolsep{0.05cm}
\renewcommand\arraystretch{1.2}
\begin{table*}[htbp]
	\caption{The radiative decay widths for two cases in units of keV. Here, case \Rmnum{1} refers to results with the pure molecule assumption, and case \Rmnum{2} is for the physical states as the mixings of bare cores and molecules.}\label{table:RWCompare}
	\centering
	\begin{tabular*}{\textwidth}{@{\extracolsep{\fill}}ccccccccccccc}\toprule[1.0pt]
	%\toprule[1.0pt]
%	&\multicolumn{12}{c}{Results from our articles}\\
	\midrule[1.0pt]
	   Process&\multicolumn{2}{c}{$\Gamma[{D_{s0}^*\to D_s^*\gamma}]$}&\multicolumn{2}{c}{$\Gamma[D_{s1}^{\prime}\to D_s^*\gamma]$}&\multicolumn{2}{c}{$\Gamma[D_{s1}^{\prime}\to D_s\gamma]$}&\multicolumn{2}{c}{$\Gamma[B_{s0}^{*}\to B_s^*\gamma]$}&\multicolumn{2}{c}{$\Gamma[B_{s1}^{\prime}\to B_s^*\gamma]$}&\multicolumn{2}{c}{$\Gamma[B_{s1}^{\prime}\to B_s\gamma]$}\\
	   \Xcline{1-1}{1.0pt}
	   \Xcline{2-3}{1.0pt}
	   \Xcline{4-5}{1.0pt}
	   \Xcline{6-7}{1.0pt}
	   \Xcline{8-9}{1.0pt}
	   \Xcline{10-11}{1.0pt}
	   \Xcline{12-13}{1.0pt}
	   $\Lambda^{\prime}$ (GeV)&\Rmnum{1}&\Rmnum{2}&\Rmnum{1}&\Rmnum{2}&\Rmnum{1}&\Rmnum{2}&\Rmnum{1}&\Rmnum{2}&\Rmnum{1}&\Rmnum{2}&\Rmnum{1}&\Rmnum{2}\\\midrule[1.0pt]
	   0.5&10.79&1.41&1.65&1.74&5.58&0.98&29.37&0.01&0.68&13.08&3.5&2.42\\
	   1.0&32.06&4.27&1.60&2.19&15.36&6.41&153.69&17.56&1.31&10.62&107.83&17.27\\
	   1.5&38.97&3.12&1.17&2.75&16.26&6.99&216.48&27.61&1.37&10.26&210.77&37.27\\\midrule[1.0pt]\midrule[1.0pt]
	\end{tabular*}
\end{table*}

To highlight the role of the bare quark-antiquark cores, we discuss two cases here: case \Rmnum{1} with the pure molecule assumption and case \Rmnum{2} with the mixture picture. We summarize the numerical results for these two cases in Table \ref{table:RWCompare}. For $D_{s0}^*(2317)\to D_s^*\gamma$, the width in the case~\Rmnum{1} is about eight times more than that of case \Rmnum{2}. In the $D_{s1}^{\prime}(2460)\to D_s\gamma$ process, there is about a threefold change. In the $D_{s1}^{\prime}(2460)\to D_s^*\gamma$ process, the radiative decay width varies little with $\Lambda^{\prime}$ for both cases. In addition, the widths for the bottom sector change dozens of times with the introduction of bare $\bar{b}s$ cores, which indicates that the bore core components have a great influence.  

Experimental data on these processes are still lacking, and different theoretical predictions with parity doubling model \cite{Bardeen:2003kt}, effective Lagrangian approach \cite{Faessler:2007gv,Faessler:2008vc,Fu:2021wde}, light-cone sum rules \cite{Colangelo:2005hv,Wang:2008wz,Wang:2006mf}, and so on, vary in a very wide range. Ref. \cite{Wang:2019ehs} treated $B_{s1}^{\prime}$ state as a molecular state and provided the widths of $B_{s1}^{\prime}\to B_s^*\gamma$ and $B_{s1}^{\prime}\to B_s\gamma$ as 0.4$-$2.6 and 45.2$-$79.8 keV, respectively, which agree relatively well with our results in case \Rmnum{1}.

The difference between different models is important for us to understand the inner structure of the $D_{s0}^{*}(2317)$ and $D_{s1}^{\prime}(2460)$ states and the predictions of their bottom analogs. Thus, we expect that the future experimental measurements on the radiative decays of $D_{s0}^{*}(2317)\to D_s^{*}\gamma$ and $D_{s1}^{\prime}(2460)\to D_s^{(*)}\gamma$, and their bottom-strange partners, are crucial for elucidating their inner structures.

\section{Summary}\label{sec6}

In this work, we systematically study the $D_{s0}^*(2317)$, $D_{s1}^{\prime}(2460)$, and their bottom analogs within a sophisticated theoretical framework. We take the unquenched effect into account to investigate the low mass problem of $D_{s0}^*(2317)$ and $D_{s1}^{\prime}(2460)$. In our framework, the experiment masses of $D_{s0}^*(2317)$ and $D_{s1}^{\prime}(2460)$ can be precisely reproduced, which indicates the unquenched effect plays an important role in the establishment of $D_{s0}^*(2317)$ and $D_{s1}^{\prime}(2460)$.

In addition to the coupled channel effect between the bare $c\bar{s}$ core and $D^{(*)}K$ channel, we also consider the $D^{(*)}K$-$D^{(*)}K$ interaction. We notice that the further introduction of the latter interaction can cause a visible mass shift that can better accommodate the experimental measurements. Our results show that $D_{s0}^*(2317)$ and $D_{s1}^{\prime}(2460)$ are the mixtures of bare $c\bar{s}$ cores and $D^{(*)}K$ components, while the latter have more weights in the compositions. Furthermore, we predict the spectra of their bottom analogs using the same coupling constants and cutoff parameters. The predictions of their mass are $M_{B_{s0}^*}=5714$ MeV and $M_{B_{s1}^{\prime}}=5762$ MeV, respectively. They are the mixture of $\bar{b}s$ and $B^{(*)}\bar{K}$.  

Due to the introduction of bare state core, the hadron-hadron channel binds tighter than the pure hadron-hadron system, which will affect their other properties. To decode the inner structure of physical states, we systematically investigate the radiative decay of $D_{s0}^*(2317)\to D_s^*\gamma$, $D_{s1}^{\prime}(2460) \to D_s^*\gamma$, $D_{s1}^{\prime}(2460) \to D_s\gamma$, $B_{s0}^{*}(5714) \to B_s^*\gamma$, $B_{s1}^{\prime}(5762) \to B_s^*\gamma$, and $B_{s1}^{\prime}(5762) \to B_s\gamma$ processes. Different components in the compositions give clearly different contributions to the radiative decay widths. Therefore we expect that future  experimental measurements of these radiative decay processes will be definitely helpful to reveal their underlying structures. 

Our exploration of the $D_{s0}^*(2317)$, $D_{s1}^{\prime}(2460)$, and their bottom analogs states provides a more profound understanding of the heavy-light meson system and nonperturbative QCD. We expect that our results can be further examined by future experiments. 

\section*{ACKNOWLEDGMENTS}
This project is partially supported by the National Natural Science Foundation of China under Grants No. 12175091, No. 12335001, No. 12247101, No. 12405098, No. 11975090, the 111 Project under Grant No. B20063, the innovation project for young science and technology
talents of Lanzhou city under Grant No. 2023-QN-107, and the Science Foundation of Hebei Normal University with contract No.~L2023B09.


\begin{thebibliography}{99}

%\cite{Chen:2016spr}
\bibitem{Chen:2016spr}
H.~X.~Chen, W.~Chen, X.~Liu, Y.~R.~Liu, and S.~L.~Zhu,
A review of the open charm and open bottom systems,
Rep. Prog. Phys. \textbf{80}, 076201 (2017).
%doi:10.1088/1361-6633/aa6420
%[arXiv:1609.08928 [hep-ph]].
%317 citations counted in INSPIRE as of 26 Jun 2023

%\cite{Chen:2022asf}
\bibitem{Chen:2022asf}
H.~X.~Chen, W.~Chen, X.~Liu, Y.~R.~Liu, and S.~L.~Zhu,
An updated review of the new hadron states,
Rep. Prog. Phys. \textbf{86}, 026201 (2023).
%doi:10.1088/1361-6633/aca3b6
%[arXiv:2204.02649 [hep-ph]].
%162 citations counted in INSPIRE as of 26 Jun 2023

%\cite{Meng:2022ozq}
\bibitem{Meng:2022ozq}
L.~Meng, B.~Wang, G.~J.~Wang, and S.~L.~Zhu,
Chiral perturbation theory for heavy hadrons and chiral effective field theory for heavy hadronic molecules,
Phys. Rep. \textbf{1019}, 1 (2023).
%doi:10.1016/j.physrep.2023.04.003
%[arXiv:2204.08716 [hep-ph]].
%62 citations counted in INSPIRE as of 26 Jun 2023

%\cite{Dong:2021juy}
\bibitem{Dong:2021juy}
X.~K.~Dong, F.~K.~Guo, and B.~S.~Zou,
A survey of heavy-antiheavy hadronic molecules,
Progr. Phys. \textbf{41}, 65 (2021).
%doi:10.13725/j.cnki.pip.2021.02.001
%[arXiv:2101.01021 [hep-ph]].
%144 citations counted in INSPIRE as of 05 Aug 2024

%\cite{Liu:2024uxn}
\bibitem{Liu:2024uxn}
M.~Z.~Liu, Y.~W.~Pan, Z.~W.~Liu, T.~W.~Wu, J.~X.~Lu, and L.~S.~Geng,
Three ways to decipher the nature of exotic hadrons: Multiplets, three-body hadronic molecules, and correlation functions,
arXiv:2404.06399.
%0 citations counted in INSPIRE as of 10 Apr 2024

%\cite{BaBar:2003oey}
\bibitem{BaBar:2003oey}
B.~Aubert \textit{et al.} ({\it BABAR }Collaboration),
Observation of a narrow meson decaying to $D_s^+ \pi^0$ at a mass of 2.32 GeV/c$^2$,
Phys. Rev. Lett. \textbf{90}, 242001 (2003).
%doi:10.1103/PhysRevLett.90.242001
%[arXiv:hep-ex/0304021 [hep-ex]].
%974 citations counted in INSPIRE as of 20 Jun 2023

%\cite{CLEO:2003ggt}
\bibitem{CLEO:2003ggt}
D.~Besson \textit{et al.} (CLEO Collaboration),
Observation of a narrow resonance of mass $2.46$~GeV/$c^2$ decaying to $D_s^+\pi$ and confirmation of the $D_{SJ}^*(2317)$ state,
Phys. Rev. D \textbf{68}, 032002 (2003); \textbf{75}, 119908(E) (2007).
%doi:10.1103/PhysRevD.68.032002
%[arXiv:hep-ex/0305100 [hep-ex]].
%692 citations counted in INSPIRE as of 20 Jun 2023

%\cite{Belle:2003kup}
\bibitem{Belle:2003kup}
Y.~Mikami \textit{et al.} (Belle Collaboration),
Measurements of the $D_{sJ}$ resonance properties,
Phys. Rev. Lett. \textbf{92}, 012002 (2004).
%doi:10.1103/PhysRevLett.92.012002
%[arXiv:hep-ex/0307052 [hep-ex]].
%294 citations counted in INSPIRE as of 21 Mar 2024

%\cite{Belle:2003guh}
\bibitem{Belle:2003guh}
P.~Krokovny \textit{et al.} (Belle Collaboration),
Observation of the $D_{sJ}(2317)$ and $D_{sJ}(2457)$ in $B$ decays,
Phys. Rev. Lett. \textbf{91}, 262002 (2003).
%doi:10.1103/PhysRevLett.91.262002
%[arXiv:hep-ex/0308019 [hep-ex]].
%393 citations counted in INSPIRE as of 21 Mar 2024

%\cite{BaBar:2004yux}
\bibitem{BaBar:2004yux}
B.~Aubert \textit{et al.} ({\it BABAR} Collaboration),
Study of $B \to D_{sJ}^{(*)+} \bar{D}^{(*)}$ decays,
Phys. Rev. Lett. \textbf{93}, 181801 (2004).
%doi:10.1103/PhysRevLett.93.181801
%[arXiv:hep-ex/0408041 [hep-ex]].
%102 citations counted in INSPIRE as of 21 Mar 2024

%\cite{BaBar:2003cdx}
\bibitem{BaBar:2003cdx}
B.~Aubert \textit{et al.} ({\it BABAR} Collaboration),
Observation of a narrow meson decaying to $D_s^+ \pi^0 \gamma$ at a mass of 2.458 GeV/c$^2$,
Phys. Rev. D \textbf{69}, 031101 (2004).
%doi:10.1103/PhysRevD.69.031101
%[arXiv:hep-ex/0310050 [hep-ex]].
%159 citations counted in INSPIRE as of 21 Mar 2024

%\cite{BaBar:2006eep}
\bibitem{BaBar:2006eep}
B.~Aubert \textit{et al.} ({\it BABAR} Collaboration),
A study of the $D^*_{sJ}(2317)^+$ and $D_{sJ}(2460)^+$ mesons in inclusive $c\bar{c}$ production near $\sqrt{s}= 10.6$ GeV,
Phys. Rev. D \textbf{74}, 032007 (2006).
%doi:10.1103/PhysRevD.74.032007
%[arXiv:hep-ex/0604030 [hep-ex]].
%98 citations counted in INSPIRE as of 21 Mar 2024

%\cite{ParticleDataGroup:2024cfk}
\bibitem{ParticleDataGroup:2024cfk}
S.~Navas \textit{et al.} (Particle Data Group),
Review of particle physics,
Phys. Rev. D \textbf{110}, 030001 (2024).
%doi:10.1103/PhysRevD.110.030001
%87 citations counted in INSPIRE as of 05 Sep 2024

%\cite{Godfrey:1985xj}
\bibitem{Godfrey:1985xj}
S.~Godfrey and N.~Isgur,
Mesons in a relativized quark model with chromodynamics,
Phys. Rev. D \textbf{32}, 189 (1985).
%doi:10.1103/PhysRevD.32.189
%3124 citations counted in INSPIRE as of 08 Jul 2023

%\cite{Godfrey:1986wj}
\bibitem{Godfrey:1986wj}
S.~Godfrey and R.~Kokoski,
The properties of $P$ wave mesons with one heavy quark,
Phys. Rev. D \textbf{43}, 1679 (1991).
%doi:10.1103/PhysRevD.43.1679
%499 citations counted in INSPIRE as of 26 Jun 2023

%\cite{Godfrey:2015dva}
\bibitem{Godfrey:2015dva}
S.~Godfrey and K.~Moats,
Properties of excited charm and charm-strange mesons,
Phys. Rev. D \textbf{93}, 034035 (2016).
%doi:10.1103/PhysRevD.93.034035
%[arXiv:1510.08305 [hep-ph]].
%145 citations counted in INSPIRE as of 26 Jun 2023

%\cite{Ebert:2009ua}
\bibitem{Ebert:2009ua}
D.~Ebert, R.~N.~Faustov, and V.~O.~Galkin,
Heavy-light meson spectroscopy and Regge trajectories in the relativistic quark model,
Eur. Phys. J. C \textbf{66}, 197 (2010).
%doi:10.1140/epjc/s10052-010-1233-6
%[arXiv:0910.5612 [hep-ph]].
%237 citations counted in INSPIRE as of 20 Jun 2023

%\cite{vanBeveren:2003kd}
\bibitem{vanBeveren:2003kd}
E.~van Beveren and G.~Rupp,
Observed $D_s(2317)$ and tentative $D(2100\text{--}2300)$ as the
charmed cousins of the light scalar nonet,
Phys. Rev. Lett. \textbf{91}, 012003 (2003).
%doi:10.1103/PhysRevLett.91.012003
%[arXiv:hep-ph/0305035 [hep-ph]].
%375 citations counted in INSPIRE as of 26 Jun 2023

%\cite{Kalashnikova:2005ui}
\bibitem{Kalashnikova:2005ui}
Y.~S.~Kalashnikova,
Coupled-channel model for charmonium levels and an option for $X(3872)$,
Phys. Rev. D \textbf{72}, 034010 (2005).
%doi:10.1103/PhysRevD.72.034010
%[arXiv:hep-ph/0506270 [hep-ph]].
%182 citations counted in INSPIRE as of 01 Jul 2023

%\cite{Silvestre-Brac:1991qqx}
\bibitem{Silvestre-Brac:1991qqx}
B.~Silvestre- Brac and C.~Gignoux,
Unitary effects in spin orbit splitting of $P$ wave baryons,
Phys. Rev. D \textbf{43}, 3699 (1991).
%doi:10.1103/PhysRevD.43.3699
%47 citations counted in INSPIRE as of 01 Jul 2023

%\cite{Li:2009ad}
\bibitem{Li:2009ad}
B.~Q.~Li, C.~Meng, and K.~T.~Chao,
Coupled-channel and screening effects in charmonium spectrum,
Phys. Rev. D \textbf{80}, 014012 (2009).
%doi:10.1103/PhysRevD.80.014012
%[arXiv:0904.4068 [hep-ph]].
%116 citations counted in INSPIRE as of 01 Jul 2023

%\cite{Danilkin:2010cc}
\bibitem{Danilkin:2010cc}
I.~V.~Danilkin and Y.~A.~Simonov,
Dynamical origin and the pole structure of $X(3872)$,
Phys. Rev. Lett. \textbf{105}, 102002 (2010).
%doi:10.1103/PhysRevLett.105.102002
%[arXiv:1006.0211 [hep-ph]].
%92 citations counted in INSPIRE as of 01 Jul 2023

%\cite{Luo:2019qkm}
\bibitem{Luo:2019qkm}
S.~Q.~Luo, B.~Chen, Z.~W.~Liu, and X.~Liu,
Resolving the low mass puzzle of $\Lambda_c(2940)^+$,
Eur. Phys. J. C \textbf{80}, 301 (2020).
%doi:10.1140/epjc/s10052-020-7874-1
%[arXiv:1910.14545 [hep-ph]].
%29 citations counted in INSPIRE as of 05 Aug 2024

%\cite{Song:2015nia}
\bibitem{Song:2015nia}
Q.~T.~Song, D.~Y.~Chen, X.~Liu, and T.~Matsuki,
Charmed-strange mesons revisited: Mass spectra and strong decays,
Phys. Rev. D \textbf{91}, 054031 (2015).
%doi:10.1103/PhysRevD.91.054031
%[arXiv:1501.03575 [hep-ph]].
%69 citations counted in INSPIRE as of 26 Jun 2023

%\cite{Gao:2022bsb}
\bibitem{Gao:2022bsb}
Z.~Gao, G.~Y.~Wang, Q.~F.~L\"u, J.~Zhu, and G.~F.~Zhao,
Canonical interpretation of the $D_{s0}^+(2590)$ resonance,
Phys. Rev. D \textbf{105}, 074037 (2022).
%doi:10.1103/PhysRevD.105.074037
%[arXiv:2201.00552 [hep-ph]].
%7 citations counted in INSPIRE as of 26 Jun 2023

%\cite{Cheng:2003kg}
\bibitem{Cheng:2003kg}
H.~Y.~Cheng and W.~S.~Hou,
$B$ decays as spectroscope for charmed four quark states,
Phys. Lett. B \textbf{566}, 193 (2003).
%doi:10.1016/S0370-2693(03)00834-7
%[arXiv:hep-ph/0305038 [hep-ph]].
%232 citations counted in INSPIRE as of 27 Jun 2023

%\cite{Chen:2004dy}
\bibitem{Chen:2004dy}
Y.~Q.~Chen and X.~Q.~Li,
A comprehensive four-quark interpretation of $D_s(2317)$, $D_s(2457)$ and $D_s(2632)$,
Phys. Rev. Lett. \textbf{93}, 232001 (2004).
%doi:10.1103/PhysRevLett.93.232001
%[arXiv:hep-ph/0407062 [hep-ph]].
%129 citations counted in INSPIRE as of 27 Jun 2023

%\cite{Kim:2005gt}
\bibitem{Kim:2005gt}
H.~Kim and Y.~Oh,
$D_s(2317)$ as a four-quark state in QCD sum rules,
Phys. Rev. D \textbf{72}, 074012 (2005).
%doi:10.1103/PhysRevD.72.074012
%[arXiv:hep-ph/0508251 [hep-ph]].
%55 citations counted in INSPIRE as of 27 Jun 2023

%\cite{Dmitrasinovic:2005gc}
\bibitem{Dmitrasinovic:2005gc}
V.~Dmitrasinovic,
$D_{s0}^+(2317)-D_0(2308)$ mass difference as evidence for tetraquarks,
Phys. Rev. Lett. \textbf{94}, 162002 (2005).
%doi:10.1103/PhysRevLett.94.162002
%80 citations counted in INSPIRE as of 17 Jun 2023

%\cite{Nielsen:2005ia}
\bibitem{Nielsen:2005ia}
M.~Nielsen, R.~D.~Matheus, F.~S.~Navarra, M.~E.~Bracco, and A.~Lozea,
Diquark-antidiquark with open charm in QCD sum rules,
Nucl. Phys. B Proc. Suppl. \textbf{161}, 193 (2006).
%doi:10.1016/j.nuclphysbps.2006.08.045
%[arXiv:hep-ph/0509131 [hep-ph]].
%25 citations counted in INSPIRE as of 27 Jun 2023

%\cite{Maiani:2004vq}
\bibitem{Maiani:2004vq}
L.~Maiani, F.~Piccinini, A.~D.~Polosa, and V.~Riquer,
Diquark-antidiquarks with hidden or open charm and the nature of $X(3872)$,
Phys. Rev. D \textbf{71}, 014028 (2005).
%doi:10.1103/PhysRevD.71.014028
%[arXiv:hep-ph/0412098 [hep-ph]].
%855 citations counted in INSPIRE as of 18 Jun 2023

%\cite{Wang:2006uba}
\bibitem{Wang:2006uba}
Z.~G.~Wang and S.~L.~Wan,
$D_{s0}(2317)$ as a tetraquark state with QCD sum rules in heavy quark limit,
Nucl. Phys. \textbf{A778}, 22 (2006).
%doi:10.1016/j.nuclphysa.2006.07.041
%[arXiv:hep-ph/0602080 [hep-ph]].
%40 citations counted in INSPIRE as of 18 Jun 2023

%\cite{Vijande:2006hj}
\bibitem{Vijande:2006hj}
J.~Vijande, F.~Fernandez, and A.~Valcarce,
Open-charm meson spectroscopy,
Phys. Rev. D \textbf{73}, 034002 (2006); \textbf{74}, 059903(E) (2006).
%doi:10.1103/PhysRevD.73.034002
%[arXiv:hep-ph/0601143 [hep-ph]].
%74 citations counted in INSPIRE as of 22 Mar 2024

%\cite{Kolomeitsev:2003ac}
\bibitem{Kolomeitsev:2003ac}
E.~E.~Kolomeitsev and M.~F.~M.~Lutz,
On heavy light meson resonances and chiral symmetry,
Phys. Lett. B \textbf{582}, 39 (2004).
%doi:10.1016/j.physletb.2003.10.118
%[arXiv:hep-ph/0307133 [hep-ph]].
%295 citations counted in INSPIRE as of 26 Jun 2023

%\cite{Szczepaniak:2003vy}
\bibitem{Szczepaniak:2003vy}
A.~P.~Szczepaniak,
Description of the $D_s^*(2320)$ resonance as the $D\pi$ atom,
Phys. Lett. B \textbf{567}, 23 (2003).
%doi:10.1016/S0370-2693(03)00865-7
%[arXiv:hep-ph/0305060 [hep-ph]].
%169 citations counted in INSPIRE as of 26 Jun 2023

%\cite{Barnes:2003dj}
\bibitem{Barnes:2003dj}
T.~Barnes, F.~E.~Close, and H.~J.~Lipkin,
Implications of a $DK$ molecule at 2.32 GeV,
Phys. Rev. D \textbf{68}, 054006 (2003).
%doi:10.1103/PhysRevD.68.054006
%[arXiv:hep-ph/0305025 [hep-ph]].
%417 citations counted in INSPIRE as of 26 Jun 2023

%\cite{Hofmann:2003je}
\bibitem{Hofmann:2003je}
J.~Hofmann and M.~F.~M.~Lutz,
Open charm meson resonances with negative strangeness,
Nucl. Phys. \textbf{A733}, 142 (2004).
%doi:10.1016/j.nuclphysa.2003.12.013
%[arXiv:hep-ph/0308263 [hep-ph]].
%164 citations counted in INSPIRE as of 26 Jun 2023

%\cite{Gamermann:2006nm}
\bibitem{Gamermann:2006nm}
D.~Gamermann, E.~Oset, D.~Strottman, and M.~J.~Vicente Vacas,
Dynamically generated open and hidden charm meson systems,
Phys. Rev. D \textbf{76}, 074016 (2007).
%doi:10.1103/PhysRevD.76.074016
%[arXiv:hep-ph/0612179 [hep-ph]].
%307 citations counted in INSPIRE as of 26 Jun 2023

%\cite{Guo:2006fu}
\bibitem{Guo:2006fu}
F.~K.~Guo, P.~N.~Shen, H.~C.~Chiang, R.~G.~Ping, and B.~S.~Zou,
Dynamically generated $0^+$ heavy mesons in a heavy chiral unitary approach,
Phys. Lett. B \textbf{641}, 278 (2006).
%doi:10.1016/j.physletb.2006.08.064
%[arXiv:hep-ph/0603072 [hep-ph]].
%258 citations counted in INSPIRE as of 26 Jun 2023

%\cite{Guo:2006rp}
\bibitem{Guo:2006rp}
F.~K.~Guo, P.~N.~Shen, and H.~C.~Chiang,
Dynamically generated $1^+$ heavy mesons,
Phys. Lett. B \textbf{647}, 133 (2007).
%doi:10.1016/j.physletb.2007.01.050
%[arXiv:hep-ph/0610008 [hep-ph]].
%142 citations counted in INSPIRE as of 26 Jun 2023

%\cite{Flynn:2007ki}
\bibitem{Flynn:2007ki}
J.~M.~Flynn and J.~Nieves,
Elastic s-wave $B\pi$, $D\pi$, $DK$ and $K\pi$ scattering from lattice calculations of scalar form-factors in semileptonic decays,
Phys. Rev. D \textbf{75}, 074024 (2007).
%doi:10.1103/PhysRevD.75.074024
%[arXiv:hep-ph/0703047 [hep-ph]].
%57 citations counted in INSPIRE as of 26 Jun 2023

%\cite{Faessler:2007gv}
\bibitem{Faessler:2007gv}
A.~Faessler, T.~Gutsche, V.~E.~Lyubovitskij, and Y.~L.~Ma,
Strong and radiative decays of the $D_{s0}^*(2317)$ meson in the $DK$-molecule picture,
Phys. Rev. D \textbf{76}, 014005 (2007).
%doi:10.1103/PhysRevD.76.014005
%[arXiv:0705.0254 [hep-ph]].
%197 citations counted in INSPIRE as of 27 Jun 2023

%\cite{Guo:2009ct}
\bibitem{Guo:2009ct}
F.~K.~Guo, C.~Hanhart, and U.~G.~Meissner,
Interactions between heavy mesons and Goldstone bosons from chiral dynamics,
Eur. Phys. J. A \textbf{40}, 171 (2009).
%doi:10.1140/epja/i2009-10762-1
%[arXiv:0901.1597 [hep-ph]].
%145 citations counted in INSPIRE as of 27 Jun 2023

%\cite{Xie:2010zza}
\bibitem{Xie:2010zza}
Z.~X.~Xie, G.~Q.~Feng, and X.~H.~Guo,
Analyzing $D_{s0}^*(2317)$ in the $DK$ molecule picture in the Beth-Salpeter approach,
Phys. Rev. D \textbf{81}, 036014 (2010).
%doi:10.1103/PhysRevD.81.036014
%26 citations counted in INSPIRE as of 27 Jun 2023

%\cite{Cleven:2010aw}
\bibitem{Cleven:2010aw}
M.~Cleven, F.~K.~Guo, C.~Hanhart, and U.~G.~Meissner,
Light meson mass dependence of the positive parity heavy-strange mesons,
Eur. Phys. J. A \textbf{47}, 19 (2011).
%doi:10.1140/epja/i2011-11019-2
%[arXiv:1009.3804 [hep-ph]].
%75 citations counted in INSPIRE as of 27 Jun 2023

%\cite{Wu:2011yb}
\bibitem{Wu:2011yb}
X.~G.~Wu and Q.~Zhao,
The mixing of $D_{s1}(2460)$ and $D_{s1}(2536)$,
Phys. Rev. D \textbf{85}, 034040 (2012).
%doi:10.1103/PhysRevD.85.034040
%[arXiv:1111.4002 [hep-ph]].
%12 citations counted in INSPIRE as of 27 Jun 2023

%\cite{Guo:2015dha}
\bibitem{Guo:2015dha}
Z.~H.~Guo, U.~G.~Mei\ss{}ner, and D.~L.~Yao,
New insights into the $D^{*}_{s0}(2317)$ and other charm scalar mesons,
Phys. Rev. D \textbf{92}, 094008 (2015).
%doi:10.1103/PhysRevD.92.094008
%[arXiv:1507.03123 [hep-ph]].
%61 citations counted in INSPIRE as of 27 Jun 2023

%\cite{Albaladejo:2016hae}
\bibitem{Albaladejo:2016hae}
M.~Albaladejo, D.~Jido, J.~Nieves, and E.~Oset,
$D^*_{s0}(2317)$ and $\textit{DK}$ scattering in $B$ decays from $BaBar$ and LHCb data,
Eur. Phys. J. C \textbf{76}, 300 (2016).
%doi:10.1140/epjc/s10052-016-4144-3
%[arXiv:1604.01193 [hep-ph]].
%26 citations counted in INSPIRE as of 27 Jun 2023

%\cite{Du:2017ttu}
\bibitem{Du:2017ttu}
M.~L.~Du, F.~K.~Guo, U.~G.~Mei\ss{}ner, and D.~L.~Yao,
Study of open-charm $0^+$ states in unitarized chiral effective theory with one-loop potentials,
Eur. Phys. J. C \textbf{77}, 728 (2017).
%doi:10.1140/epjc/s10052-017-5287-6
%[arXiv:1703.10836 [hep-ph]].
%43 citations counted in INSPIRE as of 27 Jun 2023

%\cite{Guo:2018tjx}
\bibitem{Guo:2018tjx}
Z.~H.~Guo, L.~Liu, U.~G.~Mei\ss{}ner, J.~A.~Oller, and A.~Rusetsky,
Towards a precise determination of the scattering amplitudes of the charmed and light-flavor pseudoscalar mesons,
Eur. Phys. J. C \textbf{79}, 13 (2019).
%doi:10.1140/epjc/s10052-018-6518-1
%[arXiv:1811.05585 [hep-ph]].
%41 citations counted in INSPIRE as of 27 Jun 2023

%\cite{Albaladejo:2018mhb}
\bibitem{Albaladejo:2018mhb}
M.~Albaladejo, P.~Fernandez-Soler, J.~Nieves, and P.~G.~Ortega,
Contribution of constituent quark model $c\bar{s}$ states to the dynamics of the $D_{s0}^*(2317)$ and $D_{s1}(2460)$ resonances,
Eur. Phys. J. C \textbf{78}, 722 (2018).
%doi:10.1140/epjc/s10052-018-6176-3
%[arXiv:1805.07104 [hep-ph]].
%39 citations counted in INSPIRE as of 27 Jun 2023

%\cite{Wu:2019vsy}
\bibitem{Wu:2019vsy}
T.~W.~Wu, M.~Z.~Liu, L.~S.~Geng, E.~Hiyama, and M.~P.~Valderrama,
$DK$, $DDK$, and $DDDK$ molecules\textendash{}understanding the nature of the $D_{s0}^*(2317)$,
Phys. Rev. D \textbf{100}, 034029 (2019).
%doi:10.1103/PhysRevD.100.034029
%[arXiv:1906.11995 [hep-ph]].
%37 citations counted in INSPIRE as of 27 Jun 2023

%\cite{Kong:2021ohg}
\bibitem{Kong:2021ohg}
S.~Y.~Kong, J.~T.~Zhu, D.~Song, and J.~He,
Heavy-strange meson molecules and possible candidates $D_{s0}^*(2317)$, $D_{s1}(2460)$, and $X_0(2900)$,
Phys. Rev. D \textbf{104}, 094012 (2021).
%doi:10.1103/PhysRevD.104.094012
%[arXiv:2106.07272 [hep-ph]].
%22 citations counted in INSPIRE as of 27 Jun 2023

%\cite{Wang:2012bu}
\bibitem{Wang:2012bu}
P.~Wang and X.~G.~Wang,
Study on $0^+$ states with open charm in unitarized heavy meson chiral approach,
Phys. Rev. D \textbf{86}, 014030 (2012).
%doi:10.1103/PhysRevD.86.014030
%[arXiv:1204.5553 [hep-ph]].
%34 citations counted in INSPIRE as of 27 Jun 2023

%\cite{Liu:2022dmm}
\bibitem{Liu:2022dmm}
M.~Z.~Liu, X.~Z.~Ling, L.~S.~Geng, En-Wang, and J.~J.~Xie,
Production of $D_{s0}^*(2317)$ and $D_{s1}(2460)$ in $B$ decays as $D^{(*)}K$ and $D_s^*$\ensuremath{\eta} molecules,
Phys. Rev. D \textbf{106}, 114011 (2022).
%doi:10.1103/PhysRevD.106.114011
%[arXiv:2209.01103 [hep-ph]].
%9 citations counted in INSPIRE as of 27 Jun 2023

%\cite{Huang:2021fdt}
\bibitem{Huang:2021fdt}
B.~L.~Huang, Z.~Y.~Lin, and S.~L.~Zhu,
Light pseudoscalar meson and heavy meson scattering lengths to $\mathcal{O}(p^4)$ in heavy meson chiral perturbation theory,
Phys. Rev. D \textbf{105}, 036016 (2022).
%doi:10.1103/PhysRevD.105.036016
%[arXiv:2112.13702 [hep-ph]].
%7 citations counted in INSPIRE as of 27 Jun 2023

%\cite{Liu:2011mi}
\bibitem{Liu:2011mi}
Z.~W.~Liu, Y.~R.~Liu, X.~Liu, and S.~L.~Zhu,
The pseudoscalar meson and heavy vector meson scattering lengths,
Phys. Rev. D \textbf{84}, 034002 (2011).
%doi:10.1103/PhysRevD.84.034002
%[arXiv:1104.2726 [hep-ph]].
%16 citations counted in INSPIRE as of 29 Mar 2024

%\cite{Wu:2022wgn}
\bibitem{Wu:2022wgn}
T.~C.~Wu and L.~S.~Geng,
Theoretical investigation of the molecular nature of $D_{s0}^*(2317)$ and $D_{s1}(2460)$ and the possibility of observing the $D\bar{D}K$ bound state $K_{c\bar{c}}(4180)$ in inclusive $e^+e^-\to c\bar{c}$ collisions,
Phys. Rev. D \textbf{108}, 014015 (2023).
%doi:10.1103/PhysRevD.108.014015
%[arXiv:2211.01846 [hep-ph]].
%7 citations counted in INSPIRE as of 09 Aug 2024

%\cite{Liu:2023uly}
\bibitem{Liu:2023uly}
Z.~W.~Liu, J.~X.~Lu, and L.~S.~Geng,
Study of the $DK$ interaction with femtoscopic correlation functions,
Phys. Rev. D \textbf{107}, 074019 (2023).
%doi:10.1103/PhysRevD.107.074019
%[arXiv:2302.01046 [hep-ph]].
%6 citations counted in INSPIRE as of 27 Jun 2023

%\cite{Ikeno:2023ojl}
\bibitem{Ikeno:2023ojl}
N.~Ikeno, G.~Toledo, and E.~Oset,
Model independent analysis of femtoscopic correlation functions: An application to the $D_{s0}^*(2317)$,
Phys. Lett. B \textbf{847}, 138281 (2023).
%doi:10.1016/j.physletb.2023.138281
%[arXiv:2305.16431 [hep-ph]].
%14 citations counted in INSPIRE as of 09 Aug 2024

%\cite{Chen:2016ypj}
\bibitem{Chen:2016ypj}
R.~Chen and X.~Liu,
Is the newly reported $X(5568)$ a $B\bar{K}$ molecular state?,
Phys. Rev. D \textbf{94}, 034006 (2016).
%doi:10.1103/PhysRevD.94.034006
%[arXiv:1607.05566 [hep-ph]].
%23 citations counted in INSPIRE as of 03 Jul 2023

%\cite{Chen:2022svh}
\bibitem{Chen:2022svh}
R.~Chen and Q.~Huang,
From the isovector molecular explanation of the newly $T_{c\bar{s}}^{a0(++)}(2900)$ to possible charmed-strange molecular pentaquarks,
arXiv:2208.10196.
%10 citations counted in INSPIRE as of 03 Jul 2023

%\cite{Liu:2024xbw}
\bibitem{Liu:2024xbw}
M.~Z.~Liu, X.~Z.~Ling, and L.~S.~Geng,
Productions of $D_{s0}^*(2317)$ and $D_{s1}(2460)$ in $B_{(s)}$ and $\ensuremath{\Lambda}_b(\ensuremath{\Xi}_b)$ decays,
Phys. Rev. D \textbf{109}, 056014 (2024).
%doi:10.1103/PhysRevD.109.056014
%0 citations counted in INSPIRE as of 18 Mar 2024

%\cite{Kim:2023htt}
\bibitem{Kim:2023htt}
H.~J.~Kim and H.~C.~Kim,
$D^*_{s0}(2317)$ and $B^*_{s0}$ as molecular states,
Prog. Theor. Exp. Phys. \textbf{2024}, 073D01 (2024).
%doi:10.1093/ptep/ptae095
%[arXiv:2310.13370 [hep-ph]].
%1 citations counted in INSPIRE as of 09 Aug 2024

%\cite{Montesinos:2024uhq}
\bibitem{Montesinos:2024uhq}
V.~Montesinos, M.~Albaladejo, J.~Nieves, and L.~Tolos,
Charge-conjugation asymmetry and molecular content: The $D_{s0}^\ast(2317)^\pm$ in matter,
Phys. Lett. B \textbf{853}, 138656 (2024).
%doi:10.1016/j.physletb.2024.138656
%[arXiv:2403.00451 [hep-ph]].
%1 citations counted in INSPIRE as of 09 Aug 2024

%\cite{Mohler:2011ke}
\bibitem{Mohler:2011ke}
D.~Mohler and R.~M.~Woloshyn,
$D$ and $D_s$ meson spectroscopy,
Phys. Rev. D \textbf{84}, 054505 (2011).
%doi:10.1103/PhysRevD.84.054505
%[arXiv:1103.5506 [hep-lat]].
%94 citations counted in INSPIRE as of 17 Jun 2023

%\cite{Liu:2012zya}
\bibitem{Liu:2012zya}
L.~Liu, K.~Orginos, F.~K.~Guo, C.~Hanhart, and U.~G.~Meissner,
Interactions of charmed mesons with light pseudoscalar mesons from lattice QCD and implications on the nature of the $D_{s0}^*(2317)$,
Phys. Rev. D \textbf{87}, 014508 (2013).
%doi:10.1103/PhysRevD.87.014508
%[arXiv:1208.4535 [hep-lat]].
%176 citations counted in INSPIRE as of 17 Jun 2023

%\cite{Mohler:2013rwa}
\bibitem{Mohler:2013rwa}
D.~Mohler, C.~B.~Lang, L.~Leskovec, S.~Prelovsek, and R.~M.~Woloshyn,
$D_{s0}^*(2317)$ meson and $D$-meson-kaon scattering from lattice QCD,
Phys. Rev. Lett. \textbf{111}, 222001 (2013).
%doi:10.1103/PhysRevLett.111.222001
%[arXiv:1308.3175 [hep-lat]].
%185 citations counted in INSPIRE as of 17 Jun 2023

%\cite{Alexandrou:2019tmk}
\bibitem{Alexandrou:2019tmk}
C.~Alexandrou, J.~Berlin, J.~Finkenrath, T.~Leontiou, and M.~Wagner,
Tetraquark interpolating fields in a lattice QCD investigation of the $D_{s0}^\ast(2317)$ meson,
Phys. Rev. D \textbf{101}, 034502 (2020).
%doi:10.1103/PhysRevD.101.034502
%[arXiv:1911.08435 [hep-lat]].
%16 citations counted in INSPIRE as of 08 Jul 2023

%\cite{Lang:2014yfa}
\bibitem{Lang:2014yfa}
C.~B.~Lang, L.~Leskovec, D.~Mohler, S.~Prelovsek, and R.~M.~Woloshyn,
$D_s$ mesons with $DK$ and $D^*K$ scattering near threshold,
Phys. Rev. D \textbf{90}, 034510 (2014).
%doi:10.1103/PhysRevD.90.034510
%[arXiv:1403.8103 [hep-lat]].
%162 citations counted in INSPIRE as of 17 Jun 2023

%\cite{Bali:2017pdv}
\bibitem{Bali:2017pdv}
G.~S.~Bali, S.~Collins, A.~Cox, and A.~Sch\"afer,
Masses and decay constants of the $D_{s0}^*(2317)$ and $D_{s1}(2460)$ from $N_f=2$ lattice QCD close to the physical point,
Phys. Rev. D \textbf{96}, 074501 (2017).
%doi:10.1103/PhysRevD.96.074501
%[arXiv:1706.01247 [hep-lat]].
%84 citations counted in INSPIRE as of 17 Jun 2023

%\cite{Luo:2021dvj}
\bibitem{Luo:2021dvj}
S.~Q.~Luo, B.~Chen, X.~Liu, and T.~Matsuki,
Predicting a new resonance as charmed-strange baryonic analog of $D^*_{s0}$(2317),
Phys. Rev. D \textbf{103}, 074027 (2021).
%doi:10.1103/PhysRevD.103.074027
%[arXiv:2102.00679 [hep-ph]].
%15 citations counted in INSPIRE as of 19 Jun 2023

%\cite{Hao:2022vwt}
\bibitem{Hao:2022vwt}
W.~Hao, Y.~Lu, and B.~S.~Zou,
Coupled channel effects for the charmed-strange mesons,
Phys. Rev. D \textbf{106}, 074014 (2022).
%doi:10.1103/PhysRevD.106.074014
%[arXiv:2208.10915 [hep-ph]].
%5 citations counted in INSPIRE as of 19 Jun 2023

%\cite{Simonov:2004ar}
\bibitem{Simonov:2004ar}
Y.~A.~Simonov and J.~A.~Tjon,
The coupled-channel analysis of the $D$ and $D_s$ mesons,
Phys. Rev. D \textbf{70}, 114013 (2004).
%doi:10.1103/PhysRevD.70.114013
%[arXiv:hep-ph/0409361 [hep-ph]].
%67 citations counted in INSPIRE as of 19 Jun 2023

%\cite{Yang:2023tvc}
\bibitem{Yang:2023tvc}
J.~J.~Yang, W.~Hao, X.~Wang, D.~M.~Li, Y.~X.~Li, and E.~Wang,
The mass spectrum and strong decay properties of the charmed-strange mesons within Godfrey\textendash{}Isgur model considering the coupled-channel effects,
Eur. Phys. J. C \textbf{83}, 1098 (2023).
%doi:10.1140/epjc/s10052-023-12275-3
%[arXiv:2303.11815 [hep-ph]].
%0 citations counted in INSPIRE as of 21 Mar 2024

%\cite{Coito:2011qn}
\bibitem{Coito:2011qn}
S.~Coito, G.~Rupp, and E.~van Beveren,
Quasi-bound states in the continuum: A dynamical coupled-channel calculation of axial-vector charmed mesons,
Phys. Rev. D \textbf{84}, 094020 (2011).
%doi:10.1103/PhysRevD.84.094020
%[arXiv:1106.2760 [hep-ph]].
%25 citations counted in INSPIRE as of 29 Mar 2024

%\cite{Hwang:2004cd}
\bibitem{Hwang:2004cd}
D.~S.~Hwang and D.~W.~Kim,
Mass of $D^*_{(sJ)}(2317)$ and coupled channel effect,
Phys. Lett. B \textbf{601}, 137 (2004).
%doi:10.1016/j.physletb.2004.09.040
%[arXiv:hep-ph/0408154 [hep-ph]].
%88 citations counted in INSPIRE as of 29 Mar 2024

%\cite{Ni:2021pce}
\bibitem{Ni:2021pce}
R.~H.~Ni, Q.~Li, and X.~H.~Zhong,
Mass spectra and strong decays of charmed and charmed-strange mesons,
Phys. Rev. D \textbf{105}, 056006 (2022).
%doi:10.1103/PhysRevD.105.056006
%[arXiv:2110.05024 [hep-ph]].
%14 citations counted in INSPIRE as of 22 Mar 2024

%\cite{Yang:2021tvc}
\bibitem{Yang:2021tvc}
Z.~Yang, G.~J.~Wang, J.~J.~Wu, M.~Oka, and S.~L.~Zhu,
Novel coupled channel framework connecting the quark model and lattice QCD for the near-threshold $D_s$ states,
Phys. Rev. Lett. \textbf{128}, 112001 (2022).
%doi:10.1103/PhysRevLett.128.112001
%[arXiv:2107.04860 [hep-ph]].
%38 citations counted in INSPIRE as of 09 Aug 2024

%\cite{Ortega:2016mms}
\bibitem{Ortega:2016mms}
P.~G.~Ortega, J.~Segovia, D.~R.~Entem, and F.~Fernandez,
Molecular components in $P$-wave charmed-strange mesons,
Phys. Rev. D \textbf{94}, 074037 (2016).
%doi:10.1103/PhysRevD.94.074037
%[arXiv:1603.07000 [hep-ph]].
%54 citations counted in INSPIRE as of 19 Jun 2023

%\cite{Godfrey:2016nwn}
\bibitem{Godfrey:2016nwn}
S.~Godfrey, K.~Moats, and E.~S.~Swanson,
$B$ and $B_s$ meson spectroscopy,
Phys. Rev. D \textbf{94}, 054025 (2016).
%doi:10.1103/PhysRevD.94.054025
%[arXiv:1607.02169 [hep-ph]].
%79 citations counted in INSPIRE as of 06 Dec 2023

%\cite{DiPierro:2001dwf}
\bibitem{DiPierro:2001dwf}
M.~Di Pierro and E.~Eichten,
Excited heavy-light systems and hadronic transitions,
Phys. Rev. D \textbf{64}, 114004 (2001).
%doi:10.1103/PhysRevD.64.114004
%[arXiv:hep-ph/0104208 [hep-ph]].
%399 citations counted in INSPIRE as of 20 Jun 2023

%\cite{Sun:2014wea}
\bibitem{Sun:2014wea}
Y.~Sun, Q.~T.~Song, D.~Y.~Chen, X.~Liu, and S.~L.~Zhu,
Higher bottom and bottom-strange mesons,
Phys. Rev. D \textbf{89}, 054026 (2014).
%doi:10.1103/PhysRevD.89.054026
%[arXiv:1401.1595 [hep-ph]].
%67 citations counted in INSPIRE as of 20 Jun 2023

%\cite{Lu:2016bbk}
\bibitem{Lu:2016bbk}
Q.~F.~L\"u, T.~T.~Pan, Y.~Y.~Wang, E.~Wang, and D.~M.~Li,
Excited bottom and bottom-strange mesons in the quark model,
Phys. Rev. D \textbf{94}, 074012 (2016).
%doi:10.1103/PhysRevD.94.074012
%[arXiv:1607.02812 [hep-ph]].
%70 citations counted in INSPIRE as of 18 Mar 2024

%\cite{li:2021hss}
\bibitem{li:2021hss}
Q.~Li, R.~H.~Ni, and X.~H.~Zhong,
Towards establishing an abundant $B$ and $B_s$ spectrum up to the second orbital excitations,
Phys. Rev. D \textbf{103}, 116010 (2021).
%doi:10.1103/PhysRevD.103.116010
%[arXiv:2102.03694 [hep-ph]].
%11 citations counted in INSPIRE as of 20 Jun 2023

%\cite{Hao:2022ibj}
\bibitem{Hao:2022ibj}
W.~Hao, Y.~Lu, and E.~Wang,
The assignments of the $B_s$ mesons within the screened potential model and $^3P_0$ model,
Eur. Phys. J. C \textbf{83}, 520 (2023).
%doi:10.1140/epjc/s10052-023-11689-3
%[arXiv:2212.10068 [hep-ph]].
%0 citations counted in INSPIRE as of 19 Mar 2024

%\cite{Vijande:2007ke}
\bibitem{Vijande:2007ke}
J.~Vijande, A.~Valcarce, and F.~Fernandez,
$B$ meson spectroscopy,
Phys. Rev. D \textbf{77}, 017501 (2008).
%doi:10.1103/PhysRevD.77.017501
%[arXiv:0711.2359 [hep-ph]].
%25 citations counted in INSPIRE as of 22 Mar 2024

%\cite{Colangelo:2012xi}
\bibitem{Colangelo:2012xi}
P.~Colangelo, F.~De Fazio, F.~Giannuzzi, and S.~Nicotri,
New meson spectroscopy with open charm and beauty,
Phys. Rev. D \textbf{86}, 054024 (2012).
%doi:10.1103/PhysRevD.86.054024
%[arXiv:1207.6940 [hep-ph]].
%150 citations counted in INSPIRE as of 20 Jun 2023

%\cite{Altenbuchinger:2013vwa}
\bibitem{Altenbuchinger:2013vwa}
M.~Altenbuchinger, L.~S.~Geng, and W.~Weise,
Scattering lengths of Nambu-Goldstone bosons off $D$ mesons and dynamically generated heavy-light mesons,
Phys. Rev. D \textbf{89}, 014026 (2014).
%doi:10.1103/PhysRevD.89.014026
%[arXiv:1309.4743 [hep-ph]].
%97 citations counted in INSPIRE as of 20 Jun 2023

%\cite{Sun:2018zqs}
\bibitem{Sun:2018zqs}
Z.~F.~Sun, J.~J.~Xie, and E.~Oset,
Bottom strange molecules with isospin 0,
Phys. Rev. D \textbf{97}, 094031 (2018).
%doi:10.1103/PhysRevD.97.094031
%[arXiv:1801.04367 [hep-ph]].
%3 citations counted in INSPIRE as of 20 Jun 2023

%\cite{Yang:2022vdb}
\bibitem{Yang:2022vdb}
Z.~Yang, G.~J.~Wang, J.~J.~Wu, M.~Oka, and S.~L.~Zhu,
The investigations of the P-wave $B_{s}$ states combining quark model and lattice QCD in the coupled channel framework,
J. High Energy Phys. \textbf{01} (2023) 058.
%doi:10.1007/JHEP01(2023)058
%[arXiv:2207.07320 [hep-lat]].
%4 citations counted in INSPIRE as of 27 Jun 2023

%\cite{Albaladejo:2016ztm}
\bibitem{Albaladejo:2016ztm}
M.~Albaladejo, P.~Fernandez-Soler, J.~Nieves, and P.~G.~Ortega,
Lowest-lying even-parity ${\bar{B}}_s$ mesons: Heavy-quark spin-flavor symmetry, chiral dynamics, and constituent quark-model bare masses,
Eur. Phys. J. C \textbf{77}, 170 (2017).
%doi:10.1140/epjc/s10052-017-4735-7
%[arXiv:1612.07782 [hep-ph]].
%15 citations counted in INSPIRE as of 01 Jul 2023

%\cite{Ortega:2016pgg}
\bibitem{Ortega:2016pgg}
P.~G.~Ortega, J.~Segovia, D.~R.~Entem, and F.~Fern\'andez,
Threshold effects in $P$-wave bottom-strange mesons,
Phys. Rev. D \textbf{95}, 034010 (2017).
%doi:10.1103/PhysRevD.95.034010
%[arXiv:1612.04826 [hep-ph]].
%24 citations counted in INSPIRE as of 20 Oct 2022

%\cite{Gregory:2010gm}
\bibitem{Gregory:2010gm}
E.~B.~Gregory, C.~T.~H.~Davies, I.~D.~Kendall, J.~Koponen, K.~Wong, E.~Follana, E.~Gamiz, G.~P.~Lepage, E.~H.~Muller, H.~Na \textit{et al.},
Precise $B$, $B_s$ and $B_c$ meson spectroscopy from full lattice QCD,
Phys. Rev. D \textbf{83}, 014506 (2011).
%doi:10.1103/PhysRevD.83.014506
%[arXiv:1010.3848 [hep-lat]].
%89 citations counted in INSPIRE as of 17 Jun 2023

%\cite{Lang:2015hza}
\bibitem{Lang:2015hza}
C.~B.~Lang, D.~Mohler, S.~Prelovsek, and R.~M.~Woloshyn,
Predicting positive parity $B_s$ mesons from lattice QCD,
Phys. Lett. B \textbf{750}, 17 (2015).
%doi:10.1016/j.physletb.2015.08.038
%[arXiv:1501.01646 [hep-lat]].
%92 citations counted in INSPIRE as of 17 Jun 2023

%\cite{Hudspith:2023loy}
\bibitem{Hudspith:2023loy}
R.~J.~Hudspith and D.~Mohler,
Exotic tetraquark states with two $\bar{b}$ quarks and $J^P=0^+$ and $1^+$ $B_s$ states in a nonperturbatively tuned lattice NRQCD setup,
Phys. Rev. D \textbf{107}, 114510 (2023).
%doi:10.1103/PhysRevD.107.114510
%[arXiv:2303.17295 [hep-lat]].
%10 citations counted in INSPIRE as of 06 Dec 2023

%\cite{Belle:2011wdj}
\bibitem{Belle:2011wdj}
V.~Bhardwaj \textit{et al.} (Belle Collaboration),
Observation of $X(3872)\to J/\psi \gamma$ and search for $X(3872)\to\psi'\gamma$ in $B$ decays,
Phys. Rev. Lett. \textbf{107}, 091803 (2011).
%doi:10.1103/PhysRevLett.107.091803
%[arXiv:1105.0177 [hep-ex]].
%208 citations counted in INSPIRE as of 19 Jun 2024

%\cite{BaBar:2008flx}
\bibitem{BaBar:2008flx}
B.~Aubert \textit{et al.} ({\it BaBar} Collaboration),
Evidence for $X(3872) \to \psi_{2S} \gamma$ in $B^\pm \to X_{3872} K^\pm$ decays, and a study of $B \to c \bar{c} \gamma K$,
Phys. Rev. Lett. \textbf{102}, 132001 (2009).
%doi:10.1103/PhysRevLett.102.132001
%[arXiv:0809.0042 [hep-ex]].
%318 citations counted in INSPIRE as of 19 Jun 2024

%\cite{LHCb:2014jvf}
\bibitem{LHCb:2014jvf}
R.~Aaij \textit{et al.} (LHCb Collaboration),
Evidence for the decay $X(3872)\rightarrow\psi(2S)\gamma$,
Nucl. Phys. \textbf{B886}, 665 (2014).
%doi:10.1016/j.nuclphysb.2014.06.011
%[arXiv:1404.0275 [hep-ex]].
%179 citations counted in INSPIRE as of 19 Jun 2024

%\cite{BESIII:2020nbj}
\bibitem{BESIII:2020nbj}
M.~Ablikim \textit{et al.} (BESIII Collaboration),
Study of open-charm decays and radiative transitions of the $X(3872)$,
Phys. Rev. Lett. \textbf{124}, 242001 (2020).
%doi:10.1103/PhysRevLett.124.242001
%[arXiv:2001.01156 [hep-ex]].
%40 citations counted in INSPIRE as of 19 Jun 2024

%\cite{LHCb:2024tpv}
\bibitem{LHCb:2024tpv}
R.~Aaij \textit{et al.} (LHCb Collaboration),
Probing the nature of the $\chi_{c1}(3872)$ state using radiative decays,
arXiv:2406.17006.
%0 citations counted in INSPIRE as of 05 Jul 2024

%\cite{Badalian:2012jz}
\bibitem{Badalian:2012jz}
A.~M.~Badalian, V.~D.~Orlovsky, Y.~A.~Simonov, and B.~L.~G.~Bakker,
The ratio of decay widths of X(3872) to $ \psi^{\prime}\gamma $ and $ J/\psi\gamma$ as a test of the $X(3872)$ dynamical structure,
Phys. Rev. D \textbf{85}, 114002 (2012).
%doi:10.1103/PhysRevD.85.114002
%[arXiv:1202.4882 [hep-ph]].
%42 citations counted in INSPIRE as of 19 Jun 2024

%\cite{Cincioglu:2016fkm}
\bibitem{Cincioglu:2016fkm}
E.~Cincioglu, J.~Nieves, A.~Ozpineci, and A.~U.~Yilmazer,
Quarkonium contribution to meson molecules,
Eur. Phys. J. C \textbf{76}, 576 (2016).
%doi:10.1140/epjc/s10052-016-4413-1
%[arXiv:1606.03239 [hep-ph]].
%69 citations counted in INSPIRE as of 19 Jun 2024

%\cite{Dong:2009uf}
\bibitem{Dong:2009uf}
Y.~Dong, A.~Faessler, T.~Gutsche, and V.~E.~Lyubovitskij,
$J/\psi \gamma$ and $\psi(2S) \gamma$ decay modes of the $X(3872)$,
J. Phys. G \textbf{38}, 015001 (2011).
%doi:10.1088/0954-3899/38/1/015001
%[arXiv:0909.0380 [hep-ph]].
%108 citations counted in INSPIRE as of 19 Jun 2024

%\cite{Chen:2024xlw}
\bibitem{Chen:2024xlw}
P.~Chen, Z.~W.~Liu, Z.~L.~Zhang, S.~Q.~Luo, F.~L.~Wang, J.~Z.~Wang, and X.~Liu,
Role of electromagnetic interactions in the $X(3872)$ and its analogs,
Phys. Rev. D \textbf{109}, 094002 (2024).
%doi:10.1103/PhysRevD.109.094002
%[arXiv:2401.05989 [hep-ph]].
%1 citations counted in INSPIRE as of 09 Aug 2024

%\cite{Godfrey:2015dia}
\bibitem{Godfrey:2015dia}
S.~Godfrey and K.~Moats,
Bottomonium mesons and strategies for their observation,
Phys. Rev. D \textbf{92}, 054034 (2015).
%doi:10.1103/PhysRevD.92.054034
%[arXiv:1507.00024 [hep-ph]].
%140 citations counted in INSPIRE as of 18 Jun 2024

%\cite{Segovia:2016xqb}
\bibitem{Segovia:2016xqb}
J.~Segovia, P.~G.~Ortega, D.~R.~Entem, and F.~Fern\'andez,
Bottomonium spectrum revisited,
Phys. Rev. D \textbf{93}, 074027 (2016).
%doi:10.1103/PhysRevD.93.074027
%[arXiv:1601.05093 [hep-ph]].
%117 citations counted in INSPIRE as of 18 Jun 2024

%\cite{Wang:2018rjg}
\bibitem{Wang:2018rjg}
J.~Z.~Wang, Z.~F.~Sun, X.~Liu, and T.~Matsuki,
Higher bottomonium zoo,
Eur. Phys. J. C \textbf{78}, 915 (2018).
%doi:10.1140/epjc/s10052-018-6372-1
%[arXiv:1802.04938 [hep-ph]].
%43 citations counted in INSPIRE as of 18 Jun 2024

%\cite{Li:2012vc}
\bibitem{Li:2012vc}
B.~Q.~Li, C.~Meng, and K.~T.~Chao,
Search for $\chi_{c_J}(2P)$ from higher charmonim E1 transitions and $X$, $Y$, $Z$ states,
arXiv:1201.4155.
%10 citations counted in INSPIRE as of 18 Jun 2024

%\cite{Deng:2016ktl}
\bibitem{Deng:2016ktl}
W.~J.~Deng, H.~Liu, L.~C.~Gui, and X.~H.~Zhong,
Spectrum and electromagnetic transitions of bottomonium,
Phys. Rev. D \textbf{95}, 074002 (2017).
%doi:10.1103/PhysRevD.95.074002
%[arXiv:1607.04696 [hep-ph]].
%67 citations counted in INSPIRE as of 18 Jun 2024

%\cite{Deng:2016stx}
\bibitem{Deng:2016stx}
W.~J.~Deng, H.~Liu, L.~C.~Gui, and X.~H.~Zhong,
Charmonium spectrum and their electromagnetic transitions with higher multipole contributions,
Phys. Rev. D \textbf{95}, 034026 (2017).
%doi:10.1103/PhysRevD.95.034026
%[arXiv:1608.00287 [hep-ph]].
%101 citations counted in INSPIRE as of 18 Jun 2024

%\cite{CLEO:1992xqa}
\bibitem{CLEO:1992xqa}
F.~Butler \textit{et al.} (CLEO Collaboration),
Measurement of the $D^* (2010)$ branching fractions,
Phys. Rev. Lett. \textbf{69}, 2041 (1992).
%doi:10.1103/PhysRevLett.69.2041
%157 citations counted in INSPIRE as of 21 Jun 2024

%\cite{BESIII:2014rqs}
\bibitem{BESIII:2014rqs}
M.~Ablikim \textit{et al.} (BESIII Collaboration),
Precision measurement of the $D^{*0}$ decay branching fractions,
Phys. Rev. D \textbf{91}, 031101 (2015).
%doi:10.1103/PhysRevD.91.031101
%[arXiv:1412.4566 [hep-ex]].
%16 citations counted in INSPIRE as of 21 Jun 2024

%\cite{Li:2019tbn}
\bibitem{Li:2019tbn}
Q.~Li, M.~S.~Liu, L.~S.~Lu, Q.~F.~L\"u, L.~C.~Gui, and X.~H.~Zhong,
Excited bottom-charmed mesons in a nonrelativistic quark model,
Phys. Rev. D \textbf{99}, 096020 (2019).
%doi:10.1103/PhysRevD.99.096020
%[arXiv:1903.11927 [hep-ph]].
%44 citations counted in INSPIRE as of 18 Mar 2024

%\cite{Godfrey:2003kg}
\bibitem{Godfrey:2003kg}
S.~Godfrey,
Testing the nature of the $D_{sJ}^*(2317)^+$ and $D_{sJ}(2463)^+$ states using radiative transitions,
Phys. Lett. B \textbf{568}, 254 (2003).
%doi:10.1016/j.physletb.2003.06.049
%[arXiv:hep-ph/0305122 [hep-ph]].
%186 citations counted in INSPIRE as of 18 Mar 2024

%\cite{Radford:2009bs}
\bibitem{Radford:2009bs}
S.~F.~Radford, W.~W.~Repko, and M.~J.~Saelim,
Potential model calculations and predictions for $c$ anti-$s$ quarkonia,
Phys. Rev. D \textbf{80}, 034012 (2009).
%doi:10.1103/PhysRevD.80.034012
%[arXiv:0903.0551 [hep-ph]].
%32 citations counted in INSPIRE as of 18 Mar 2024

%\cite{Chen:2020jku}
\bibitem{Chen:2020jku}
S.~F.~Chen, J.~Liu, H.~Q.~Zhou, and D.~Y.~Chen,
Electric transitions of the charmed-strange mesons in a relativistic quark model,
Eur. Phys. J. C \textbf{80}, 290 (2020).
%doi:10.1140/epjc/s10052-020-7852-7
%[arXiv:2003.07988 [hep-ph]].
%2 citations counted in INSPIRE as of 18 Mar 2024

%\cite{Green:2016occ}
\bibitem{Green:2016occ}
N.~Green, W.~W.~Repko, and S.~F.~Radford,
Note on predictions for $c\bar{s}$ quarkonia using a three-loop static potential,
Nucl. Phys. \textbf{A958}, 71 (2017).
%doi:10.1016/j.nuclphysa.2016.11.006
%[arXiv:1605.06393 [hep-ph]].
%4 citations counted in INSPIRE as of 18 Mar 2024

%\cite{Godfrey:2005ww}
\bibitem{Godfrey:2005ww}
S.~Godfrey,
Properties of the charmed $P$-wave mesons,
Phys. Rev. D \textbf{72}, 054029 (2005).
%doi:10.1103/PhysRevD.72.054029
%[arXiv:hep-ph/0508078 [hep-ph]].
%75 citations counted in INSPIRE as of 18 Mar 2024

%\cite{Lutz:2007sk}
\bibitem{Lutz:2007sk}
M.~F.~M.~Lutz and M.~Soyeur,
Radiative and isospin-violating decays of $D_{s}$-mesons in the hadrogenesis conjecture,
Nucl. Phys. \textbf{A813}, 14 (2008).
%doi:10.1016/j.nuclphysa.2008.09.003
%[arXiv:0710.1545 [hep-ph]].
%112 citations counted in INSPIRE as of 20 Mar 2024

%\cite{Colangelo:2005hv}
\bibitem{Colangelo:2005hv}
P.~Colangelo, F.~De Fazio, and A.~Ozpineci,
Radiative transitions of $D^*_{sJ}(2317)$ and $D_{sJ}(2460)$,
Phys. Rev. D \textbf{72}, 074004 (2005).
%doi:10.1103/PhysRevD.72.074004
%[arXiv:hep-ph/0505195 [hep-ph]].
%124 citations counted in INSPIRE as of 20 Mar 2024

%\cite{Cleven:2014oka}
\bibitem{Cleven:2014oka}
M.~Cleven, H.~W.~Grie\ss{}hammer, F.~K.~Guo, C.~Hanhart, and U.~G.~Mei\ss{}ner,
Strong and radiative decays of the $D^*_{s0}(2317)$ and $D_{s1}(2460)$,
Eur. Phys. J. A \textbf{50}, 149 (2014).
%doi:10.1140/epja/i2014-14149-y
%[arXiv:1405.2242 [hep-ph]].
%63 citations counted in INSPIRE as of 18 Mar 2024

%\cite{Fu:2021wde}
\bibitem{Fu:2021wde}
H.~L.~Fu, H.~W.~Grie\ss{}hammer, F.~K.~Guo, C.~Hanhart, and U.~G.~Mei\ss{}ner,
Update on strong and radiative decays of the $D_{s0}^*(2317)$ and $D_{s1}(2460)$ and their bottom cousins,
Eur. Phys. J. A \textbf{58}, 70 (2022).
%doi:10.1140/epja/s10050-022-00724-8
%[arXiv:2111.09481 [hep-ph]].
%12 citations counted in INSPIRE as of 08 Jul 2023

%\cite{Faessler:2007us}
\bibitem{Faessler:2007us}
A.~Faessler, T.~Gutsche, V.~E.~Lyubovitskij, and Y.~L.~Ma,
$D^* K$ molecular structure of the $D_{s1}(2460)$ meson,
Phys. Rev. D \textbf{76}, 114008 (2007).
%doi:10.1103/PhysRevD.76.114008
%[arXiv:0709.3946 [hep-ph]].
%114 citations counted in INSPIRE as of 18 Mar 2024

%\cite{Xiao:2016hoa}
\bibitem{Xiao:2016hoa}
C.~J.~Xiao, D.~Y.~Chen, and Y.~L.~Ma,
Radiative and pionic transitions from the $D_{s1}(2460)$ to the $D_{s0}^\ast(2317)$,
Phys. Rev. D \textbf{93}, 094011 (2016).
%doi:10.1103/PhysRevD.93.094011
%[arXiv:1601.06399 [hep-ph]].
%26 citations counted in INSPIRE as of 18 Mar 2024

%\cite{Feng:2012zzf}
\bibitem{Feng:2012zzf}
G.~Q.~Feng and X.~H.~Guo,
$DK$ molecule in the Bethe-Salpeter equation approach in the heavy quark limit,
Phys. Rev. D \textbf{86}, 036004 (2012).
%doi:10.1103/PhysRevD.86.036004
%7 citations counted in INSPIRE as of 18 Mar 2024

%\cite{Feng:2012zze}
\bibitem{Feng:2012zze}
G.~Q.~Feng, X.~H.~Guo, and Z.~H.~Zhang,
Studying the $D^*K$ molecular structure of $D_{s1}(2460)^+$ in the Bethe-Salpeter approach,
Eur. Phys. J. C \textbf{72}, 2033 (2012).
%doi:10.1140/epjc/s10052-012-2033-y
%8 citations counted in INSPIRE as of 18 Mar 2024

%\cite{Faessler:2008vc}
\bibitem{Faessler:2008vc}
A.~Faessler, T.~Gutsche, V.~E.~Lyubovitskij, and Y.~L.~Ma,
Molecular structure of the $B^*_{sl}(5725)$ and $B_{s1}(5778)$ bottom-strange mesons,
Phys. Rev. D \textbf{77}, 114013 (2008).
%doi:10.1103/PhysRevD.77.114013
%[arXiv:0801.2232 [hep-ph]].
%58 citations counted in INSPIRE as of 18 Mar 2024

%\cite{Wang:2019ehs}
\bibitem{Wang:2019ehs}
Z.~Y.~Wang, J.~J.~Qi, Q.~X.~Yu, and X.~H.~Guo,
$B_{s1}(5778)$ as a $B^*\bar{K}$ molecule in the Bethe-Salpeter equation approach,
Phys. Rev. D \textbf{100}, 096009 (2019).
%doi:10.1103/PhysRevD.100.096009
%[arXiv:1906.09002 [hep-ph]].
%1 citations counted in INSPIRE as of 18 Mar 2024

%\cite{Zhang:2022pxc}
\bibitem{Zhang:2022pxc}
Z.~L.~Zhang, Z.~W.~Liu, S.~Q.~Luo, F.~L.~Wang, B.~Wang, and H.~Xu,
\ensuremath{\Lambda_c(2910)} and \ensuremath{\Lambda_c(2940)} as conventional baryons dressed with the $D^*N$ channel,
Phys. Rev. D \textbf{107}, 034036 (2023).
%doi:10.1103/PhysRevD.107.034036
%[arXiv:2210.17188 [hep-ph]].
%4 citations counted in INSPIRE as of 26 Jun 2023

%\cite{Guo:2017jvc}
\bibitem{Guo:2017jvc}
F.~K.~Guo, C.~Hanhart, U.~G.~Mei\ss{}ner, Q.~Wang, Q.~Zhao, and B.~S.~Zou,
Hadronic molecules,
Rev. Mod. Phys. \textbf{90}, 015004 (2018); \textbf{94}, 029901(E) (2022).
%doi:10.1103/RevModPhys.90.015004
%[arXiv:1705.00141 [hep-ph]].
%931 citations counted in INSPIRE as of 26 Jun 2023

%\cite{Weinberg:1965zz}
\bibitem{Weinberg:1965zz}
S.~Weinberg,
Evidence that the deuteron is not an elementary particle,
Phys. Rev. \textbf{137}, B672-B678 (1965).
%doi:10.1103/PhysRev.137.B672
%347 citations counted in INSPIRE as of 04 Oct 2022

%\cite{Tornqvist:1995kr}
\bibitem{Tornqvist:1995kr}
N.~A.~Tornqvist,
Understanding the scalar meson $q\bar{q}$ nonet,
Z. Phys. C \textbf{68}, 647 (1995).
%doi:10.1007/BF01565264
%[arXiv:hep-ph/9504372 [hep-ph]].
%480 citations counted in INSPIRE as of 03 Oct 2022

%\cite{Eichten:1978tg}
\bibitem{Eichten:1978tg}
E.~Eichten, K.~Gottfried, T.~Kinoshita, K.~D.~Lane, and T.~M.~Yan,
Charmonium: The model,
Phys.\ Rev.\ D \textbf{17}, 3090 (1978).
%[erratum: Phys. Rev. D \textbf{21} (1980), 313]
%doi:10.1103/PhysRevD.17.3090
%1616 citations counted in INSPIRE as of 16 Apr 2022

%\cite{Danilkin:2009hr}
\bibitem{Danilkin:2009hr}
I.~V.~Danilkin and Y.~A.~Simonov,
Channel coupling in heavy quarkonia: Energy levels, mixing, widths and new states,
Phys. Rev. D \textbf{81}, 074027 (2010).
%doi:10.1103/PhysRevD.81.074027
%[arXiv:0907.1088 [hep-ph]].
%63 citations counted in INSPIRE as of 02 Sep 2022

%\cite{Lu:2017hma}
\bibitem{Lu:2017hma}
Y.~Lu, M.~N.~Anwar, and B.~S.~Zou,
How large is the contribution of excited mesons in coupled-channel effects?,
Phys. Rev. D \textbf{95}, 034018 (2017).
%doi:10.1103/PhysRevD.95.034018
%[arXiv:1701.00692 [hep-ph]].
%19 citations counted in INSPIRE as of 18 Oct 2022

%\cite{Anwar:2018yqm}
\bibitem{Anwar:2018yqm}
M.~N.~Anwar, Y.~Lu, and B.~S.~Zou,
$\chi_{b}(3P)$ multiplet revisited: Hyperfine mass splitting and radiative transitions,
Phys. Rev. D \textbf{99}, 094005 (2019).
%doi:10.1103/PhysRevD.99.094005
%[arXiv:1806.01155 [hep-ph]].
%10 citations counted in INSPIRE as of 18 Oct 2022

%\cite{Ortega:2021fem}
\bibitem{Ortega:2021fem}
P.~G.~Ortega, J.~Segovia, D.~R.~Entem, and F.~Fernandez,
The $D_{s0}(2590)^+$ as the dressed $c\bar{s}(2^1S_0)$ meson in a coupled-channels calculation,
Phys. Lett. B \textbf{827}, 136998 (2022).
%doi:10.1016/j.physletb.2022.136998
%[arXiv:2111.00023 [hep-ph]].
%6 citations counted in INSPIRE as of 20 Oct 2022

%\cite{Ortega:2021yis}
\bibitem{Ortega:2021yis}
P.~G.~Ortega, D.~R.~Entem, and F.~Fern\'andez,
Does the $J^{PC}=1^{+\ensuremath{-}}$ counterpart of the $X(3872)$ exist?
Phys. Lett. B \textbf{829}, 137083 (2022).
%doi:10.1016/j.physletb.2022.137083
%[arXiv:2111.02475 [hep-ph]].
%1 citations counted in INSPIRE as of 20 Oct 2022

%\cite{Ortega:2009hj}
\bibitem{Ortega:2009hj}
P.~G.~Ortega, J.~Segovia, D.~R.~Entem, and F.~Fernandez,
Coupled channel approach to the structure of the $X(3872)$,
Phys. Rev. D \textbf{81}, 054023 (2010).
%doi:10.1103/PhysRevD.81.054023
%[arXiv:0907.3997 [hep-ph]].
%108 citations counted in INSPIRE as of 02 Sep 2022

%\cite{Hiyama:2003cu}
\bibitem{Hiyama:2003cu}
E.~Hiyama, Y.~Kino, and M.~Kamimura,
Gaussian expansion method for few-body systems,
Prog. Part. Nucl. Phys. \textbf{51}, 223 (2003).
%doi:10.1016/S0146-6410(03)90015-9
%482 citations counted in INSPIRE as of 02 Sep 2022

%\cite{Hiyama:2012sma}
\bibitem{Hiyama:2012sma}
E.~Hiyama,
Gaussian expansion method for few-body systems and its applications to atomic and nuclear physics,
Prog. Theor. Exp. Phys. \textbf{2012}, 01A204 (2012).
%doi:10.1093/ptep/pts015
%29 citations counted in INSPIRE as of 02 Sep 2022

%\cite{Wise:1992hn}
\bibitem{Wise:1992hn}
M.~B.~Wise,
Chiral perturbation theory for hadrons containing a heavy quark,
Phys. Rev. D \textbf{45}, R2188 (1992).
%doi:10.1103/PhysRevD.45.R2188
%831 citations counted in INSPIRE as of 02 Oct 2022

%\cite{Manohar:2000dt}
\bibitem{Manohar:2000dt}
A.~V.~Manohar and M.~B.~Wise,
Heavy quark physics,
Camb. Monogr. Part. Phys. Nucl. Phys. Cosmol. \textbf{10}, 1 (2000).
%481 citations counted in INSPIRE as of 02 Oct 2022

%\cite{Yan:1992gz}
\bibitem{Yan:1992gz}
T.~M.~Yan, H.~Y.~Cheng, C.~Y.~Cheung, G.~L.~Lin, Y.~C.~Lin, and H.~L.~Yu,
Heavy quark symmetry and chiral dynamics,
Phys. Rev. D \textbf{46}, 1148 (1992); \textbf{55}(E), 5851 (1997).
%doi:10.1103/PhysRevD.46.1148
%754 citations counted in INSPIRE as of 03 Jul 2023

%\cite{Burdman:1992gh}
\bibitem{Burdman:1992gh}
G.~Burdman and J.~F.~Donoghue,
Union of chiral and heavy quark symmetries,
Phys. Lett. B \textbf{280}, 287 (1992).
%doi:10.1016/0370-2693(92)90068-F
%676 citations counted in INSPIRE as of 03 Jul 2023

%\cite{Falk:1992cx}
\bibitem{Falk:1992cx}
A.~F.~Falk and M.~E.~Luke,
Strong decays of excited heavy mesons in chiral perturbation theory,
Phys. Lett. B \textbf{292}, 119 (1992).
%doi:10.1016/0370-2693(92)90618-E
%[arXiv:hep-ph/9206241 [hep-ph]].
%260 citations counted in INSPIRE as of 03 Jul 2023

%\cite{Casalbuoni:1996pg}
\bibitem{Casalbuoni:1996pg}
R.~Casalbuoni, A.~Deandrea, N.~Di Bartolomeo, R.~Gatto, F.~Feruglio, and G.~Nardulli,
Phenomenology of heavy meson chiral Lagrangians,
Phys. Rep. \textbf{281}, 145 (1997).
%doi:10.1016/S0370-1573(96)00027-0
%[arXiv:hep-ph/9605342 [hep-ph]].
%658 citations counted in INSPIRE as of 03 Jul 2023

%\cite{Lin:1999ad}
\bibitem{Lin:1999ad}
Z.~w.~Lin and C.~M.~Ko,
A Model for $J / \psi$ absorption in hadronic matter,
Phys. Rev. C \textbf{62}, 034903 (2000).
%doi:10.1103/PhysRevC.62.034903
%[arXiv:nucl-th/9912046 [nucl-th]].
%241 citations counted in INSPIRE as of 03 Jul 2023

%\cite{Nagahiro:2008mn}
\bibitem{Nagahiro:2008mn}
H.~Nagahiro, L.~Roca, and E.~Oset,
Meson loops in the $f_0(980)$ and $a_0(980)$ radiative decays into $\rho$, $\omega$,
Eur. Phys. J. A \textbf{36}, 73 (2008).
%doi:10.1140/epja/i2008-10567-8
%[arXiv:0802.0455 [hep-ph]].
%38 citations counted in INSPIRE as of 03 Jul 2023

%\cite{Wang:2019aoc}
\bibitem{Wang:2019aoc}
F.~L.~Wang, R.~Chen, Z.~W.~Liu, and X.~Liu,
Possible triple-charm molecular pentaquarks from $\Xi_{cc}D_1/\Xi_{cc}D_2^*$ interactions,
Phys. Rev. D \textbf{99}, 054021 (2019).
%doi:10.1103/PhysRevD.99.054021
%[arXiv:1901.01542 [hep-ph]].
%23 citations counted in INSPIRE as of 03 Jul 2023

%\cite{Isola:2003fh}
\bibitem{Isola:2003fh}
C.~Isola, M.~Ladisa, G.~Nardulli, and P.~Santorelli,
Charming penguins in $B \to K^* \pi$, $K(\rho, \omega, \phi)$ decays,
Phys. Rev. D \textbf{68}, 114001 (2003).
%doi:10.1103/PhysRevD.68.114001
%[arXiv:hep-ph/0307367 [hep-ph]].
%168 citations counted in INSPIRE as of 03 Jul 2023

%\cite{Chen:2011cj}
\bibitem{Chen:2011cj}
D.~Y.~Chen, X.~Liu, and T.~Matsuki,
Two charged strangeonium-like structures observable in the $Y(2175) \to \phi(1020)\pi^{+} \pi^{-}$ process,
Eur. Phys. J. C \textbf{72}, 2008 (2012).
%doi:10.1140/epjc/s10052-012-2008-z
%[arXiv:1112.3773 [hep-ph]].
%33 citations counted in INSPIRE as of 03 Jul 2023

%\cite{Sun:2011uh}
\bibitem{Sun:2011uh}
Z.~F.~Sun, J.~He, X.~Liu, Z.~G.~Luo, and S.~L.~Zhu,
$Z_b(10610)^\pm$ and $Z_b(10650)^\pm$ as the $B^*\bar{B}$ and $B^*\bar{B}^{*}$ molecular states,
Phys. Rev. D \textbf{84}, 054002 (2011).
%doi:10.1103/PhysRevD.84.054002
%[arXiv:1106.2968 [hep-ph]].
%190 citations counted in INSPIRE as of 21 Jun 2024

%\cite{Breit:1929zz}
\bibitem{Breit:1929zz}
G.~Breit,
The effect of retardation on the interaction of two electrons,
Phys. Rev. \textbf{34}, 553 (1929).
%doi:10.1103/PhysRev.34.553
%245 citations counted in INSPIRE as of 03 Jul 2023

%\cite{Xu:2017tsr}
\bibitem{Xu:2017tsr}
H.~Xu, B.~Wang, Z.~W.~Liu, and X.~Liu,
$D D^{*}$ potentials in chiral perturbation theory and possible molecular states,
Phys. Rev. D \textbf{99}, 014027 (2019); \textbf{104}, 119903(E) (2021).
%doi:10.1103/PhysRevD.99.014027
%[arXiv:1708.06918 [hep-ph]].
%49 citations counted in INSPIRE as of 26 Jun 2023

%\cite{Micu:1968mk}
\bibitem{Micu:1968mk}
L.~Micu,
Decay rates of meson resonances in a quark model,
Nucl. Phys. \textbf{B10}, 521 (1969).
%doi:10.1016/0550-3213(69)90039-X
%542 citations counted in INSPIRE as of 26 Jun 2023

%\cite{LeYaouanc:1972vsx}
\bibitem{LeYaouanc:1972vsx}
A.~Le Yaouanc, L.~Oliver, O.~Pene, and J.~C.~Raynal,
Naive quark pair creation model of strong interaction vertices,
Phys. Rev. D \textbf{8}, 2223 (1973).
%doi:10.1103/PhysRevD.8.2223
%706 citations counted in INSPIRE as of 26 Jun 2023

%\cite{Chen:2016iyi}
\bibitem{Chen:2016iyi}
B.~Chen, K.~W.~Wei, X.~Liu, and T.~Matsuki,
Low-lying charmed and charmed-strange baryon states,
Eur. Phys. J. C \textbf{77}, 154 (2017).
%doi:10.1140/epjc/s10052-017-4708-x
%[arXiv:1609.07967 [hep-ph]].
%78 citations counted in INSPIRE as of 26 Jun 2023

%\cite{MartinezTorres:2014kpc}
\bibitem{MartinezTorres:2014kpc}
A.~Mart\'\i{}nez Torres, E.~Oset, S.~Prelovsek, and A.~Ramos,
Reanalysis of lattice QCD spectra leading to the $D_{s0}^*(2317)$ and $D_{s1}^*(2460)$,
J. High Energy Phys. \textbf{05}, (2015) 153.
%doi:10.1007/JHEP05(2015)153
%[arXiv:1412.1706 [hep-lat]].
%91 citations counted in INSPIRE as of 20 Jun 2023

%\cite{Tan:2021bvl}
\bibitem{Tan:2021bvl}
Y.~Tan and J.~Ping,
$D^*_{s0}(2317)$ and $D_{s1}(2460)$ in an unquenched quark model,
arXiv:2111.04677.
%6 citations counted in INSPIRE as of 20 Jun 2023

%\cite{Albaladejo:2016lbb}
\bibitem{Albaladejo:2016lbb}
M.~Albaladejo, P.~Fernandez-Soler, F.~K.~Guo, and J.~Nieves,
Two-pole structure of the $D^\ast_0(2400)$,
Phys. Lett. B \textbf{767}, 465 (2017).
%doi:10.1016/j.physletb.2017.02.036
%[arXiv:1610.06727 [hep-ph]].
%85 citations counted in INSPIRE as of 21 Jun 2023

%\cite{Du:2017zvv}
\bibitem{Du:2017zvv}
M.~L.~Du, M.~Albaladejo, P.~Fern\'andez-Soler, F.~K.~Guo, C.~Hanhart, U.~G.~Mei\ss{}ner, J.~Nieves, and D.~L.~Yao,
Towards a new paradigm for heavy-light meson spectroscopy,
Phys. Rev. D \textbf{98}, 094018 (2018).
%doi:10.1103/PhysRevD.98.094018
%[arXiv:1712.07957 [hep-ph]].
%60 citations counted in INSPIRE as of 21 Jun 2023

%\cite{Brodsky:1968ea}
\bibitem{Brodsky:1968ea}
S.~J.~Brodsky and J.~R.~Primack,
The electromagnetic interactions of composite systems,
Ann. Phys. (N.Y.) \textbf{52}, 315 (1969).
%doi:10.1016/0003-4916(69)90264-4
%251 citations counted in INSPIRE as of 26 Mar 2023

%\cite{Kwong:1988ae}
\bibitem{Kwong:1988ae}
W.~Kwong and J.~L.~Rosner,
$D$ wave quarkonium levels of the $\Upsilon$ family,
Phys. Rev. D \textbf{38}, 279 (1988).
%doi:10.1103/PhysRevD.38.279
%219 citations counted in INSPIRE as of 26 Mar 2023

%\cite{Luo:2023hnp}
\bibitem{Luo:2023hnp}
S.~Q.~Luo, Z.~W.~Liu, and X.~Liu,
New type of hydrogenlike charm-pion or charm-kaon matter,
Phys. Rev. D \textbf{107}, 054022 (2023).
%doi:10.1103/PhysRevD.107.054022
%[arXiv:2302.13202 [hep-ph]].
%1 citations counted in INSPIRE as of 21 Mar 2024

%\cite{Zhang:2006ix}
\bibitem{Zhang:2006ix}
Y.~J.~Zhang, H.~C.~Chiang, P.~N.~Shen, and B.~S.~Zou,
Possible S-wave bound-states of two pseudoscalar mesons,
Phys. Rev. D \textbf{74}, 014013 (2006).
%doi:10.1103/PhysRevD.74.014013
%[arXiv:hep-ph/0604271 [hep-ph]].
%85 citations counted in INSPIRE as of 21 Mar 2024

%\cite{Chen:2010re}
\bibitem{Chen:2010re}
D.~Y.~Chen, Y.~B.~Dong, and X.~Liu,
Long-distant contribution and $\chi_{c1}$ radiative decays to light vector meson,
Eur. Phys. J. C \textbf{70}, 177 (2010).
%doi:10.1140/epjc/s10052-010-1449-5
%[arXiv:1005.0066 [hep-ph]].
%25 citations counted in INSPIRE as of 05 Dec 2023

%\cite{Chen:2014sra}
\bibitem{Chen:2014sra}
D.~Y.~Chen, X.~Liu, and T.~Matsuki,
Observation of $e^+e^-\to \chi_{c0}\omega$ and missing higher charmonium $\psi(4S)$,
Phys. Rev. D \textbf{91}, 094023 (2015).
%doi:10.1103/PhysRevD.91.094023
%[arXiv:1411.5136 [hep-ph]].
%33 citations counted in INSPIRE as of 05 Dec 2023

%\cite{Kaymakcalan:1983qq}
\bibitem{Kaymakcalan:1983qq}
O.~Kaymakcalan, S.~Rajeev, and J.~Schechter,
NonAbelian anomaly and vector meson decays,
Phys. Rev. D \textbf{30}, 594 (1984).
%doi:10.1103/PhysRevD.30.594
%478 citations counted in INSPIRE as of 21 Mar 2024

%\cite{Oh:2000qr}
\bibitem{Oh:2000qr}
Y.~S.~Oh, T.~Song, and S.~H.~Lee,
$J / \psi$ absorption by $\pi$ and $\rho$ mesons in meson exchange model with anomalous parity interactions,
Phys. Rev. C \textbf{63}, 034901 (2001).
%doi:10.1103/PhysRevC.63.034901
%[arXiv:nucl-th/0010064 [nucl-th]].
%217 citations counted in INSPIRE as of 21 Mar 2024

%\cite{Wang:2006ida}
\bibitem{Wang:2006ida}
Z.~G.~Wang and S.~L.~Wan,
Analysis of the vertices $D^* D_{s}K$, $D_{s}^* D K$, $D_{0} D_s K$ and $D_{s0} DK$ with the light-cone QCD sum rules,
Phys. Rev. D \textbf{74}, 014017 (2006).
%doi:10.1103/PhysRevD.74.014017
%[arXiv:hep-ph/0606002 [hep-ph]].
%69 citations counted in INSPIRE as of 28 May 2024

%\cite{Dong:2008gb}
\bibitem{Dong:2008gb}
Y.~B.~Dong, A.~Faessler, T.~Gutsche, and V.~E.~Lyubovitskij,
Estimate for the $X(3872)\to \gamma J/\psi$ decay width,
Phys. Rev. D \textbf{77}, 094013 (2008).
%doi:10.1103/PhysRevD.77.094013
%[arXiv:0802.3610 [hep-ph]].
%147 citations counted in INSPIRE as of 05 Dec 2023

%\cite{Xiao:2016mho}
\bibitem{Xiao:2016mho}
C.~J.~Xiao and D.~Y.~Chen,
Possible $B^{(\ast)} \bar{K}$ hadronic molecule state,
Eur. Phys. J. A \textbf{53}, 127 (2017).
%doi:10.1140/epja/i2017-12310-x
%[arXiv:1603.00228 [hep-ph]].
%42 citations counted in INSPIRE as of 21 Mar 2024

%\cite{Becirevic:2009yb}
\bibitem{Becirevic:2009yb}
D.~Becirevic, B.~Blossier, E.~Chang, and B.~Haas,
$g_{B^*B\pi}$-coupling in the static heavy quark limit,
Phys. Lett. B \textbf{679}, 231 (2009).
%doi:10.1016/j.physletb.2009.07.031
%[arXiv:0905.3355 [hep-ph]].
%69 citations counted in INSPIRE as of 21 Mar 2024

%\cite{Bardeen:2003kt}
\bibitem{Bardeen:2003kt}
W.~A.~Bardeen, E.~J.~Eichten, and C.~T.~Hill,
Chiral multiplets of heavy - light mesons,
Phys. Rev. D \textbf{68}, 054024 (2003).
%doi:10.1103/PhysRevD.68.054024
%[arXiv:hep-ph/0305049 [hep-ph]].
%521 citations counted in INSPIRE as of 20 Jan 2024

%\cite{Wang:2006mf}
\bibitem{Wang:2006mf}
Z.~G.~Wang,
Radiative decays of the $D_{s0}(2317)$, $D_{s1}(2460)$ and the related strong coupling constants,
Phys. Rev. D \textbf{75}, 034013 (2007).
%doi:10.1103/PhysRevD.75.034013
%[arXiv:hep-ph/0612225 [hep-ph]].
%21 citations counted in INSPIRE as of 08 Apr 2024

%\cite{Wang:2008wz}
\bibitem{Wang:2008wz}
Z.~G.~Wang,
Radiative decays of the strange-bottomed ($0^+$,$1^+$) mesons,
Commun. Theor. Phys. \textbf{52}, 91 (2009).
%doi:10.1088/0253-6102/52/1/21
%[arXiv:0803.1223 [hep-ph]].
%6 citations counted in INSPIRE as of 03 Apr 2024`   `   

\end{thebibliography}
\end{document}